%% file: abcelArxiv.tex
\documentclass{amsart}
\usepackage{amsmath}
\usepackage[dvips]{graphicx}
\usepackage{enumerate}
\usepackage{natbib}
\usepackage{url} 

\usepackage{helvet}
\usepackage{amsmath}
\usepackage{amsfonts}
\usepackage{mathtools}

\usepackage{helvet}
\usepackage{natbib}
\usepackage{subcaption}
\usepackage{color}
\usepackage{enumitem}
\usepackage{xr}
\usepackage{fullpage}
\usepackage{lscape}

\DeclareMathOperator*{\argmax}{arg\,max}
\DeclareMathOperator*{\argmin}{arg\,min}

\newcommand{\s}[1]{s(#1)}
\newcommand{\sa}{s}
\newcommand{\sm}{\mathfrak{s}}
\newcommand{\gt}{s_1}
\newcommand{\gn}{s_o}
\newcommand{\gi}{s_i}
\newcommand{\et}{E^0}
\newcommand{\htr}{H^0}

\newtheorem{lemma}{Lemma}
\newtheorem{theorem}{Theorem}

\newtheorem{remark}{Remark}

\title[ABC Empirical Likelihood]{On an Empirical Likelihood based Solution to the Approximate Bayesian Computation Problem}

\author[Chaudhuri]{Sanjay Chaudhuri} 
\address{Department of Statistics, University of Nebraska-Lincoln, Nebraska 68583}
\email{schaudhuri2@unl.edu}
\author[Ghosh]{Subhroshekhar Ghosh}
\address{Department of Mathematics, National University of Singapore, Singapore 119076.}
\email{subhrowork@gmail.com}
\author[Pham]{Kim Cuc Pham}
\address{Department of Statistics and Applied Probability, National University of Singapore, \quad Singapore 117546.}
\email{staptkc@u.nus.edu}

\begin{document}


\begin{abstract}
Approximate Bayesian Computation (ABC) methods are applicable to statistical models specified by generative processes with analytically intractable likelihoods.  These methods try to approximate the posterior density of a model  parameter by comparing the observed data with additional process-generated simulated datasets. 
  For computational benefit, only the values of certain well-chosen summary statistics are usually compared, instead of the whole dataset.
    Most ABC procedures are computationally expensive, justified only heuristically, and have poor asymptotic properties.
    In this article, we introduce a new empirical likelihood-based approach to the ABC paradigm called ABCel.  
The proposed procedure is computationally tractable and approximates the target log posterior of the parameter as a sum of two functions of the data --- namely, the mean of the optimal log-empirical likelihood weights and the estimated differential entropy of the summary functions.
We rigorously justify the procedure via direct and reverse information projections onto appropriate classes of probability densities.  Past applications of empirical likelihood in ABC demanded constraints based on analytically tractable estimating functions that involve both the data and the parameter; although by the nature of the ABC problem such functions may not be available in general. In contrast, we use constraints that are functions of the summary statistics only. Equally importantly, we show that our construction directly connects to the reverse information projection and estimate the relevant differential entropy by a k-NN estimator.
  We show that ABCel is posterior consistent and has highly favourable asymptotic properties.  
Its construction justifies the use of simple summary statistics like moments, quantiles, etc, which in practice produce accurate approximation of the posterior density.  We illustrate the performance of the proposed procedure in a range of  applications.


\end{abstract}

\maketitle

\section{Introduction}
The concept of likelihood is central to parametric statistical inference. However, for many models encountered in natural, 
engineering, and environmental sciences, tractable analytic forms of their likelihoods are not available.  These models are often specified by a generative process,
in the sense that independent samples can be generated from them for any input value of the model parameters.
Approximate Bayesian computation (ABC) methods \citep{rubin1984,tavare+bgd97,beaumont+zb02,wood10,marin+prr11,fearnhead+p12,blum+nps13,mengersen+pr13} are useful for Bayesian inference for models
like these.
  Given the observed data, their objective is to estimate the posterior density of the parameters associated with the data generating process without specifying a functional relationship between those parameters and the data.

  In this article, we introduce a new modified empirical likelihood-based approach to the ABC problem that we call \textit{ABCel}.
  Along the line of the traditional ABC procedures, it assumes the availability of the observed data and the ability to generate independent and identically distributed data sets of the same size from the generating process for any given value of the parameter of interest.  
  In particular compared to the empirical likelihood-based $BC_{el}-AMIS$ algorithm \citep{mengersen+pr13} our method does not require specifications of estimating equations depending both on the observed data and the parameters, which are typically unavailable.
  The estimating equations are specified by the differences in the values of the appropriate summary statistics of the observed and the replicated data sets.  These equations form natural constraints for our proposed modified empirical likelihood without directly involving the parameters.
ABCel can be rigorously justified using various information projections and basic principles of Bayesian statistics. 
Furthermore, ABCel exhibits many favourable asymptotic properties and is computationally tractable.

Because of their potential application to complex models, ABC methods have generated immense interest in statistics. The procedures proposed in the literature can be classified into two broad groups.  The \emph{Simple rejection ABC} procedures try to sample from the parameter posterior directly.
A proposed value of the parameter is taken as a valid draw from the posterior if the corresponding replication generated from the process is closer to the observation than a pre-specified tolerance.


Even though attractive at first glance, and in spite of the availability of sophisticated and efficient sampling
algorithms \citep{marjoram+mpt03,sisson+ft07,beaumont+rmc09} with improved efficiency, {\it the curse of dimensionality} makes the simple rejection algorithm computationally expensive in high-dimensions.
More crucially, the accuracy of the posterior approximation depends heavily on the value of the pre-specified tolerance. Intuitively, small tolerances are preferred, but they are computationally prohibitive.  
Available results (e.g. \citet{frazier+mrr18}, \citet{li+f18a}, \citet{li+f18b}, \citet{millerDunson2019}, \citet{berntonEtAl2019}) show that, unless the pre-specified tolerance satisfies certain conditions which depend both on the summary statistics as well as the specified distance function (\citet{millerDunson2019, berntonEtAl2019}),
the resulting rejection ABC posteriors may not have desirable asymptotic properties (e.g. Bayesian consistency, correct asymptotic frequentist coverage of credible intervals).
We refer to \citet{robertSurvey2016} for a more detailed and succinct discussion of the possible pitfalls of the rejection ABC procedures.

Alternatives to the rejection-based ABC are provided by the so-called {\it pseudo-likelihood methods}.  For each value of the parameter, these methods attempt to estimate the likelihood of the observed summaries, from observations simulated from the data generating process.
One of the most popular pseudo-likelihood method is the {\it synthetic likelihood} introduced by \citet{wood10}.  Here, in order to compute the likelihood, the summary statistics are assumed to be approximately jointly distributed as a multivariate normal random vector.  Their mean and the covariance matrix vary with the parameter and are estimated using the summaries simulated from the data generating process (see \cite{price+dln16}).
Synthetic likelihood does not perform well when the normal approximations of the summary statistics are inaccurate. This happens, e.g. when extreme values of the observations are used as summaries, or often when the process generates data vectors with dependent components, e.g. from a time series, etc.
Extensions that relax the requirement of normality have been a continuous topic of interest for many researchers in this area.
\citet{Fasiolo2016} consider an extended saddle-point approximation, whereas \citet{Dutta2016} proposes
a method based on logistic regression.
By making use of various transformations \citet{anNottDrovandi2020} and \citet{priddleDrovandi2020} consider semi-parametric extensions of synthetic likelihood.  \citet{drovandi+pl15} describe an encompassing framework for many of the above suggestions, which they call parametric Bayesian indirect inference.  
\citet{frazier2020robust} have recently proposed a robustified version of synthetic likelihood that is able to detect and provide some degree of robustness to misspecification.  

The $BC_{el}-AMIS$ procedure introduced by \citet{mengersen+pr13} is pseudo-likelihood based, where the intractable data likelihood is replaced by a non-parametric empirical likelihood \citep{owen01}.  This procedure follows the traditional Bayesian empirical likelihood (BayesEL) procedures \citep{lazar03,chaudhuri+g11} and specifies the likelihood from the jumps of the joint empirical distribution function of the data computed under appropriate constraints.
Empirical likelihood does not require the summaries to be approximately normal. However, the $BC_{el}-AMIS$ procedure typically requires constraints based on analytically tractable estimating functions of both the data and the parameters.
By the nature of the ABC problem, such functions are not readily available, and thus, the proposed $BC_{el}-AMIS$ algorithm is not always easy to implement in practise.  The exponentially tilted empirical likelihood \citep{schennach2005} based ABC proposed by \citet{grazianLiseo2017} suffers from similar problems. 



The proposed paradigm of ABCel, which is essentially a modified empirical likelihood-based method, neither uses any tolerance parameter nor assume any specific form of a pseudo-likelihood. It first finds an analytic expression of an approximation of the target posterior.
This expression is then used to approximate the data density and obtain an optimal approximate of the target.  

\subsection{Notations}
In order to provide an overview of our arguments detailed in the rest of this article, we lay out some notations. Suppose $\sa$ is a pre-specified possibly vector valued summary statistic and $\gn$ is its observed value obtained with an unknown input $\theta_o$ of the parameter $\theta$.  Like most ABC procedures, we would like to approximate the posterior density $\Pi(\theta\mid \gn)$.
Let $f_0(\sa(\theta)\mid \theta)$ is the true density the statistics $\sa$ inherits from the data generating mechanism.  This density is unknown, which prevents direct computation of the required posterior.

For an input value $\theta$, let $\gt$ be one replicated summary generated from the data generating process.  For reasons described below in detail, it is more convenient to focus on the set of all joint densities $\mathcal{F}$ defined on $(\theta,\gt,\gn)$.  For an assumed prior density $\pi(\theta)$ over the parameters, the set $\mathcal{F}$ contains the true joint density
\[
f_0(\theta,\gt,\gn)=f_0(\gt\mid\theta)f_0(\gn\mid\theta)\pi(\theta).
\]
We briefly explain the ABCel in following steps.

\subsection{A closed form expression of the approximate posterior.} Suppose $f$ is any density in $\mathcal{F}$.  We first find an approximation of $\Pi(\theta\mid\gn)$ corresponding to $f$.  This is achieved by projecting the conditional density $f(\theta,\gt\mid\gn)$ on an appropriate subset of joint densities defined on $(\theta,\gt)$, and subsequently marginalising out $\gt$ (see Section \ref{sec:just}).  We show that (See Theorem \ref{thm:postAprx}) this approximate posterior can be written in a closed form as: 
\[
f^{\prime}(\theta\mid\gn)=\frac{e^{\et_{\gt\mid\theta}[\log f(\theta,\gt,\gn)]+\htr_{\gt\mid\theta}(\theta)}}{\int_{t\in\Theta} e^{\et_{\gt\mid t}[\log f(t,\gt,\gn)]+\htr_{\gt\mid t}(t)}dt},
\]
where $\htr_{\gt\mid\theta}(\theta)$ is the differential entropy of $f_0(\gt\mid\theta)$ at $\theta$ and the expectation is computed with respect to $f_0(\gt\mid\theta)$.  Note that, in the expression of the approximate posterior the effect of the replicate summaries are integrated out.  
Furthermore, if one happens to choose $f(\theta,\gt,\gn)=f_o(\theta,\gt,\gn)$, then it turns out that $f^{\prime}(\theta\mid\gn)=\Pi(\theta\mid\gn)$. In other words, the approximation is exact.  Thus our strategy here is to approximate the true joint density $f_0(\theta,\gt,\gn)$ and compute the corresponding approximate posterior using the expression described above. 

\subsection{Approximation of the target densities.} The approximation of the target joint $f_0(\theta,\gt,\gn)$ is achieved by first noting that for any joint density $f(\theta,\gt,\gn)\in\mathcal{F}$ and for any $\theta$, and $\gn$  (see Theorem \ref{thm:error}):
\begin{equation}\label{eq:2ashrt}
\log f\left(\theta,\gn\right)-\left\{\et_{\gt\mid \theta}\left[\log f\left(\theta,\gt,\gn\right)\right]+\htr_{\gt\mid \theta}(\theta)\right\}\ge 0.
\end{equation}
Here again the equality holds if $f(\theta,\gt,\gn)=f_0(\theta,\gt,\gn)$, however, L.H.S. of \eqref{eq:2ashrt} may have multiple minima over $\mathcal{F}$.
Thus, in the second step we specify a subset $\mathcal{F}^{\prime}\subseteq\mathcal{F}$,
such that the L.H.S. of \eqref{eq:2ashrt} has a unique minimum at $f_0$ over $\mathcal{F}^{\prime}$  (see Theorem \ref{thm:fp}).  Finding the minimum of the L.H.S. of \eqref{eq:2ashrt} involves minimising the Kullback-Leibler divergence between the conditional densities $f_0(\gt\mid\theta)$
and $f(\gt\mid\gn,\theta)$ over $f(\theta,\gt,\gn)\in\mathcal{F}^{\prime}$.  The optimum $f(\gt\mid\gn,\theta)$ approximates $f_0(\gt\mid\theta)$, the true density of the data generating process (see Theorem \ref{thm:fp}).

\subsection{Connection to the proposed modified empirical likelihood.} Finally, in Section \ref{sec:conEl} we argue that the proposed modified empirical likelihood-based procedure follows the above recipe in the sample.  In particular, for various natural summary statistics, the proposed procedure automatically minimises the divergence in \eqref{eq:2ashrt} approximately over the set $\mathcal{F}^{\prime}$.  
This is achieved first of all, by the constraints under which the empirical likelihood is computed and secondly, by the flexibility of the estimated likelihood .  
The proposed summary based constraints approximately translate to an equality constraint on the conditional marginal density of the observed and replicated data and the summaries.  The densities in $\mathcal{F}^{\prime}$ need to satisfy this condition (see Section \ref{sec:optAprx}).  
Additionally, by only requiring positive weights on the observations, the procedure can minimise the above divergence over a large set of densities without explicitly specifying their parametric forms.  
That is, together with the constraints, the proposed empirical likelihood is optimal approximately over the whole of $\mathcal{F}^{\prime}$, which strongly justifies the proposed procedure.

Once the optimal weights are computed, the corresponding $f^{\prime}_0(\theta\mid \gn)$ is computed using the mean of log weights and by estimating the differential entropy of the replicate summaries using an estimator 
due to \citet{berrettSamworthMing2019} which is a weighted version of the k-nearest neighbour \citet{kozLeo87} (see Section \ref{sec:diffent}).




The proposed procedure estimates the posterior by approximating joint densities. 
Thus, simple summaries like quantiles, moments, etc. which approximately specify the underlying density perform justifiably well (see Section \ref{sec:expl}).  Such summaries abound in the literature of the goodness of fit tests \citep{agostinoBook}, moment problem \citep{gut2002} etc.
Furthermore, this density approximation approach allows a variety of summaries like those based on e.g. the Fourier transform of the data.  These are useful when the observations are dependent (see Section \ref{sec:bb}), which is often the case.


The proposed posterior has many favourable properties.  Asymptotically, under mild conditions, the proposed posterior is Bayesian and posterior consistent 
for the true value of the parameter when both the sample size and the number of replications grow unbounded (Section \ref{sec:asymp}).  We invoke the results from \citet{ghosh2019empirical}, to further explore its properties when the number of replications increases, but the sample size is held fixed (Section \ref{sec:m}).  

Finally, perhaps the biggest advantage of our procedure is its easy implementation.  In order to compute the likelihood, a user only needs to specify an appropriate set of summary statistics and the number of replications to be simulated for each value of the parameter. For the estimation of the differential entropy, the order of the percentile (see Section \ref{sec:diffent}) has to be chosen.  Unlike many ABC procedures, no other parameters tuning or otherwise are either need to be specified or estimated. 
Moreover, both the empirical likelihood and the proposed differential entropy estimator can be computed using a fast algorithm implemented in multiple software.  An easy adaptive Markov chain Monte Carlo procedure due to \citet{haarioEtAl2001} can be adapted to efficiently sample from the resulting posterior (see Section \ref{Ssec:impl2} of the supplement).

\section{ABC Empirical Likelihood Posterior}\label{sec:multsamp}

In this section, we introduce the basic ideas of the proposed ABC Empirical Likelihood (ABCel) posterior.  A  modified empirical likelihood-based method which only depends on the observed data and replicated data simulated from the generating process is described.  One part of the ABCel posterior is constructed using this likelihood.  The other part is an estimate of a differential entropy, which is computed using a non-parametric Euclidean likelihood.
We only describe the motivation and computation of ABCel posterior in this section.  The justification of the procedure is presented in the subsequent sections.  

\subsection{Setup}
Let $\theta$ be the input parameter of the data generating process.  We assume $\theta$ takes values in a set $\Theta$ and assign a prior $\pi(\theta)$ to it.  For any given value $\theta\in\Theta$, the process generates i.i.d. n-dimensional random vectors from an unknown density depending on parameter $\theta$.  Since the density of the same random variable would change with the value of the parameter, we make their connection explicit in the notation.  As for example, the observed data $x_o$ is a realisation of the random variable $X_o(\theta_o)$, i.e. the random variable $X_o(\theta)$ generated from the process with $\theta=\theta_o$.  
Additionally, for each $\theta\in\Theta$, realisations from $m$ i.i.d. replicated random variables $X_i(\theta)$, $i=1$, $2$, $\ldots$, $m$ are drawn from the process with input parameter value $\theta$.  That is in total, we consider a set of $n-$dimensional random vectors 
$\left\{X_i(\theta), i\in\mathbb{M}_o,\theta\in\Theta\right\}$, where $\mathbb{M}_o=\{o\}\cup\mathbb{N}$, i.e. the set of positive integers appended with the symbol $o$.  By construction, conditional on $\theta$, $\left\{X_i(\theta), i\in\mathbb{M}_{o}\right\}$ are independent and identically distributed. 

The true density $f_0(X_o(\theta)\mid\theta)$ is unknown, which prevents computation of the \emph{exact} posterior $\Pi(\theta\mid X_o)\coloneqq\Pi(\theta\mid X_o(\theta)=x_o)\propto f_0(x_o\mid \theta)\pi(\theta)$.  The problem is to approximate the posterior using the observation $x_o$ and the replications $X_i(\theta)$, $i=1$, $2$, $\ldots$, $m$. 

Direct approximation of $\Pi(\theta\mid X_o)$ may be computationally cumbersome \citep{drovandi_frazier_2022} and in most ABC applications inference on $\theta$ is made using a posterior conditional on $r\times 1$ vector $\s{\cdot}$ of summary statistics of the observations.

Suppose that for a given $\theta\in\Theta$, $\s{X(\theta)}$ inherits a density $f_0(\s{X(\theta)}\mid\theta)$ from $X(\theta)$. Using the summaries $\s{X_i(\theta)}$, $i\in\mathbb{M}_o$, most ABC procedures estimate the \emph{target} posterior
\begin{equation}\label{eq:truepost}
\Pi(\theta\mid\s{X_o})\coloneqq\Pi(\theta\mid \s{X_o(\theta)}=\s{x_o})=\frac{f_0(\s{x_o}\mid\theta)\pi(\theta)}{\int f_0(\s{x_o}\mid t)\pi(t)dt}.
\end{equation}

\subsection{Construction of ABCel Posterior}
The ABCel posterior is based on the following observation.  Suppose $\theta=\theta_o$. Then by construction, the random variables $\s{X_o(\theta_o)}$, $\s{X_1(\theta)}$, $\ldots$, $\s{X_m(\theta)}$ are identically distributed.  Now if $\et_{\sa\mid\theta_o}$ denotes the expectation w.r.t $f_0(\s{X_{i}(\theta_o)}\mid\theta_o)$, then for any $i=1,\dots, m$,
\begin{equation}
  \et_{\sa\mid\theta_o}\left[\s{X_{i}(\theta_o)}-\s{X_{o}(\theta_o)}\right]=0. \label{eq:ex}
\end{equation}

The proposed empirical likelihood estimator of the posterior consists of two parts.  The first is an empirical likelihood which is constructed using constraints based on the expectation in \eqref{eq:ex}.  For any $\theta\in\Theta$ and for each $i=1$, $2$, $\ldots$, $m$, define 
\begin{equation}\label{eq:h}
  h_i(\theta)=\s{X_i(\theta)}-\s{X_o(\theta_o)},
  \end{equation}
and the random set:
\begin{equation}\label{eq:w1} 
\mathcal{W}_{\theta}
=\left\{w~:~\sum^m_{i=1}w_i\left[\s{X_{i}(\theta)}-\s{X_{o}(\theta_o)}\right]=0\right\}\cap\Delta_{m-1},
\end{equation}  
where $\Delta_{m-1}$ is the $m-1$ dimensional simplex.  



We define the optimal weights $\hat{w}$ as: 
\begin{equation}
  \hat{w}\coloneqq\hat{w}(\theta)\coloneqq\argmax_{w\in\mathcal{W}_{\theta}}\left(\prod^m_{i=1}mw_i\right). \label{eq:w2}
\end{equation}
If the problem in \eqref{eq:w2} is infeasible, $\hat{w}$ is defined to be zero. These optimal weights are used in the first part of the posterior estimate.

The second part requires an estimate of the differential entropy $\htr_{\sa\mid\theta}(\theta)$ of $f_0(\cdot\mid\theta)$ at the input $\theta\in\Theta$, which is defined by, $\htr_{\sa\mid\theta}(\theta)=-\int f_0(\sa\mid\theta)\log f_0(\sa\mid\theta)d\sa$.  Let, $\hat{H}^{0}_{\sa\mid\theta}(\theta)$ is an estimate of $\htr_{\sa\mid\theta}(\theta)$ (see Section \ref{sec:diffent} below for details).

By using this estimate and the optimal $\hat{w}$ we define \emph{ABC empirical likelihood} (\emph{ABCel}) estimate of the required posterior as,
\begin{align}
  \hat{\Pi}(\theta\mid \s{X_o})&=\frac{\left[e^{\left(\frac{1}{m}\sum^m_{i=1}\log\left(\hat{w}_i(\theta)\right)+\hat{H}^0_{\sa\mid\theta}(\theta)\right)}\right]\pi(\theta)}{\int_{t\in\Theta}\left[e^{\left(\frac{1}{m}\sum^m_{i=1}\log\left(\hat{w}_i(t)\right)+\hat{H}^0_{\sa\mid t}(t)\right)}\right]\pi(t)dt}
\label{eq:mpost}
\end{align}
When $\prod^m_{i=1}\hat{w}_i=0$, we define $\hat{\Pi}(\theta\mid \s{X_o})=0$.


The empirical likelihood used in \eqref{eq:mpost} is different from the original Bayesian empirical likelihood (BayesEL) posterior \citep{lazar03, chaudhuri+g11} and the previous use of Bayesian empirical likelihood in an ABC setting \citep{mengersen+pr13} in two ways.  
First, instead of the sum, it uses the mean of the log-weights.  This is significant in several ways (see below) and can be justified by an information projection argument described in Section \ref{sec:fnform}.  

The second aspect is our choice of the constraints, which is probably more significant.  Usual BayesEL formulations (as in \citet{mengersen+pr13}) would have used constraints which are functions of $\sa(X_o)$ and $\theta$.  Such estimating equations are not necessarily known in an ABC problem.
In our formulation, we avoid such specifications using constraints based on $\sa(X_o)$ and the replicated summaries $\sa(X_i)$, $i=1$, $2$, $\ldots$, $m$.  The summaries in \eqref{eq:h} are routinely used in \emph{Exponential Random Graph Models (ERGM)} literature 
\citep{horvat_czabarke_toroczkai_2015}, however the weights are obtained by maximising the entropy \citep{jaynes_1957a,jaynes_1957b} instead of a likelihood as in \eqref{eq:w2} above.  This is equivalent to maximising a cross-entropy term (see \eqref{eq:crent}).  Unlike the rejection ABC, we do not need to specify any distance function or any tolerance parameter.        

From the formulation of the constraints, the optimal weights in \eqref{eq:w2} define a constrained joint-conditional empirical distribution function supported on  $m$ observations $\left(\s{X_i(\theta)},\s{x_o}\right)$ given $\theta$. 
This is somewhat similar to the data-replication methods, discussed in \citet{lele+dl07} and \citet{doucet+gr02} (see also \citet{gourieroux1996}).   More importantly, as we argue in Section \ref{sec:conEl} below, for simple choices of summary our constraints ensure that 
the above joint-conditional $f\left(\s{X_i(\theta)},\s{X_o(\theta)}\mid \theta\right)$ is estimated by approximately equating the underlying marginal conditional densities $f\left(X_i(\theta)\mid \theta\right)$ and $f\left(X_o(\theta)\mid \theta\right)$ of $X_i(\theta)$ and $X_o(\theta)$ respectively, which provides an argument in favour of the optimality of our procedure.




No analytic expression for the proposed ABCel posterior exists in general.  By construction, each $\hat{w}_i$ is bounded for all values $\theta$.  All components of $\hat{w}$ in \eqref{eq:w2} and the ABCel posterior are strictly positive iff the origin of $\mathbb{R}^r$ is in the interior of the convex hull defined by the vectors $h_1$, $h_2$, $\ldots$, $h_m$.  
Otherwise the ABCel posterior would be zero (even though in the boundary of the above convex hull, the constrained optimisation in \eqref{eq:w2} is still feasible). 
It is well-known (see e.g. \citet{chaudhuri+my17}) that the supports of the Bayesian empirical likelihood (BayesEL) posteriors are in general non-convex.  It is expected that the proposed ABCel posterior will suffer from the same deficiency as well. 
However, as we discuss below (see Section \ref{sec:impl}) the non-convexity of the support does not make the proposed ABCel posterior computationally expensive.  One can device easy Markov chain Monte Carlo (MCMC) techniques to draw samples from this posterior at a reasonable computational cost.  Such samples are enough for making posterior inference.

Finally, the proposed method is more general than the synthetic likelihood \citep{wood10}.  
The latter assumes normality of the joint distribution of the summary statistics.  
Even though many summary statistics are asymptotically normally distributed, this is not always the case.  This is especially true if the process generates dependent data sets eg. a time series, spatial data etc.  In such cases, the synthetic likelihood can perform quite poorly (see e.g. Section \ref{sec:arch} below).  
Some relaxation of normality has been proposed by various authors, but many of these procedures require specification or estimation of additional tuning parameters. In our empirical likelihood approximation, we only require the observed data and simulated data from the generating process for a given $\theta$.

\subsection{Differential Entropy Estimation}\label{sec:diffent}
Several estimators of differential entropy have been studied in the literature.  The oracle estimator is given by $-\sum^m_{i=1}\log f_0(\s{X_i(\theta)})/m$.  In this article we implement a weighted k-nearest neighbour based estimator due to \citet{kozLeo87} described in \citet{berrettSamworthMing2019}. 
This estimator is easy to compute and has better asymptotic properties than histogram or kernel-based estimators \citep{hallMorton1993}, specially for high dimensional summaries.  

The nearest-neighbour estimator requires us to specify $k$, the order of the nearest neighbour.  Ideally, $k$ should depend on $m$.  Our experiments suggests any value of $k$ as long as it is not very small or not very large, makes little difference.
Note that, other than the summary statistics and the number of replications $m$, this $k$ is the only parameter an user needs to specify in order to compute the proposed posterior.  No other parameters tuning or otherwise are required.

\subsection{Example} 
\begin{figure}[t]
\begin{center}
  \begin{subfigure}{.5\columnwidth}
{\resizebox{3in}{2.5in}{\includegraphics{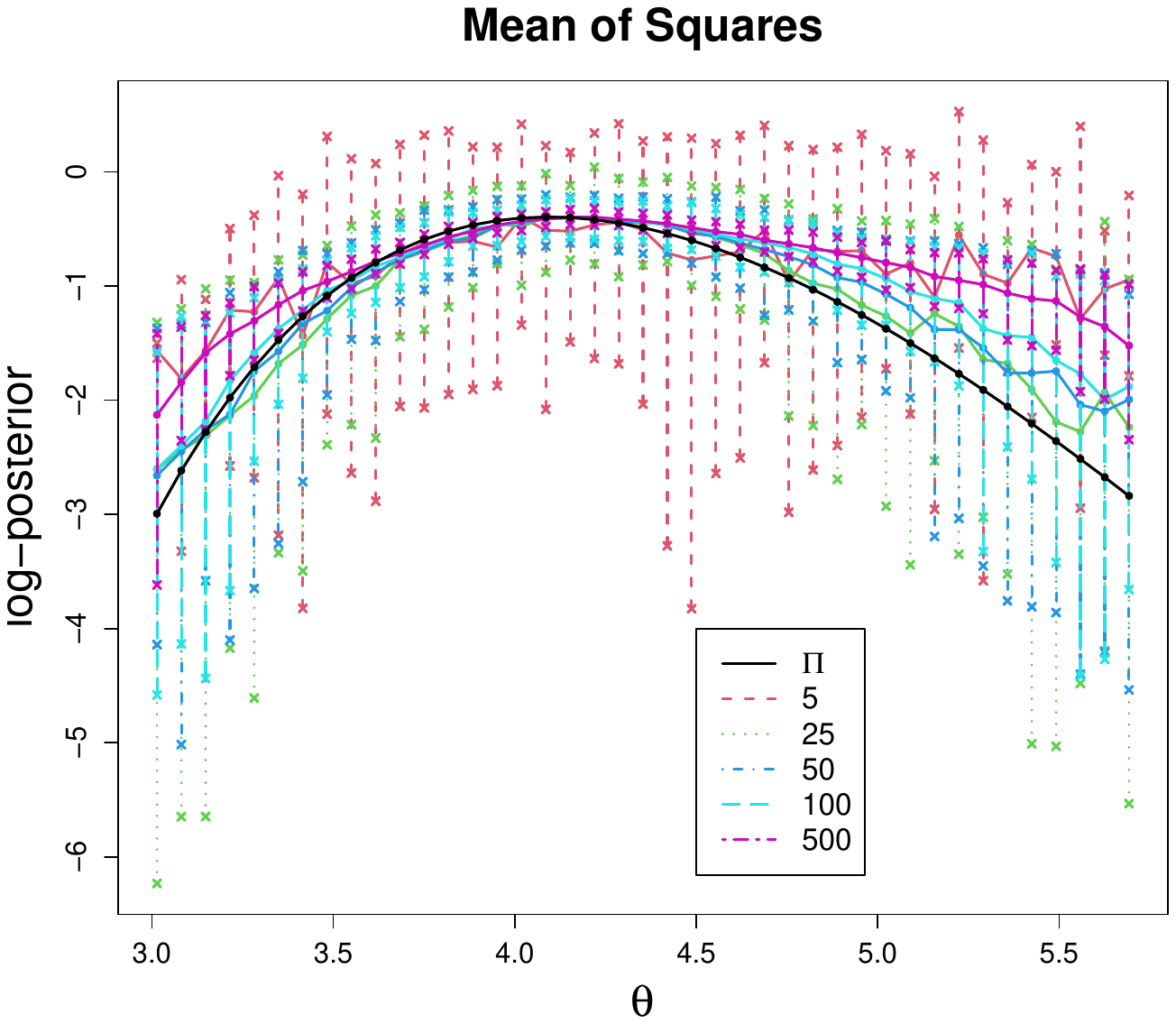}}}
\caption{$\hat{\Pi}(\theta\mid s^{(1)}(X_o))$}
\label{fig:hatSigma}
\end{subfigure}\begin{subfigure}{.5\columnwidth}
{\resizebox{3in}{2.5in}{\includegraphics{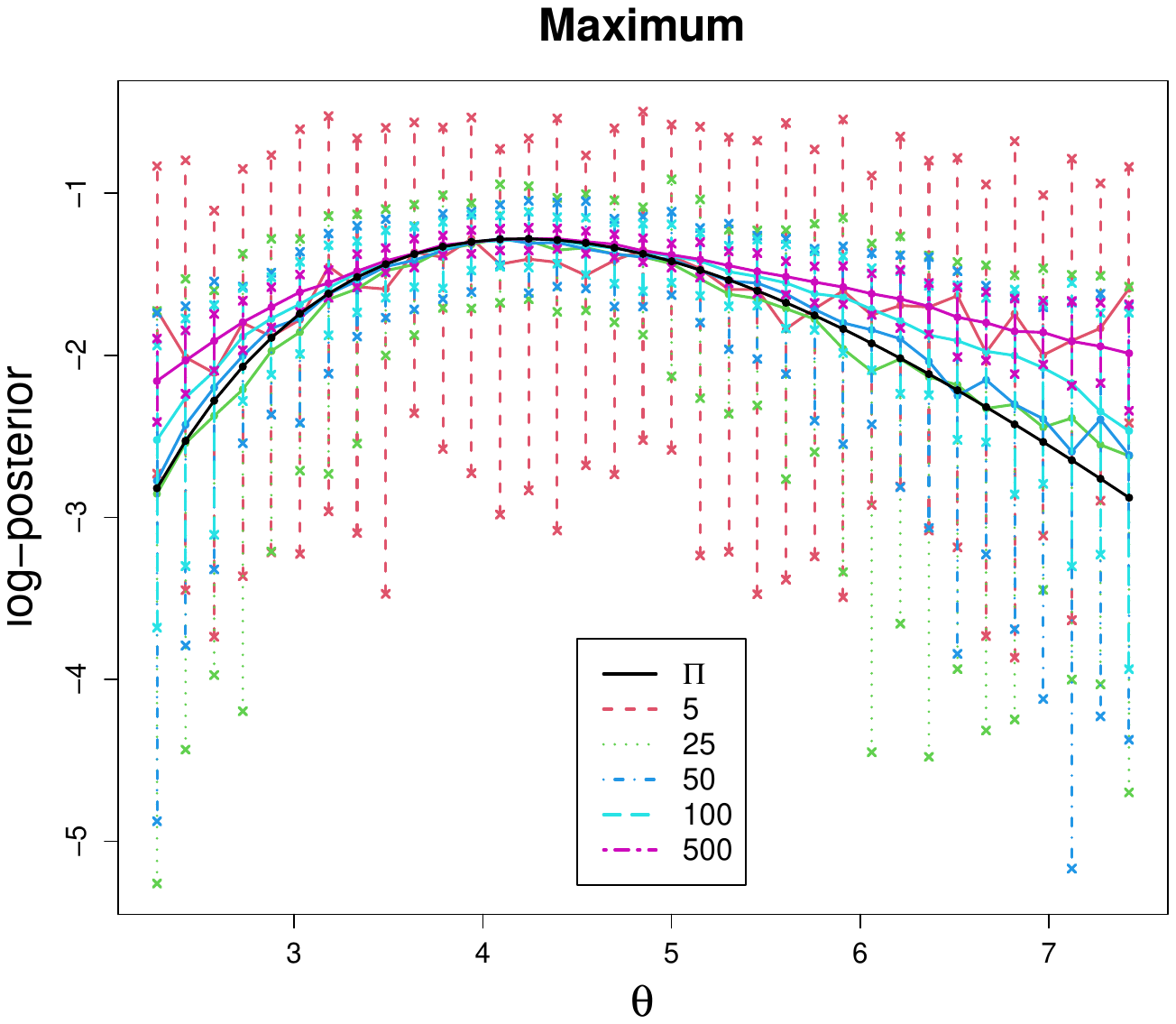}}}
\caption{$\hat{\Pi}(\theta\mid s^{(2)}(X_o))$}
\label{fig:hatMax}
  \end{subfigure}
  \end{center}
\caption{Comparison of the true log-posterior with the logarithm of the proposed estimator for different values of $m$.  The samples of size $n=100$ were drawn from $N(0,\theta)$ distribution with $\theta_o=4$.  We chose (a) $s^{(1)}(X_i)=\sum_j X^2_{ij}/n$ and (b) $s^{(2)}(X_i)=\max_j(X_{ij})$ and a $U(0,10)$ prior on $\theta$.  
The true log-posterior is in black.  For each value of $\theta$ and $m$ the means and the $95\%$ confidence intervals of the estimated log-posterior based on $100$ repetitions are shown.}
\label{fig:post}
\end{figure}

In Figure \ref{fig:post} we compare the shape of the ABCel log-posteriors with the true log-posteriors $\Pi$ for the variance of a Normal distribution with zero mean conditional on (a) $s^{(1)}(X_i)=\sum_jX^2_{ij}/n$ (Figure \ref{fig:hatSigma}) and (b) $s^{(2)}(X_i)=\max_j(X_{ij})$ (Figure \ref{fig:hatMax}).  
Here, for each $i=1$, $2$, $\ldots$, $m$, and $j=1$, $2$, $\ldots$, $100$, the observation $X_{ij}$ is drawn from a $N\left(0,\theta\right)$, with $\theta_o=4$.  We assume that the parameter $\theta$ follows a $U(0,10)$ prior.
 
The log-posteriors were compared on a grid of parameters whose true posterior values were larger than the $.05$.  
Based on $100$ repetitions, At each value of $\theta$ and $m$, the mean and the endpoints of the symmetric $95\%$ confidence intervals are shown in the figure.
To make the comparison of the shapes easier, for each $m$, the maximum of the mean of ABCel log-posterior was matched with the maximum value of the true log-posterior.  

From Figure \ref{fig:post} it follows that for $m=25$ and $m=50$, for each value of $\theta$ the means of the estimated log-posteriors (solid coloured lines) are very close to the true log-posterior (solid black line) for both $s^{(1)}(X_o)$ and $s^{(2)}(X_o)$.  
Furthermore, the $95\%$ confidence bands always cover the corresponding true value of the log-posterior.  
It is evident that the proposed ABCel posterior is a good approximation of the true posterior up to a scaling constant.   This is even true for the summary function $s^{(2)}(X_o)$, which unlike $s^{(1)}(X_o)$, asymptotically does not converge to a normal random variable under any centring or scaling.  



As the number of replicates i.e. $m$ increases (see $m=500$), in Figure \ref{fig:post} the log-posterior, tends to get more flat in shape.  However, the confidence bands get narrower.  This is somewhat expected. We have kept the number of summaries fixed here.  However, statistical intuition mandates that the number of summaries used should increase with the number of replications.  Using results from \citep{ghosh2019empirical}, we discuss such phenomena in more detail in Section \ref{sec:m} below.  Furthermore, an example illustrating the inter-relationship between the number and the nature of the summary statistics with the number of replications can be found in Section \ref{normal}.




\section{Justification For the ABC Empirical Likelihood Posterior}\label{sec:just}
In this section we provide a rigorous justification for the proposed modified empirical likelihood-based posterior estimate $\hat{\Pi}(\theta\mid \s{X_o})$ introduced in Section \ref{sec:multsamp}.  Our arguments use direct and reverse information projections of appropriate conditional densities on judiciously chosen density sets.
We first discuss a general functional form of a posterior approximation, then use this functional form to find an accurate approximation of the true posterior.  Both of these approximations are obtained in the population. Finally, it is argued that our posterior estimate $\hat{\Pi}(\theta\mid\s{X_o})$ approximates the above recipe in the sample.    

Recall that the observed summary random variable $\gn\!\coloneqq\s{X_o(\theta)}$ and for some input $\theta\in\Theta$, $\gt\!\coloneqq\s{X_1(\theta)}$ denote a replicated summary random variable corresponding to $X_1(\theta)$ obtained from the data generating process.
In order to justify the proposed empirical likelihood-based estimator, it is more convenient to work with the joint densities defined on $(\theta,\gt,\gn)$.  Let $\mathcal{F}$ be the set of all joint densities defined on $(\theta,\gt,\gn)$.  Suppose $f_0(\gt\mid\theta)$ is the unknown density $\gt$ inherits from the data-generating process.
Since $\gn$ is assumed to be generated from the same process, for the \emph{true} data-generating process, for all $\theta\in\Theta$, the conditional density of $\gn$ given $\theta$ is same as the conditional density of $\gt$ given $\theta$.  
The corresponding \emph{true} joint density of $(\theta,\gt,\gn)$ is given by $f_0\left(\theta,\gt,\gn\right)\coloneqq f_0\left(\gt\mid\theta\right)f_0\left(\gn\mid\theta\right)\pi(\theta)$.  Clearly $f_0(\theta,\gt,\gn)\in\mathcal{F}$.  Furthermore, by definition it follows that, $\Pi(\theta\mid \s{X_o})=f_0(\theta\mid\gn)$, 
where $f_0(\theta\mid\gn)$ is the conditional density of $\theta$ given $\gn$ corresponding to the joint $f_0(\theta,\gt,\gn)$.



By construction, Since under the true data generating process, $\gn$ is conditionally independent of $\gt$ given $\theta$, for all $\theta\in\Theta$, the conditional density of $\gt$ given $(\gn,\theta)$ equals $f_0(\gt\mid\theta)$ and the \emph{true} conditional density of $(\theta,\gt)$ given $\gn$ can be written as: 
\begin{equation}\label{eq:truejp}
f_0(\theta,\gt\mid \gn)=f_0(\gt\mid\gn,\theta)f_0(\theta\mid\gn)=f_0(\gt\mid\theta)\Pi(\theta\mid\s{X_o})
\end{equation}

Let $\mathcal{Q}_{\Theta}$ be the set of all densities defined on the set $\Theta$.  Clearly both the prior $\pi$ and the posterior $\Pi(\theta\mid \s{X_o})$ are in $\mathcal{Q}_{\Theta}$.  Using the motivation from \eqref{eq:truejp} above, suppose $\mathcal{Q}^{\prime}$ is the subset of densities on $(\theta,\gt)$ defined as:
\begin{equation}\label{def:qp}
  \mathcal{Q}^{\prime}=\left\{q^{\prime}(\theta)f_0(\gt\mid \theta)~:~q^{\prime}(\theta)\in\mathcal{Q}_{\Theta}\right\}.
\end{equation}

Since for any density $f(\theta,\gn,\gt)\in\mathcal{F}$, the true $f_0(\theta,\gt\mid \gn)\in\mathcal{Q}^{\prime}$,  our goal here is to first find an information projection of the corresponding conditional density of $(\theta,\gt)$ given $\gn$ (denoted $f(\theta,\gt\mid \gn)$) on the set $\mathcal{Q}^{\prime}$.  This projection is an approximation of $f_0(\theta,\gt\mid\gn)$.  The posterior $\Pi(\theta\mid\s{X_o})$ is then approximated by integrating the projection over $\gt$. 

\subsection{Functional form of the Posterior Approximation}\label{sec:fnform}
Let $f(\theta,\gt,\gn)\in\mathcal{F}$, and $f(\theta,\gt\mid \gn)$ be the corresponding conditional density of $(\theta,\gt)$ given $\gn$.  We compute the information projection of $f(\theta,\gt\mid \gn)$ on $\mathcal{Q}^{\prime}$ by minimising Kullback-Leibler divergence between the above conditional density and each density $q(\theta,\gt)\in\mathcal{Q}^{\prime}$.  
For any $\gn$, the Kullback-Leibler divergence \citep{coverThomasBook} between $q(\theta,\gt)$ and $f(\theta,\gt\mid \gn)$ is defined as $D_{KL}\left[q(\theta,\gt)\mid\mid f(\theta,\gt\mid \gn)\right]=\int q(\theta,\gt)\log\left(\frac{q(\theta,\gt)}{f(\theta,\gt\mid \gn)}\right)d\gt d\theta$.  The \emph{information projection} of $f(\theta,\gt\mid \gn)$ onto $\mathcal{Q}^{\prime}$ is given by:
\[
q^{\star}(\theta,\gt)\coloneqq\argmin_{q(\theta,\gt)\in\mathcal{Q}^{\prime}}D_{KL}\left[q(\theta,\gt)\mid\mid f(\theta,\gt\mid \gn)\right].
\]
Since the set $\mathcal{Q}^{\prime}$ is convex \citep{whittakerBook}, for any density $f(\theta,\gt\mid \gn)$ its projection is unique.  Next, we find an analytic expression of $q^{\star}(\theta,\gt)$. 
\begin{theorem} \label{thm:postAprx}
For any density $f\in\mathcal{F}$, let $\et_{\gt\mid\theta}\left[\log f(\theta,\gt,\gn)\right]=\int f_0(\gt\mid \theta)\log f(\theta,\gt,\gn) d\gt$ and $\htr_{\gt\mid\theta}(\theta)=-\int f_0(\gt\mid \theta)\log f_0(\gt\mid \theta)d\gt$ be the differential entropy of the density $f_0(\gt\mid \theta)$. Furthermore, let us define:
\begin{equation}\label{eq:fprime}
f^{\prime}(\theta\mid \gn)\coloneqq\frac{e^{\et_{\gt\mid\theta}[\log f(\theta,\gt,\gn)]+\htr_{\gt\mid\theta}(\theta)}}{\int_{t\in\Theta} e^{\et_{\gt\mid t}[\log f(t,\gt,\gn)]+\htr_{\gt\mid t}(t)}dt}.
\end{equation}
Then $q^{\star}(\theta,\gt)=f^{\prime}(\theta\mid \gn)f_0(\gt\mid\theta)$. 
\end{theorem}

The proof of above theorem is presented in the Appendix.  We show that, for any $q(\theta,\gt)=q^{\prime}(\theta)f_0(\gt\mid \theta)\in\mathcal{Q}^{\prime}$, such that $q^{\prime}\in\mathcal{Q}_{\Theta}$, the relationship $D_{KL}\left[q(\theta,\gt)\mid\mid f(\theta,\gt\mid \gn)\right]=D_{KL}\left[q^{\prime}(\theta)\mid\mid f^{\prime}(\theta\mid \gn)\right]+C$ 
holds, where $C$ is a non-negative function of $\gn$ and some hyper-parameters of the prior, and does not depend on $q$ or $q^{\prime}$.  Now the L.H.S. is minimum when $q^{\prime}(\theta)=f^{\prime}(\theta\mid \gn)$, from which the result follows.

Theorem \ref{thm:postAprx} shows that for any joint density $f(\theta,\gt,\gn)\in\mathcal{F}$, the density $f_0(\gt\mid\theta)f^{\prime}(\theta\mid \gn)$ is the best approximation of $f_0(\gt\mid\theta)\Pi(\theta\mid \s{X_o})$ over $\mathcal{Q}^{\prime}$, for all $\theta$, $\gt$ and $\gn$.
The posterior $\Pi(\theta\mid \s{X_o})$ can naturally be approximated by integrating this best approximation over $\gt$.  Since $f^{\prime}(\theta\mid \gn)$ is independent of $\gt$, the corresponding approximation of $\Pi(\theta\mid \s{X_o})$ is trivially given by
$\int f_0(\gt\mid\theta)f^{\prime}(\theta\mid \gn)d\gt=f^{\prime}(\theta\mid \gn).$ 


If $f(\theta,\gt,\gn)=f_0(\theta,\gt,\gn)$, clearly $f_0(\theta,\gt\mid \gn)\in\mathcal{Q}^{\prime}$, and by definition it is it's own information projection.  That is the approximation of $\Pi(\theta\mid \s{X_o})$ is exact.  That is $f^{\prime}_0(\theta\mid \gn)=\Pi(\theta\mid \s{X_o})$.  Furthermore, when $f_0(\gt\mid\theta)$ belongs to a location family $\htr_{\gt\mid\theta}(\theta)$ is not a function of $\theta$.  In that case $f^{\prime}_0(\theta\mid \gn)\propto exp\left\{\et_{\gt\mid\theta}[\log f_0(\theta,\gt,\gn)]\right\}$.




Note that, like it should in a Bayesian procedure, in the expression of $f^{\prime}(\theta\mid\gn)$, the effect of the replicate summary $\gt$ gets integrated out.  In the proposed empirical likelihood-based estimator, the expectation of the log-joint density is approximated by the mean of the log-optimal weights,
which approximately averages out the effect of the replicated summaries from the posterior estimate.  Furthermore, the proposed empirical likelihood estimates an optimal approximate of the true posterior, as we argue below.   

\subsection{Optimal Posterior Approximation}\label{sec:optAprx}
Theorem \ref{thm:postAprx} shows that for any joint density $f(\theta,\gt,\gn)\in\mathcal{F}$, the density $f^{\prime}(\theta\mid \gn)$ provides an approximation of $\Pi(\theta\mid \s{X_o})$ via information projection, with no other assumption required.  Furthermore, the
approximation is exact when the chosen joint density $f(\theta,\gt,\gn)$ is the true joint density $f_0(\theta,\gt,\gn)$.  It however, does not provide a way to choose the joint $f(\theta,\gt,\gn)\in\mathcal{F}$ such that $f^{\prime}(\theta\mid \gn)$ is an optimal approximation of the true posterior in any sense.
We discuss the criterion of such optimality in this section and then discuss its relationship with the proposed empirical likelihood-based procedure.

To that goal, suppose for a joint density $f(\theta,\gt,\gn)\in\mathcal{F}$, $f(\theta,\gn)$, and $f(\gt\mid\gn,\theta)$ respectively denote the corresponding marginal density of $(\theta,\gn)$ and the conditional density of $\gt$ given $\theta$ and $\gn$.
Furthermore, suppose $f(\gt\mid\theta)$ and $f(\gn\mid\theta)$ respectively denote the conditional density of $\gt$ and $\gn$ given $\theta$.  Note that, unless $f(\theta,\gt,\gn)$ is the the true joint density $f_0$, the two conditional densities of $\gt$ and $\gn$ given $\theta$ may not be equal.  The optimality criterion is based on the following result.



\begin{theorem}\label{thm:error}\begin{enumerate}[label={(\alph*)}]
  \item \label{stt:main} Let $f(\theta,\gt,\gn)\in\mathcal{F}$.  Then for all $\theta$, $\gt$ and $\gn$,
{\small
    \begin{equation}
   \log f\left(\theta,\gn\right)-\left\{\et_{\gt\mid \theta}\left[\log f\left(\theta,\gt,\gn\right)\right]+\htr_{\gt\mid \theta}(\theta)\right\}
    =D_{KL}\left[f_0\left(\gt\mid\theta\right)\mid\mid f\left(\gt\mid \gn,\theta\right)\right].\label{eq:sbst}
  \end{equation}
}
    \item If under the joint $f(\theta,\gt,\gn)$, $\gn$ is conditionally independent of $\gt$ given $\theta$, it follows that:
\begin{equation*}
  \log f\left(\theta,\gn\right)-\left\{\et_{\gt\mid \theta}\left[\log f\left(\theta,\gt,\gn\right)\right]+\htr_{\gt\mid \theta}(\theta)\right\}=D_{KL}\left[f_0\left(\gt\mid\theta\right)\mid\mid f\left(\gt\mid\theta\right)\right].
\end{equation*}
\item \label{stt:f0} If $f(\theta,\gt,\gn)=f_0(\theta,\gt,\gn)$, $\et_{\gt\mid \theta}\left[\log f\left(\theta,\gt,\gn\right)\right]+\htr_{\gt\mid \theta}(\theta)=\log f\left(\theta,\gn\right)=\log f_0\left(\theta,\gn\right)$.  Furthermore, $f^{\prime}(\theta\mid \gn)=f(\theta\mid \gn)=f_0(\theta\mid \gn)=\Pi(\theta\mid \s{X_o})$.
\end{enumerate}
  \end{theorem}

Theorem \ref{thm:error}\ref{stt:main} can be proved by a direct expansion of the left-hand side of the expression. The other two statements follow from the first.  In particular, we get $\et_{\gt\mid\theta}[\log f_0(\theta,\gt,\gn)]+\htr_{\gt\mid\theta}(\theta)=\log f_0(\theta,\gn)$.

This theorem shows that for any joint density $f(\theta,\gt,\gn)\in\mathcal{F}$, $f^{\prime}(\theta\mid \gn)$ is not same as the corresponding conditional density $f(\theta\mid \gn)$.  The log-numerator in the expression of $f^{\prime}(\theta\mid \gn)$ is a lower bound of $\log f(\theta,\gn)$.
Furthermore, their difference equals the Kullback-Leibler divergence between the true data-generating density $f_0(\gt\mid \theta)$ and the user-specified conditional density of $\gt$ given $\gn$ and $\theta$ i.e. $f(\gt\mid \gn,\theta)$. 
Clearly, $f_0$ is a minima of this divergence over $\mathcal{F}$, for which, by Theorem \ref{thm:error}\ref{stt:f0}, the approximation of $\Pi(\theta\mid \gn)$ by $f^{\prime}_0(\theta\mid \gn)$ is exact.  Thus, we minimise the above Kullback-Leibler divergence to find the optimal approximation.  

For any density $f(\theta,\gt,\gn)\in\mathcal{F}$, $D_{KL}\left[f_0\left(\gt\mid\theta\right)\mid\mid f\left(\gt\mid \gn,\theta\right)\right]=0\Leftrightarrow f_0\left(\gt\mid\theta\right)=f\left(\gt\mid \gn,\theta\right)$ for all $\gt$, $\gn$ and $\theta$.
This in turn implies that under $f(\theta,\gt,\gn)$, $\gt$ is conditionally independent of $\gn$ given $\theta$ and $f(\gt\mid \theta)=f_0(\gt\mid \theta)$ for all $\gt$, $\gn$ and $\theta$.  

Since the choice of $f(\gn\mid\theta)$ and the marginal $f(\theta)$ remains arbitrary, clearly, $f_0(\theta,\gt,\gn)$ is not the unique density in $\mathcal{F}$ minimising the above Kullback-Leibler divergence.  In order to make the minimal argument unique,
define $\mathcal{F}^{\prime}\subseteq\mathcal{F}$ be the collection of all joint-densities $f(\theta,\gt,\gn)\in\mathcal{F}$, such that for all values of $\theta\in\Theta$, 
\begin{enumerate}[label={(\alph*)}]
\item the corresponding conditional density of $\gt$ given $\theta$ is the same as the corresponding conditional density of $\gn$ given $\theta$, and 
\item the corresponding marginal density of $\theta$ is the prior $\pi$.
\end{enumerate}



The constraints that specify $\mathcal{F}^{\prime}$ are natural and comply to an user's priori belief about the data generating process.  In particular, true joint density $f_0\in\mathcal{F}^{\prime}$.  That is it minimises the divergence in \eqref{eq:sbst} over $\mathcal{F}^{\prime}$.  
However, if $f\in\mathcal{F}^{\prime}$ such that the above divergence is zero, then for all $\theta$, $\gt$ and $\gn$, $f(\theta,\gt,\gn)=f(\gt\mid \gn,\theta)f(\gn\mid\theta)f(\theta)=f_0(\gt\mid\theta)f(\gn\mid\theta)f(\theta)$.  Furthermore, by the construction 
of $\mathcal{F}^{\prime}$, it follows that $f(\gn\mid\theta)=f_0(\gn\mid\theta)$ and $f(\theta)=\pi(\theta)$ for all $\gn$ and $\theta$.  So it follows that, for all $\theta$, $\gt$ and $\gn$, $f(\theta,\gt,\gn)=f_0(\gt\mid\theta)f_0(\gn\mid\theta)\pi(\theta)=f_0(\theta,\gt,\gn)$.

From the arguments above, the following result is now evident.

\begin{theorem}\label{thm:fp}
Suppose $\mathcal{F}^{\prime}$ is the subset of densities over $(\theta,\gt,\gn)$ as defined above. Then $f_0\in\mathcal{F}^{\prime}$ uniquely minimises $D_{KL}\left[f_0\left(\gt\mid\theta\right)\mid\mid f\left(\gt\mid \gn,\theta\right)\right]$ over $\mathcal{F}^{\prime}$. 
\end{theorem}

An estimate of $f_o(\theta,\gt,\gn)$ can therefore be obtained as: 
\begin{equation}\label{eq:sbst2}
 \hat{f}_0(\theta,\gt,\gn)=\argmin_{f\in\mathcal{F}^{\prime}}D_{KL}\left[f_0\left(\gt\mid\theta\right)\mid\mid f\left(\gt\mid \gn,\theta\right)\right].
\end{equation}

The estimate $\hat{f}_0(\theta,\gt,\gn)$ in \eqref{eq:sbst2} is actually a reverse information projection of $f_0(\gt\mid\theta)$ on the set of densities $f(\gt\mid \gn,\theta)$ such that $f(\theta,\gt,\gn)\in\mathcal{F}^{\prime}$.
Furthermore, since $f_0(\gt\mid \theta)$ is fixed, we get 
\begin{align}
~&\argmin_{f\in\mathcal{F}^{\prime}}D_{KL}\left[f_0\left(\gt\mid\theta\right)\mid\mid f\left(\gt\mid \gn,\theta\right)\right]
=\argmax_{f\in\mathcal{F}^{\prime}}\int f_0(\gt\mid\theta)\log f(\gt\mid \gn,\theta)d\gt\nonumber\\ 
=&\mbox{arg}\max_{f\in\mathcal{F}^{\prime}}\left\{\int f_0(\gt\mid\theta)\log f(\gt,\gn\mid\theta)d\gt-\log\int f(\gt,\gn\mid\theta)d\gt\right\}.\label{eq:crent}
\end{align}

\noindent That is, in order to minimise our loss function, we only need to maximise the cross-entropy term over the specified $\mathcal{F}^{\prime}$. 


\subsection{Connection to the proposed ABCel posterior}\label{sec:conEl}
From the justifications presented above, for appropriate summary statistics, the task is to specify the set of joint densities $\mathcal{F}^{\prime}$, at least approximately, and minimise the divergence in \eqref{eq:sbst} over this specified set.  Once $\hat{f}_0(\theta,\gt,\gn)$ is computed, 
the corresponding approximation of $\Pi(\theta\mid \gn)$ is given by the corresponding $\hat{f}^{\prime}_0(\theta\mid\gn)$ can be obtained by substituting $f(\theta,\gt,\gn)$ by $\hat{f}^{\prime}_0(\theta,\gt,\gn)$ in \eqref{eq:fprime}.  
We now argue that with simple choices of summary statistics, the proposed modified empirical likelihood-based procedure follows the same recipe.



For simplicity, suppose the vector of summary statistics $\sa$ consists of $r$ quantiles of the observed or the replicated data vectors.
In the notations of Section \ref{sec:multsamp}, for $i=1$, $2$, $\ldots$, $m$, let $\gi=\s{X_i(\theta)}$ be the values of summary of $X_i(\theta)$ generated with input $\theta\in\Theta$.  Assuming that the problem in \eqref{eq:w2} is feasible the optimal weights $\hat{w}(\theta)$ satisfy the constraints:
\begin{equation*}
\hat{w}(\theta)\in\Delta_{m-1}\text{  and } \sum^m_{i=1}\hat{w}_i(\theta)(\gi-\gn)=0.
\end{equation*}
By our construction, the empirical estimate of the conditional joint distribution of the random vector $(\gt,\gn)$ given $\theta$ can be obtained as:
\[
\hat{F}_m(t_1,t_o\mid \theta)=\frac{1}{m}\sum^m_{i=1}\hat{w}_i(\theta)1_{\{(\gi,\gn)\le(t_1,t_o)\}}.
\] 
We first verify that the condition (a) in the definition of $\mathcal{F}^{\prime}$ is approximately satisfied.  Note that the constraints imply that:
\[
\int \gt d\hat{F}_m(t_1,t_o\mid \theta)=\gn.
\]
That is the conditional joint distribution is estimated by matching $\gn$ with the marginal conditional expectation of $\gt$ given $\theta$.

The concept of matching the expected quantiles with the observed is the key behind the goodness-of-fit plots like the Q-Q plots, probability plots, etc.  If the match is close, the densities of the corresponding random variables are approximately equal.  
Following the same argument, the proposed empirical likelihood-based procedure computes the estimate $\hat{F}_m$ by approximately equating the conditional marginal densities of the observed data $X_o$ and the replicated data $X_1$ given the input parameter value $\theta$.  
Now since the summary statistics, $\sa$ (in this case $r$ quantiles) are deterministic functions of the data, consequently, the conditional marginal densities of $\gt$ and $\gn$ given $\theta$ would be approximately equal.  That is, the condition (a) in the definition of the set $\mathcal{F}^{\prime}$ is approximately satisfied. 

The proposed empirical likelihood based method maximises the sample version of the Kullback-Leibler divergence in \eqref{eq:crent}.  This sample version is naturally obtained by 
estimating the marginal distribution of $\gt$ given $\theta$, by its empirical estimate, which puts equal weight of $1/m$ on each observation $\gi$ and by estimating conditional joint distribution of $(\gt,\gn)$ given $\theta$ by $\hat{F}_m$ described above.  In view of the constraint that $\hat{w}(\theta)\in\Delta_{m-1}$, this justifies 
the objective function $m^{-1}\sum^m_{i=1}\log w_i$, that is maximised in \eqref{eq:w2}.

Other than the constraints which defines $\mathcal{F}^{\prime}$, in the proposed method $\hat{F}_m$ is computed with minimal restrictions.  For any $\theta$,  the maximum of $m^{-1}\sum^m_{i=1}\log w_i(\theta)$ finite when there exist a $w\in\mathcal{W}_{\theta}$, such that $w_i(\theta)>0$, for all $i=1$, $2$, $\ldots$, $m$.
This is equivalent to maximising the divergence in \eqref{eq:crent} over all joint densities $f(\theta,\gt,\gn)\in\mathcal{F}^{\prime}$ such that, for all $i=1$, $2$, $\ldots$, $m$, each observation $(\gi,\gn)$ is in the support of the conditional density $f(\gt,\gn\mid\theta)$.  
The proposed empirical likelihood-based method thus maximises the sample version of \eqref{eq:crent} over a large flexible set of non-parametrically specified distributions approximately satisfying the constraints which define $\mathcal{F}^{\prime}$. No parameter tuning or otherwise need to be specified or estimated (as in e.g. \citet{anNottDrovandi2020}).



The key to the justification presented above is the existence of summary statistics which approximately specify the density of the underlying random variable.  
Such summaries have been rigorously studied in statistics. Other than the quantiles, moments, up crossing proportions, etc can be used.  We discuss various choices for the summary statistics in Section \ref{sec:impl} below.    


\section{Properties of the ABC Empirical Likelihood Posterior}\label{sec:asymp}
The asymptotic properties of conventional ABC methods have been a topic of much recent research
\citep{frazier+mrr18,li+f18a,li+f18b}.  Here 
we investigate some basic asymptotic properties of our proposed empirical likelihood method. The proofs of the results are deferred to the supplement. 



Following \citet{owen01} the weights in \eqref{eq:w2} can be expressed as $\hat{w}_i=\{m(1+\hat{\lambda}^Th_i)\}^{-1}$, where $\hat{\lambda}$ is obtained by solving the equation $\sum^m_{i=1}h_i/(1+\hat{\lambda}^Th_i)=0$.

\subsection{Posterior Consistency}\label{sec:postcons}
In what follows below, we consider limits as $n$ and $m=m(n)$ grow unbounded.  Furthermore, for convenience, we make the dependencies of $X_o$ and $X_1$, $X_2$, $\ldots$, $X_m\in\mathbb{R}^n$ on sample size $n$ as well as parameter $\theta$ explicit. 
In what follows, a sequence of events $\{E_n, n\ge 1\}$ is said to occur with high probability, if $P(E_n)\rightarrow 1$ as $n\rightarrow\infty$.

Suppose that we define 
$h^{(n)}_i\left(\theta\right)=\left\{\s{X^{(n)}_i(\theta)}-\s{X^{(n)}_o(\theta_o)}\right\}$, 
and assume that \\ $\et_{\s{X^{(n)}_i(\theta)}\mid\theta}[\s{X^{(n)}_i(\theta)}]$ is finite so that we can write $\s{X^{(n)}_i(\theta)}=\et_{\s{X^{(n)}_i(\theta)}\mid\theta}\left[\s{X^{(n)}_i(\theta)}\right]+\xi^{(n)}_i(\theta)=\sm^{(n)}(\theta)+\xi^{(n)}_i(\theta)$,
where $\et_{\s{X^{(n)}_i(\theta)}\mid\theta}[\xi^{(n)}_i(\theta)]=0$ for all $i$, $n$ and $\theta$.

We make the following assumptions.
\begin{itemize}
\item[(A1)] (Identifiability and convergence) There is a sequence of positive increasing real numbers $b_n\rightarrow\infty$, such that, 
$\sm^{(n)}(\theta)=b_n\left\{\sm(\theta)+o(1)\right\}$,
where $\sm(\theta)$ is a one-to-one function of $\theta$ that does not depend on $n$.  Furthermore, $\sm(\theta)$ is continuous at $\theta_o$ and for each $\epsilon>0$,  and for all $\theta\in\Theta$, there exists $\delta>0$, such that whenever $\mid\mid\theta-\theta_o\mid\mid>\epsilon$, $\mid\mid \sm(\theta)-\sm(\theta_o)\mid\mid>\delta$.

\item[(A2)] (Feasibility) For each $\theta$, $n$ and $i=o,1$, $\ldots$, $m(n)$, the vectors $\xi^{(n)}_i(\theta)$ are identically distributed, supported over the whole space, and their distribution puts positive mass on every orthant, $\mathcal{O}_u$ of $\mathbb{R}^r$, $u=1$, $2$, $\ldots$, $2^r$.  Furthermore, for every orthant $\mathcal{O}_u$, as $n\rightarrow\infty$, 
$\sup_{\{i~:~\xi^{(n)}_i(\theta)\in\mathcal{O}_u\}}\mid\mid \xi^{(n)}_i(\theta)\mid\mid\longrightarrow\infty$
in probability, uniformly in $\theta$.   
\item[(A3)] (Growth of extrema of Errors) As $n\rightarrow\infty$, 
$\sup_{i\in\{o,1,2,\ldots, m(n)\}}\mid\mid \xi^{(n)}_i(\theta)\mid\mid b^{-1}_n\rightarrow 0$
in probability, uniformly in $\theta\in\Theta$.
\end{itemize}


Assumption (A1) ensures identifiability and additionally implies that $\sm^{(n)}(\theta)/b_n-\sm(\theta)$ converges to zero uniformly in $\theta$.
Assumption (A2) is important for ensuring that with high probability the empirical likelihood ABC posterior is a valid
probability measure for $n$ large enough.  Assumptions (A2) and (A3) also link the number of simulations $m$ to $n$ 
and ensure concentration of the posterior with increasing $n$. 
The proofs of the results below are given in the Appendix.  The main result, Theorem 1, shows posterior consistency for the proposed empirical likelihood method.  

Let $l_n(\theta):=\exp(\sum^{m(n)}_{i=1}\log\left(\hat{w}_i(\theta)\right)/m(n))$ and for each $n$, we define\\ $\Theta_n\coloneqq\left\{\theta~:~\mid\mid\sm(\theta)-\sm(\theta_o)\mid\mid\le b^{-1}_n\right\}$.
By continuity of $\sm$ at $\theta_0$, $\Theta_n$ is nonempty for each $n$.  Furthermore, since $b_n$ is increasing in $n$, $\Theta_n$ is a decreasing sequence of sets in $n$. 

\begin{lemma}\label{lem:1}
  Under assumptions (A1) to (A3), with high probability, the likelihood $l_n(\theta)>0$ for all $\theta\in\Theta_n$.
\end{lemma}

Lemma \ref{lem:1} shows that for large $n$ the estimated likelihood is strictly positive in a neighbourhood of $\theta_0$.  Next, we show that the empirical likelihood is zero outside certain neighbourhood of $\theta_0$.

\begin{lemma}\label{lem:2}
  Under assumptions (A1) - (A3), for every $\epsilon>0$, let $B(\theta_0,\epsilon)$ be the open ball of radius $\epsilon$ centred at $\theta_o$. The empirical likelihood is zero outside $B(\theta_0,\epsilon)$, with high probability.
\end{lemma}

Now suppose we choose $\epsilon=b^{-1}_1$ and $n>n(b^{-1}_1)$ 
such that $l_n(\theta)$ is positive on $\Theta_n$ with high probability.  Furthermore, for all $n$ and for all $\theta\in\Theta_n$, $\min_{i\ne j}\mid\mid \s{X_j(\theta)}-\s{(X_i(\theta)}\mid\mid>0$ with probability $1$, which implies for an appropriate choice of $k$, (see the Supplement) the estimate of the differential entropy $\mid\hat{H}^{0(n)}_{\sa\mid \theta}(\theta)\mid<\infty$ with probability $1$ as well.  This proves that for large values of $n$, with high probability:
$\int_{\theta\in\Theta}l_n(\theta)e^{\hat{H}^{0(n)}_{\sa\mid \theta}(\theta)}\pi(\theta)d\theta\ge\int_{\theta\in\Theta_n}l_n(\theta)e^{\hat{H}^{0(n)}_{\sa\mid \theta}(\theta)}\pi(\theta)d\theta>0$,
and 
$
\hat{\Pi}_n\left(\theta\mid \s{X_o(\theta_o)}\right)=\left(l_n(\theta)e^{\hat{H}^{0(n)}_{\sa\mid \theta}(\theta)}\pi(\theta)\right)/\int_{t\in\Theta}l_n(t)e^{\hat{H}^{0(n)}_{\sa\mid t}(t)}\pi(t)dt$
is a valid probability measure (with high probability).  The main result, Theorem 1 below, establishes posterior consistency.

\begin{theorem}\label{thm:1}
As $n\rightarrow\infty$, $\hat{\Pi}_n\left(\theta\mid \s{X_o(\theta_o)}\right)$ converges in probability to $\delta_{\theta_o}$, where $\delta_{\theta_0}$ is the degenerate probability measure supported at $\theta_0$. 
\end{theorem}

\subsection{Behaviour of the Proposed Posterior with Growing Number of Replications}\label{sec:m}
We now discuss how the proposed ABCel posterior behaves with fixed sample size $n$ and observed summary and growing $m$.  Our primary goal is to is to find appropriate number of replicates i.e. $m$ for a fixed sample size $n$.  We also discuss the bias-variance trade-off as observed in Figure \ref{fig:post} in more details.

Under the setup of fixed $n$ and the observed summary, it is more appropriate to consider expectation of $h^{(n)}_i(\theta)$ conditional on $(\theta,\s{X^{(n)}_o(\theta_o)})$ .  Since each $X^{(n)}_i(\theta)$ is conditionally independent of $X^{(n)}_o(\theta_o)$ given $\theta$, for each $i=1$, $2$, $\ldots$, $m$, and $\theta\in\Theta$ we get:
\[
\et_{\s{X^{(n)}_i(\theta)}\mid(\theta,\s{X^{(n)}_o(\theta_o)})}\left[h^{(n)}_i(\theta)\right]=\et_{\s{X^{(n)}_i(\theta)}\mid\theta}\left[\s{X^{(n)}_i(\theta)}\right]-\s{X^{(n)}_o(\theta_o)}\ne 0 \text{~~~a.e.}.
\]
That is, for fixed $n$, after conditioning on $\s{X^{(n)}_o(\theta_o)}$, the constraints in the problem \eqref{eq:w2} $h^{(n)}_i(\theta)$, $i=1$, $2$, $\ldots$, $m$ are misspecified for all $\theta\in \Theta$ almost everywhere (even when $\theta=\theta_o$).
The constrained optimisation problem in \eqref{eq:w2} however could still be feasible and the resulting estimated posterior could be positive.  The properties of empirical likelihood under misspecified but feasible constraint have been studied by \citet{ghosh2019empirical}.  We now evoke their results.

  Using the notations introduced above, when $r=1$, i.e. there is only one constraint present, 
under conditions similar to those described above, it can be shown that, \citep[Theorem $3.4$]{ghosh2019empirical} for any $\theta\in\Theta$:
\begin{align}\label{eq:m}
l_m(\theta)&\coloneqq\frac{1}{m}\sum^m_{i=1}\log(\hat{w}(\theta))=-\frac{1}{\mathcal{M}_m(\theta)}\left|E_{\s{X^{(n)}_1(\theta)}\mid\theta}[\s{X^{(n)}_1(\theta)}]-\s{X^{(n)}_o(\theta_o)}]\right|(1+o_p(1)),\nonumber\\
& =-\frac{b_n}{\mathcal{M}_m(\theta)}\left|(\sm(\theta)-\sm(\theta_o)+o(1))-\frac{\xi^{(n)}_o(\theta_o)}{b_n} \right|(1+o_p(1)),  
\end{align}
where $\mathcal{M}_m(\theta)$ is a non-random $o(m)$ sequence such that, as $m\rightarrow\infty$, $\mathcal{M}_m \rightarrow\infty$ and both\\
$\mathcal{M}^{-1}_m(\theta)\max_{1\le i\le m}\left|\xi^{(n)}_i(\theta)\right|1_{\left\{\xi^{(n)}_i(\theta)>0\right\}}=1+o_p(1)$, and $\mathcal{M}^{-1}_m(\theta)\max_{1\le i\le m}\left|\xi^{(n)}_i(\theta)\right|1_{\left\{\xi^{(n)}_i(\theta)<0\right\}}=1+o_p(1)$ are satisfied.

The sequence $\mathcal{M}_m(\theta)$ is the rate at which the maximum of the $\s{X_i(\theta)}$ grows away from its mean.  The above conditions are easily satisfied. As for example, when $\xi^{(n)}_o(\theta_0)$ is a $N(0,\sigma^2_0)$ random variable, $\mathcal{M}_m\sim\sigma_0\sqrt{2\log m}$.


In the rest of this section, we assume that $r=1$.  Using the results from \citet{ghosh2019empirical} it is possible to specify bounds on the rate of growth of the number of replicates with the sample size. 
Since the differential entropy plays a relatively minor role in determining the posterior, in what follows we assume that for each $\theta$, the estimate of the differential entropy remains bounded focus on $l_m(\theta)$.  Furthermore, for brevity, we present the results as {\it Remarks} below.  More details are available in the supplement.  

\subsubsection{Bounds on the growth of the number of replications in terms of sample size}\label{sec:bounds}
We first consider the bounds of the replication size $m$ in terms of sample size $n$.  Our results follow from various advantageous properties of the posterior.  For the purposes of easier description and illustration, we would sometime assume that the errors $\xi^{(n)}_o$ follow a $N(0,\sigma^2_o)$ distribution.  

The posterior is Bayesian consistent \citep{frazier+mrr18} if with high probability two things happen:
  First, $\exp(l_m(\theta))$ would converge to zero for all $\theta\ne\theta_o$ and second, $\exp(l_m(\theta_o))$ would not collapse to zero.   

  \begin{remark}\label{rem:baycon1}
    In order to ensure the first condition it is enough to choose $m$ and $n$ such that $b_n/\mathcal{M}_m(\theta)$ diverges.  An upper bound of the rate of growth of $m$ can thus be obtained by inverting the relation $b_n>\mathcal{M}_m(\theta)$.  Depending on the distribution of $\xi^{(n)}_o$, $m$ can be much larger than $n$.
For example, if $\xi^{(n)}_o$ follows a normal distribution with mean zero and variance $\sigma^2_o$, $b_n=\sqrt{n}$ and $\mathcal{M}_m(\theta)$ is of the order $\sigma_o\sqrt{2\log(m)}$, which allows an upper bound of $m$ as large as $\exp(n/(2\sigma^2_o))$.   
  \end{remark}

  \begin{remark}\label{rem:baycon2}
    A more accurate relationship can be obtained from the second condition.  The condition implies that there exists a constant $C_1>0$ such that, $l_m(\theta_o)>-C_1$ with a high probability. 
    Assuming that $\xi^{(n)}_o(\theta_o)$ is a $N(0,\sigma_o^2)$ random variable, from \eqref{eq:m}, it follows that $Pr[l_m(\theta_o)\le -C_1]\le \exp(-C_1^2\log m)=m^{-C^2_1}$.
    Now if we fix the rate of reduction of the above probability to $n^{-\alpha}$ for some $\alpha$, we get $m=n^{\alpha/C^2_1}$.
    \end{remark}



  Other bounds can be found by controlling the rate at which the probability of a Type I error for testing the null hypothesis $\theta=\theta_o$ against the unrestricted alternative decrease to zero. By construction $l_m(\theta)$ is different from the traditional empirical likelihood, so this problem is of broad interest.

  \begin{remark}\label{rem:test}
    The log-likelihood ratio $\log LR(\theta_o)$ turns out to be $l_m(\theta_o)+\log m$.  The test rejects $H_0$ if $\log LR(\theta_o)$ is smaller than $\log C_0$, for some pre-specified $C_0\in(0,1)$.  Assuming that, $\xi^{(n)}(\theta)$ is a $N(0,\sigma^2_0)$ random variable, it follows that, the probability of rejecting the null hypothesis is given by (see the Supplement):
    \[
    Pr[\log m+l_m(\theta_o)\le \log C_0]\le exp\left\{(\log m)^3-2(\log m)^2\log C_0+(\log m)(\log C_0)^2\right\}.
    \]
    Now ensuring that the probability of rejecting the null hypothesis reduces at the rate of $p_n$, we get $p_n=exp\left\{-(\log m)^3\right\}$, which implies the number of replications $m=exp\{(-\log p_n)^{1/3}\}$.
  \end{remark}

  \subsubsection{Behaviour of the log-likelihood when $\mathcal{M}_m(\theta)/b_n$ diverges}
This scenario includes the situation when the sample size $n$ is fixed and the number of replication $m$ grows. We discuss the bias-variance trade-off or the flattening of the approximate likelihood as observed in Figure \ref{fig:post}.  
  \begin{remark}\label{rem:div}
    Let us fix $\theta\ne\theta_o$ and suppose $\xi^{(n)}_o(\theta_o)$ follows a $N(0,\sigma^2_o)$ distribution.  For a fixed $C_2>0$, it can be shown that (see the supplement):
\begin{equation}\label{eq:lbf}
Pr[l_m(\theta)\le -C_2]\le \left(\frac{1}{m}\right)^{\left\{C_2-\frac{b_n}{\sigma_o\sqrt{2\log m}}\left|\sm(\theta)-\sm(\theta_o)\right|\right\}^2}.
\end{equation}
Now, if $\mathcal{M}_m(\theta)/b_n=\sigma_o\sqrt{2\log m}/b_n$ diverges with $m$ and $n$, clearly, for large values of $m$ and $n$, $Pr[l_m(\theta)\le -C_2]\approx m^{-C^2_2}$.  That is, for any fixed $C_2>0$ and $\theta\ne\theta_o$,  $l_m(\theta)\ge -C_2$ with a high probability.
Furthermore, for a fixed $n$, R.H.S. of \eqref{eq:lbf} is a decreasing function in $m$.  That is if the sample size is kept fixed, increasing the number of replications will increase the probability of $l_m(\theta)\ge -C_2$.  As a result, the log likelihood will become increasingly flat in shape. This is exactly the phenomenon that was observed in Figure \ref{fig:post}.  Remark \ref{rem:div} provides actual justification to our observation.

Statistical intuition mandates with an increase in $m$, we should increase the number of summary statistics. Remark \ref{rem:div} does not apply to such situations.  We present evidence in favour of our intution in Example \ref{normal} below.
  \end{remark}
  


\section{Choice of Summary Statistics}\label{sec:impl}
A judicious choice of summary statistics is crucial for a good performance of any ABC procedure \citep{fearnhead+p12}.  
The proposed method does not necessarily require summaries that are sufficient for the parameter, which according to many authors (e.g. \citet{frazier+mrr18,robertSurvey2016}) are usually not available. Rather from the arguments in Section \ref{sec:just}, it mandates an use of summaries which approximately define the density of $X_i(\theta)$, for $i\in\mathbb{M}_o$.

Sample quantiles, extreme values, or proportion of samples exceeding the certain pre-specified thresholds that directly put constraints on the data density (see \citet{agostinoBook}) can be used.  Moreover, moments, if they exist, may under certain conditions (e.g. Carleman's condition) specify a density (see \citet{gut2002}). Thus, moments can be used as summaries in many cases as well.

For complex data models, with dependent components, marginal summaries may not be adequate. In such cases, constraints can be based on joint moments, joint quantiles or joint exceedances, etc can be used. Other than these generic choices, one can base the constraints on the functionals of transformed variables. Since a density is a one-to-one function of its characteristic function, for dependent data sets, constraints based on the smoothed spectral density of the data can be used.
For example, in the case of stochastic processes, summaries based on the exceedance proportions of log-amplitudes, which actually put constraints on the auto-covariance function of the process, are often beneficial (see Section \ref{sec:bb} below).




In our experience, often moments work the best.  A judicious mix and match of various forms of summaries decided after some inspection of the summaries of observed data are required.  It should also be recognised that summaries with widely different scales or an ill-conditioned covariance matrix may lead to a poor estimate of differential entropy and subsequently to slow mixing of the Markov Chain Monte Carlo procedures.  


Finally, the number of summaries required would depend partly on their nature, partly on the number of replications $m$, and partly on the sample size $n$ (see \eqref{eq:lbf}).  Even though some judgements is required, evidence shows (see Section \ref{sec:expl}) for any given problem, appropriate summaries can be found without much effort.   



\section{Illustrative Examples and Applications}\label{sec:expl}
  
We illustrate the utility of the ABCel method with four examples involving data simulated from a standard Gaussian model, an ARCH(1) model (also considered in \citet{mengersen+pr13}), The simple recruitment, boom and bust model \citep{anNottDrovandi2020}, and a rael life example modelled as an elliptical inclusion model respectively.  Here in order to address dependence we use non-Gaussian summaries based on auto-covariance function and the periodogram of the data.  
We also present a real application based on stereological extremes \citep{anderson2002largest}.  More examples on the traditional $g$-and-$k$ model and an application to Erd\"{o}s-Renyi random graphs can be found in the supplement.  


\begin{figure}[t]
\begin{center}
\begin{subfigure}{.45\columnwidth}
\resizebox{2.25in}{2.25in}{\includegraphics{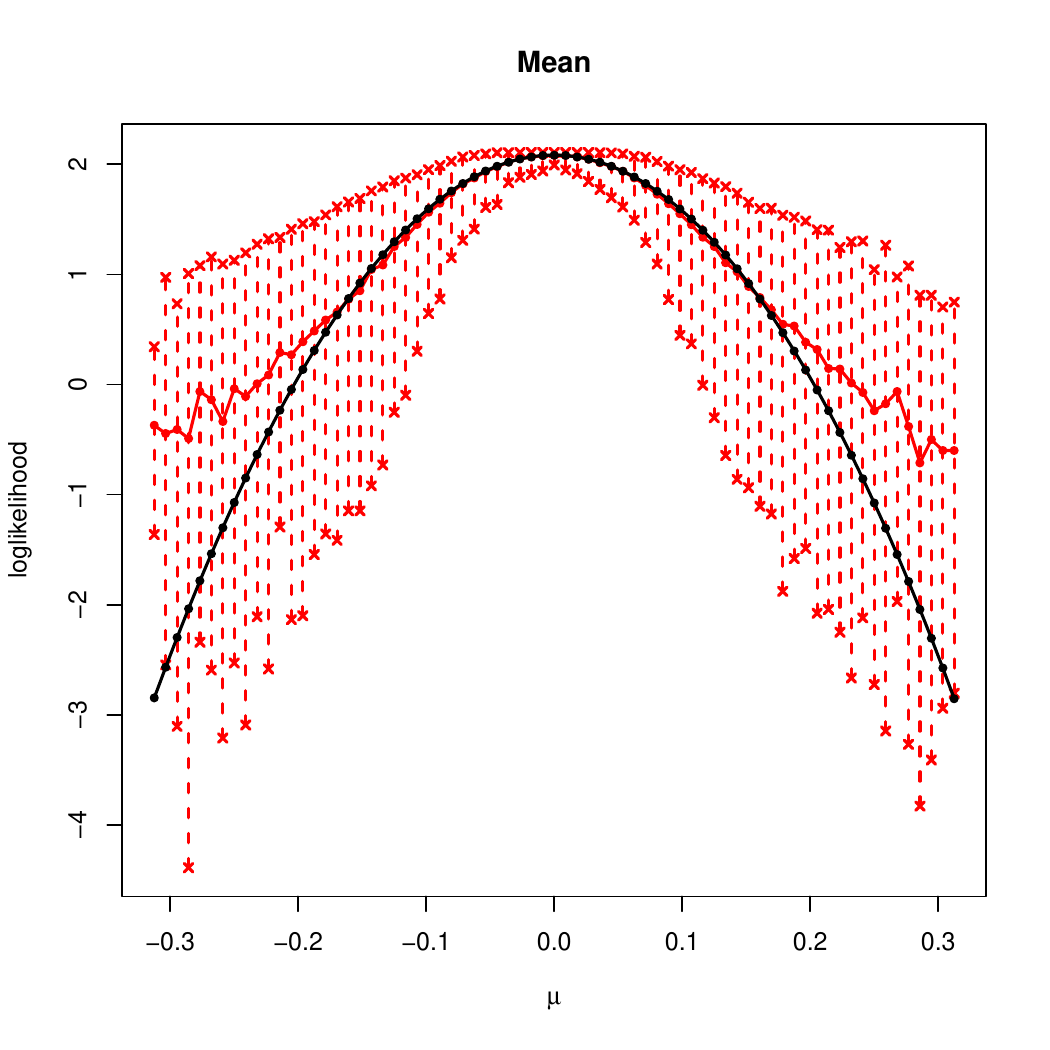}}
\caption{\label{fig:F1a}}
\end{subfigure}
\begin{subfigure}{.45\columnwidth}
\resizebox{2.25in}{2.25in}{\includegraphics{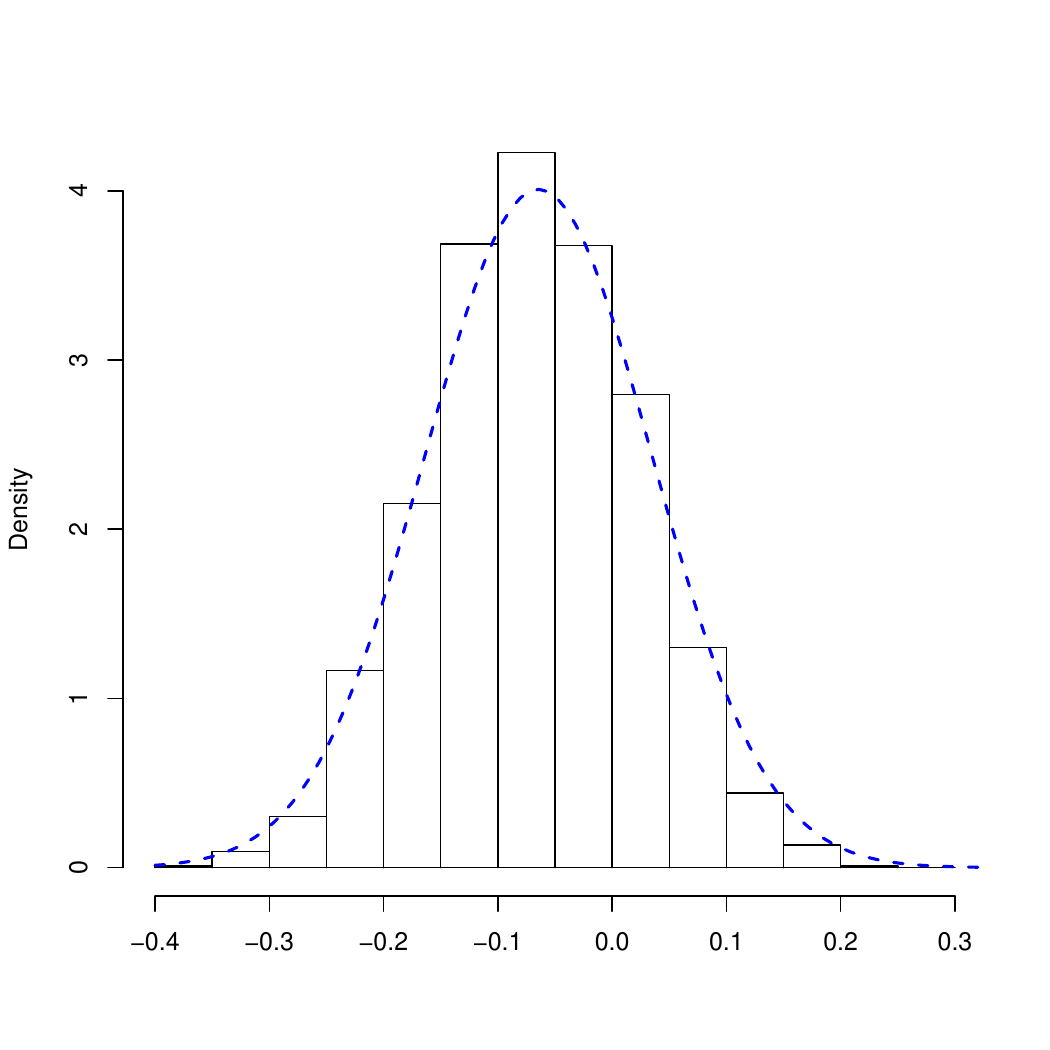}}
\caption{\label{fig:F1b}}
\end{subfigure}
\caption{Comparison of the true posterior of the mean of a Normal distribution with unit variance conditional on the sample mean with our proposed empirical likelihood based ABC posterior. Here $n=100$ and $m=25$.  
Figure \ref{fig:F1a} directly compares the true log-posterior (black curve) with the means and 95\% credible intervals of the proposed approximate posterior based on $1000$ replications for each parameter value (in red).  
Figure \ref{fig:F1b} compares the true posterior (dashed line) with the histogram of the samples drawn from the proposed empirical likelihood based ABC posterior (underlying histogram). 
}
\label{F1}
\end{center}
\end{figure}

\subsection{Normal distribution}\label{normal}

Our first example considers inference about a mean $\mu$ for a random sample of size $n=100$ from a normal density, $N(\mu,1)$.  
The prior for $\mu$ is $N(0,1)$.  The observed data $X_o$ is generated with $\mu=0$. The exact posterior for $\mu$ is  
normal, $N(\sum^n_{j=1}X_{oj}/(n+1),(n+1)^{-1})$.  The proposed empirical likelihood based method was implemented with $m=25$.  We considered several choices of constraint functions  $\sa^{(1)}$, $\ldots$, $\sa^{(r)}$.  
Specifically, for $i=o,1,\ldots,m$, we take (a) $\sa^{(1)}(X_i(\theta))=\bar{X}_{i\cdot}=\sum^n_{j=1}X_{ij}(\theta)/n$,  (b) $\sa^{(2)}(X_i(\theta))=\sum^n_{j=1}\left(X_{ij}(\theta)-\bar{X}_{i\cdot}\right)^2/n$, (c) $\sa^{(3)}(X_i(\theta))=\sum^n_{j=1}\left(X_{ij}(\theta)-\bar{X}_{i\cdot}\right)^3/n$, (d) $\sa^{(4)}(X_i(\theta))=\mbox{median of }X_i(\theta)$, (e) $\sa^{(5)}(X_i(\theta))=\mbox{first quartile of }X_i(\theta)$, (f) $\sa^{(6)}(X_i(\theta))=\mbox{third quartile of }X_i(\theta)$.  Here the constrains considered use the
first three central moments ((a)-(c)) and the three quartiles ((d)-(f)).
Combinations of these constraints are considered within the empirical likelihood procedure.  

The posteriors obtained from our proposed empirical likelihood-based ABC method with the above summaries are close to the true posterior. An illustrative example, with sample mean as a summary, is presented in Figure \ref{F1}.
Here, the true posterior density, i.e. the dashed line, is quite close to the histogram of the samples drawn from the posterior obtained from the proposed method.

In order to compare the performance of the proposed procedure for different choices of the summary statistics, we consider frequentist coverages and the average lengths of the $95\%$ credible intervals.  The results are presented in Table~\ref{Tab2s}.  The covereages are based on $100$ repeats of the procedure.  For each repetition, MCMC approximations to the posterior are based on $50,000$ samples with $50,000$ iterations discarded as burn-in.

As we have shown before (See Figure \ref{fig:post} and Remark \ref{rem:div}) the approximate posterior gets flatter if we keep the number of summary statistics fixed and increase the number of replications $m$.  That is, with increasing $m$, one should increase the number of summary statistics used.  The same argument mandates that when we increase the number of summary statistics we should also increase the number of replications.  In Table~\ref{Tab2s} we report the value of $m$ for which the Monte Carlo Frequentist coverages were close to the nominal value of $95\%$.

\begin{table}[t]
 \caption{\label{Tab2s} The coverage and the average length of $95\%$ credible intervals for $\mu$ for various choices of constraint functions when $\mu=0$ and $n=100$.  The coverage for the true posterior is $0.95$ and average length is $0.39$ (2 d.p.).}
  \begin{center}
\begin{tabular}{lccc}
Constraint Functions &$m$&Coverage & Average Length\\
Mean, (a).&$25$&$0.95$&$0.360$\\
Median, (e).&$25$&$0.95$&$0.446$\\
First two central moments, (a), (b).&$40$&$0.94$&$0.331$\\
Mean and Median, (a), (e).&$40$&$0.94$&$0.330$\\
First three central moments, (a), (b), (c).&$70$&$0.91$&$0.307$\\
Three quartiles, (e), (f), (g).&$75$&$0.93$&$0.329$\\
  \end{tabular}
  \end{center}
\end{table}


From Table \ref{Tab2s}, it is clear that the proposed method performs quite well for various sets of summary statistics.  For mean and the median the frequentist coverage is matches exactly the nominal value.  Note that the sample mean is minimal sufficient for $\mu$, and would be an ideal choice of summary statistic
  in conventional likelihood-free procedures such as ABC.  However, median is not sufficient for the mean, but still produces the exact coverage.  Table \ref{Tab2s} also shows that when multiple summary statistics are used, by increasing the number of replicates it is possible to obtain an approximate posterior with frequentist coverage close to the nominal value of $.95\%$.


\subsection{An ARCH(1) model}\label{sec:arch}  
We now present examples where summary statistics are not close to normal, so that 
the assumptions behind the synthetic likelihood are not satisfied.  
We consider an autoregressive conditional heteroskedastic or ARCH(1) model, where for each $i=o,1,2,\ldots,m$, the components $X_{i1}(\theta),X_{i2}(\theta),\ldots,X_{in}(\theta)$ are dependent for all $\theta=(\alpha_0,\alpha_1)\in\Theta$.  This model was also considered in \citet{mengersen+pr13}. 
For each $i$, the time series ${X_{ij}}_{1\leq j \leq n}$ is generated by 
$X_{ij}(\theta)=\sigma_{ij}(\theta)\epsilon_{ij}$, $\sigma_{ij}^2(\theta)=\alpha_0+\alpha_1X_{i(j-1)}^2(\theta)$,
where $\epsilon_{ij}$ are i.i.d. $N(0,1)$ random variables, $\alpha_0>0$, and $0<\alpha_1<1$. 
We assume a uniform prior over $(0,5)\times(0,1)$ for $(\alpha_0,\alpha_1)$. 

\begin{figure}[t]
  \begin{center}
    \begin{subfigure}{.45\columnwidth}
      \resizebox{2in}{2in}{\rotatebox{90}{\includegraphics{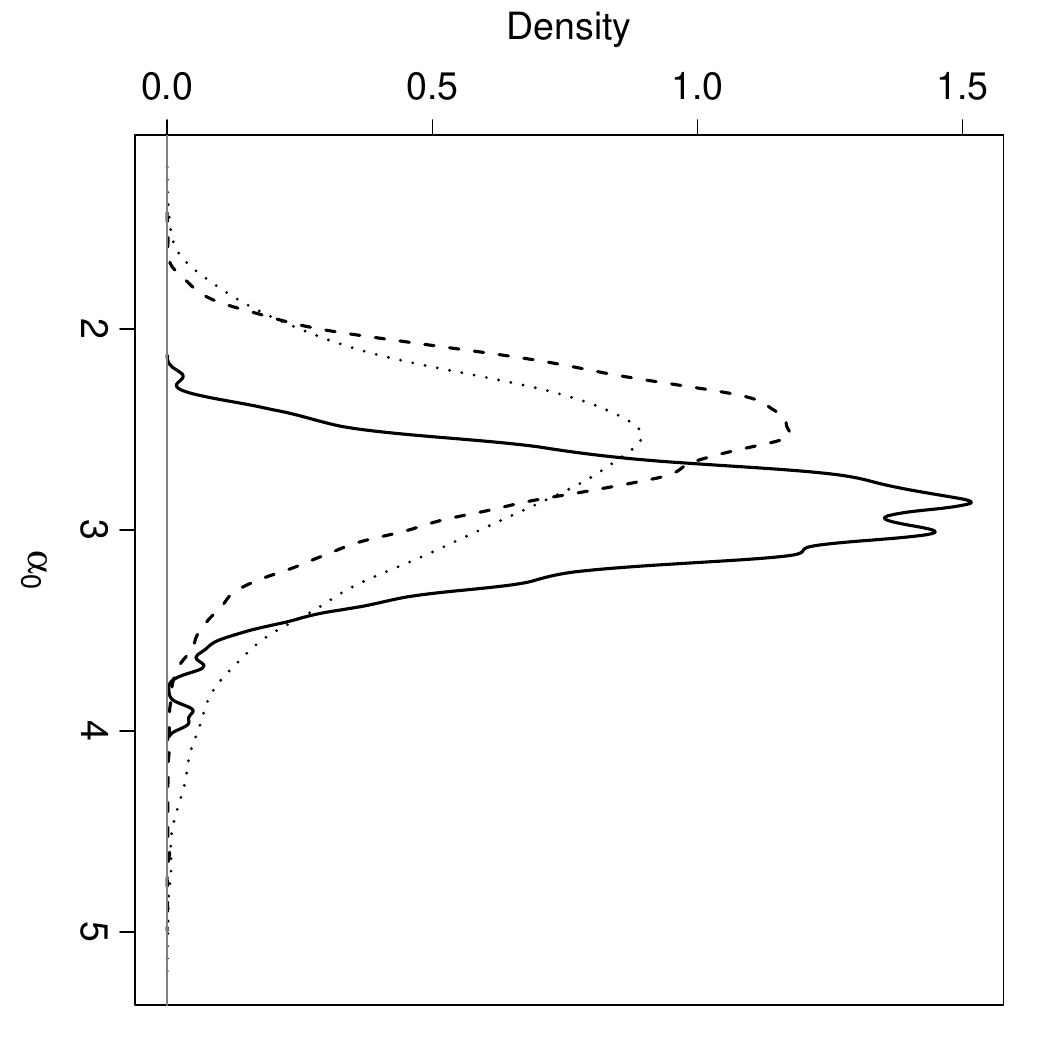}}}
      \end{subfigure}~\begin{subfigure}{.45\columnwidth}
    \resizebox{2in}{2in}{\includegraphics{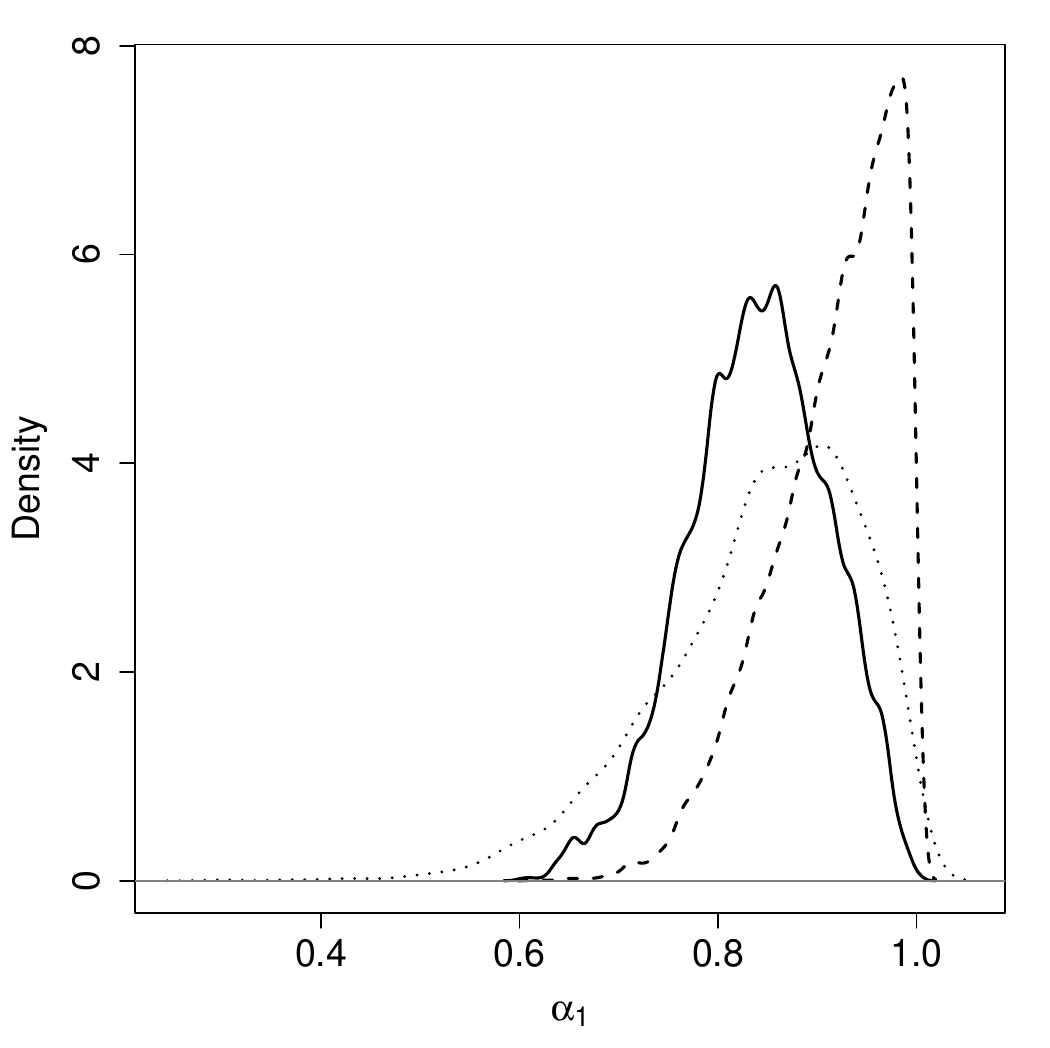}}
    \end{subfigure}\\
    \begin{subfigure}{.45\columnwidth}
      \resizebox{2in}{2in}{\rotatebox{90}{\includegraphics{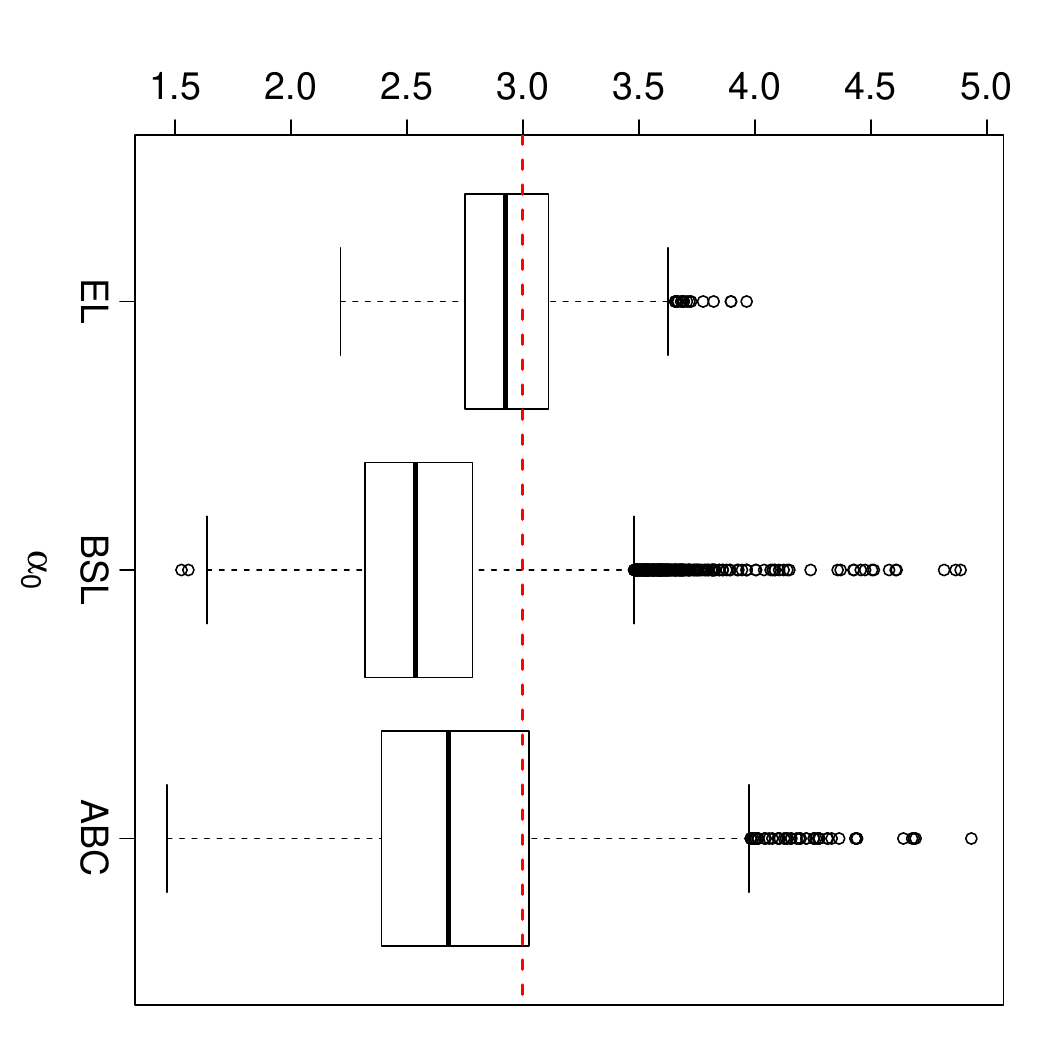}}}
      \end{subfigure}~\begin{subfigure}{.45\columnwidth}
    \resizebox{2in}{2in}{\rotatebox{90}{\includegraphics{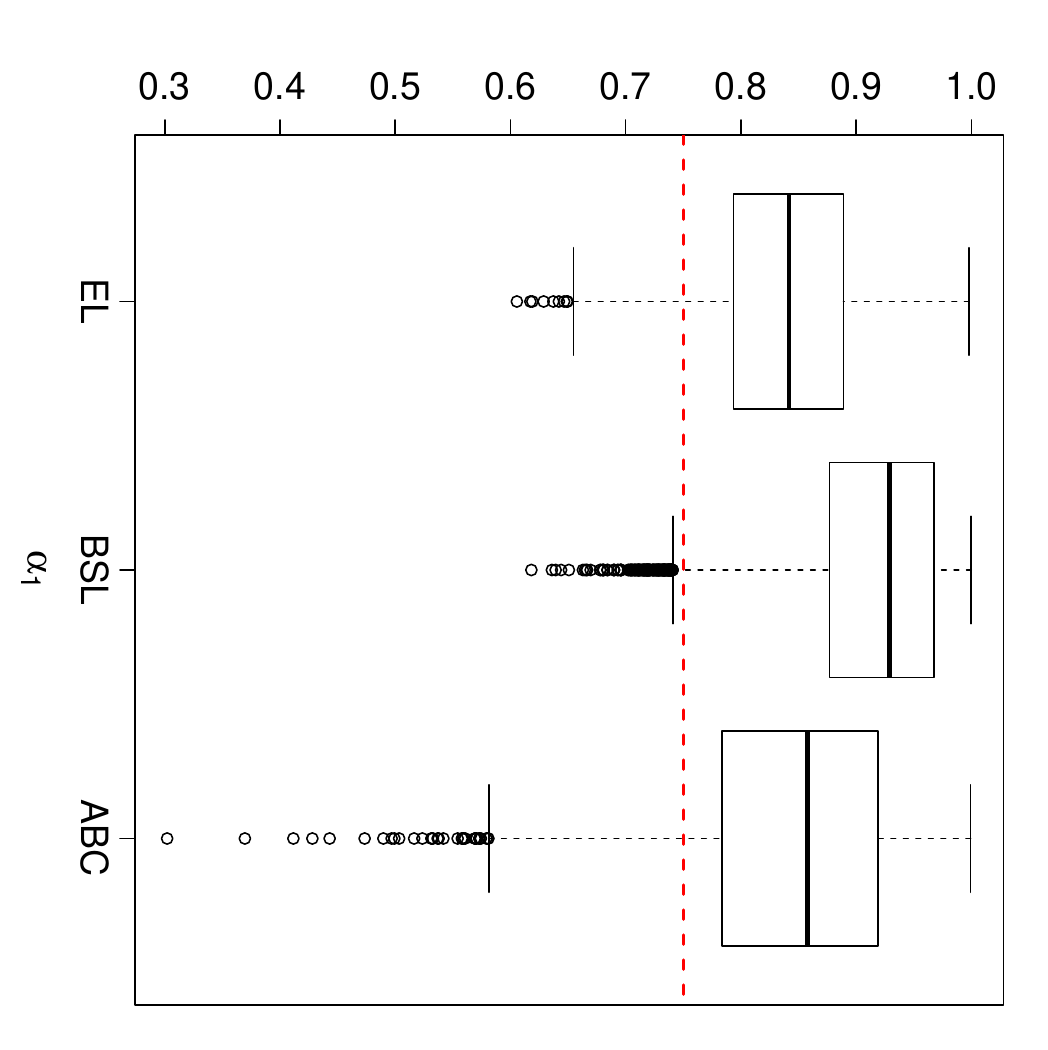}}}
    \end{subfigure}
\caption{Estimated marginal posterior densities of parameters $\alpha_0$ and $\alpha_1$ in the ARCH(1) model.  The top row
shows kernel density estimates (empirical likelihood ABC (solid), synthetic likelihood (dashed), rejection ABC (dotted)), while the bottom row shows boxplots of posterior samples.  In the boxplots, the horizontal
dotted lines show the true parameter values.}
\label{F3}
\end{center}
\end{figure}

Our summary statistics include the three quartiles of the absolute values of the data.  Since the data is dependent we also use the following summary statistic.  Let, for a fixed $i$ and for each $j$, $Y_{ij}(\theta)=X^2_{ij}(\theta)-\sum^n_{j=1}X^2_{ij}(\theta)/n$. 
Then for each $i=1$, $2$, $\ldots$, $m$, we define,
\[
\sa^{(4)}(X_i(\theta))=\frac{1}{n}\sum^n_{j=2}\left(1_{\{(Y_{ij}(\theta)\cdot Y_{i(j-1)}(\theta))\ge0\}}-1_{\{(Y_{ij}(\theta)\cdot Y_{i(j-1)}(\theta))<0\}}\right).
\] 
That is, $\sa^{(4)}$ is the difference between the proportion of the concordant and that of the discordant pairs between series $Y_i$ with its lag-$1$ version. Empirical evidence suggests that $\sa_4$ performs better than the usual lag-$1$ autocovariance of the series $X^2_i$.  The quartiles of the absolute values of the data provide some information about the marginal distribution.
 
Our observed data were of size $n=1000$, with $(\alpha_0,\alpha_1)=(3,0.75)$ and we used $m=50$ replicates for each
likelihood approximation for both empirical and synthetic likelihoods in Bayesian computations.  Marginal posterior densities were estimated for the parameters based on $50,000$ sampling iterations with $50,000$ iterations burn-in for both the synthetic likelihood and proposed empirical likelihood.
We compare these methods with the posterior obtained using rejection ABC with $1,000,000$ samples, a tolerance of $0.0025$, and linear regression adjustment.  
The estimated marginal posterior densities in Figure \ref{F3} for the proposed method are quite close to those obtained from the rejection ABC.  The synthetic likelihood produces quite different marginal posterior densities, 
especially for $\alpha_1$.  In this example 
the $\sa_4$ statistic is highly non-Gaussian, so the assumptions of the synthetic likelihood are not satisfied.  

\begin{figure}[t]
  \begin{center}
    \begin{subfigure}{.45\columnwidth}
      \resizebox{2in}{2in}{\includegraphics{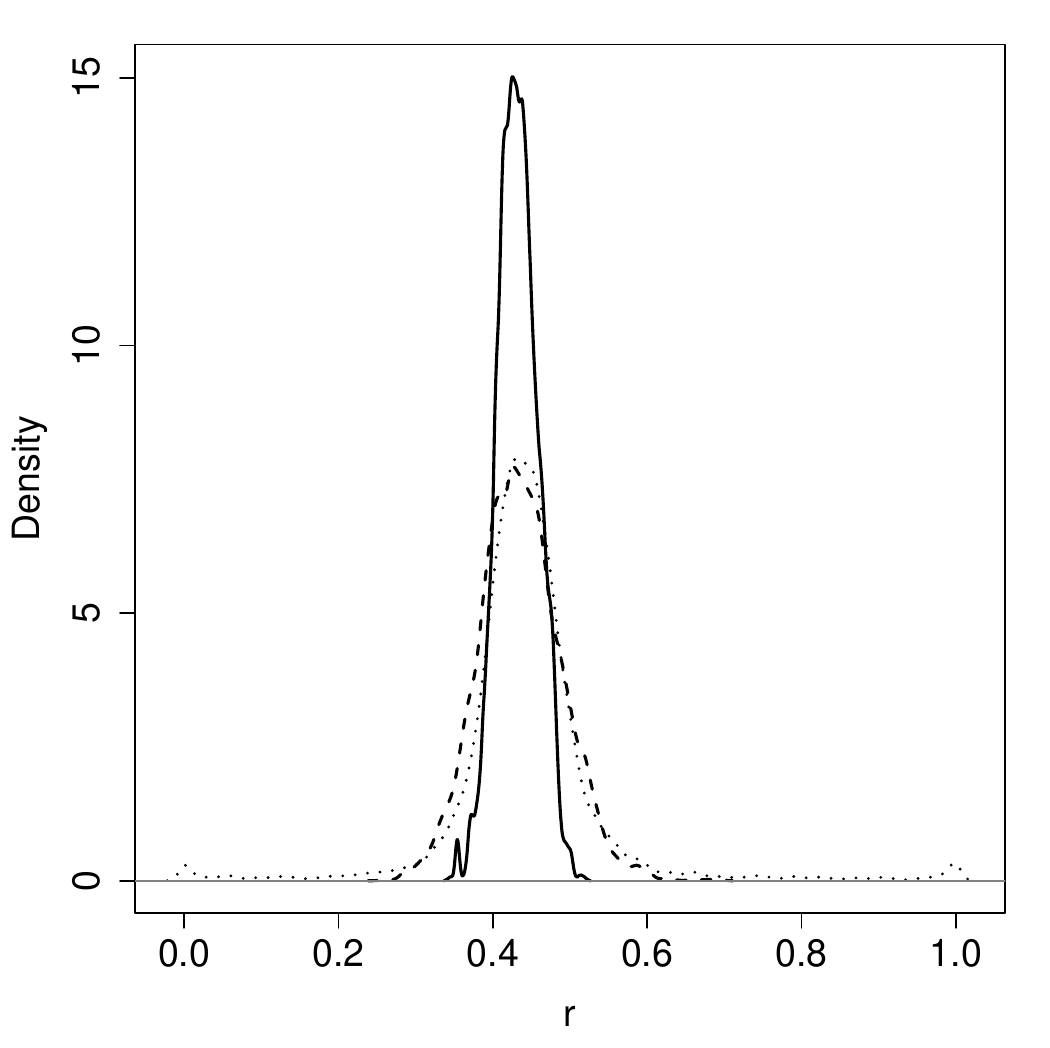}}
      \end{subfigure}~\begin{subfigure}{.45\columnwidth}
    \resizebox{2in}{2in}{\rotatebox{90}{\includegraphics{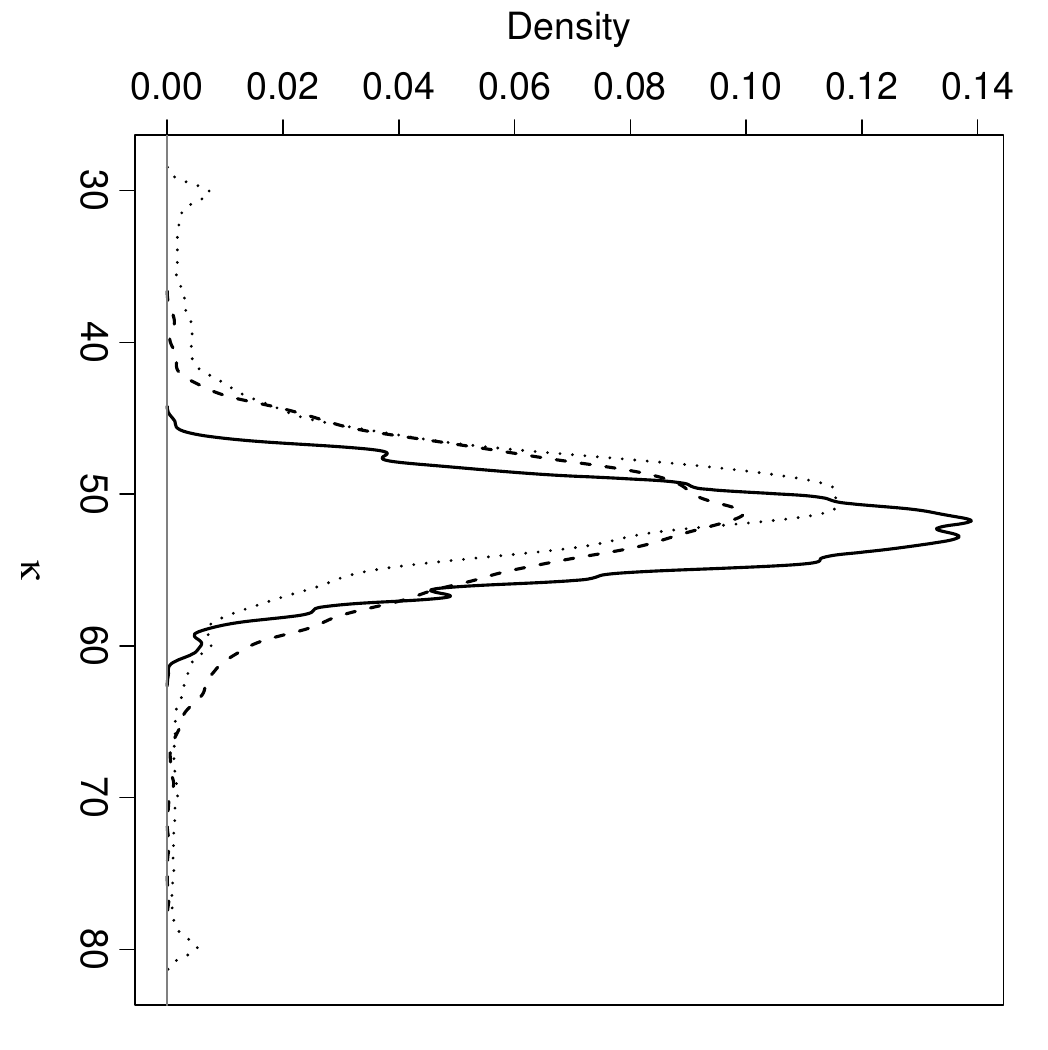}}}
    \end{subfigure}\\
    \begin{subfigure}{.45\columnwidth}
      \resizebox{2in}{2in}{\includegraphics{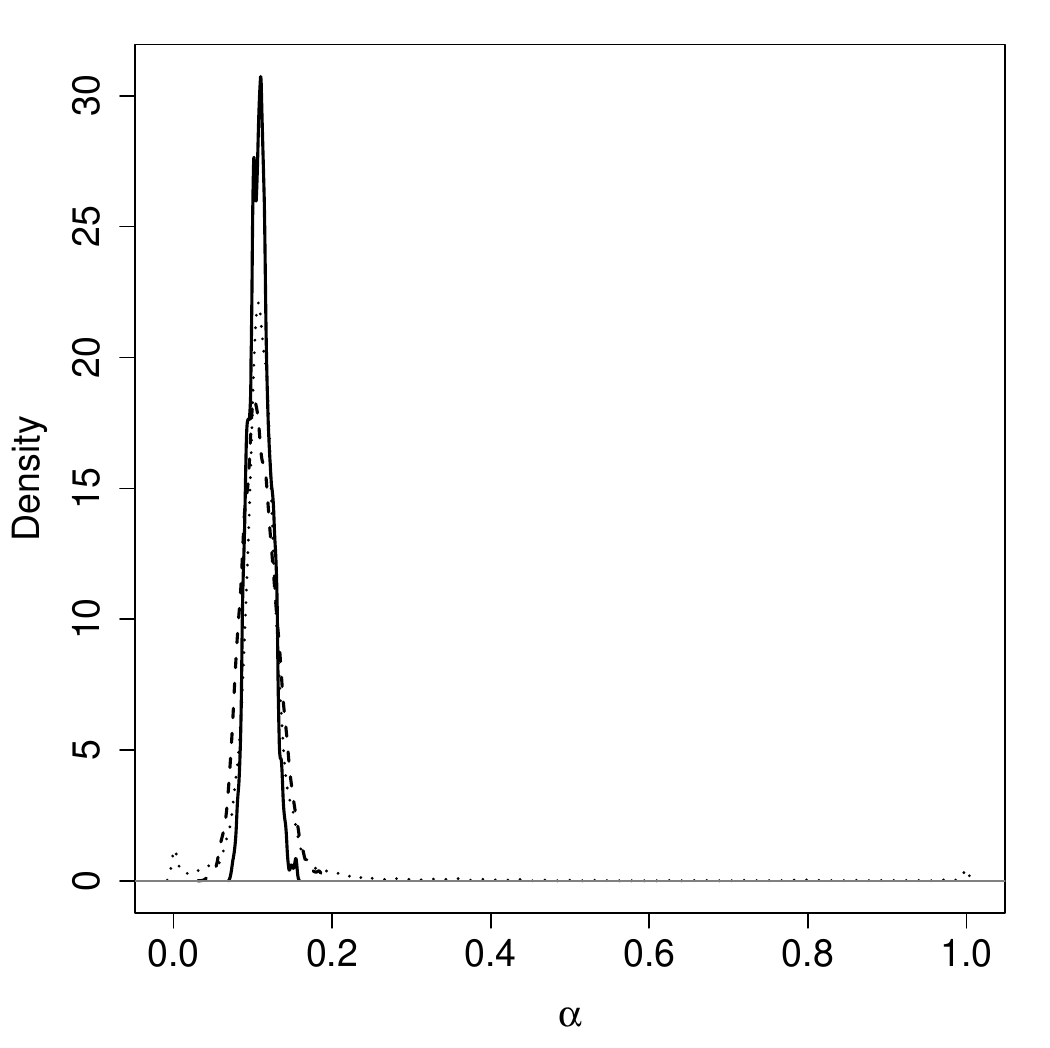}}
      \end{subfigure}~\begin{subfigure}{.45\columnwidth}
    \resizebox{2in}{2in}{\includegraphics{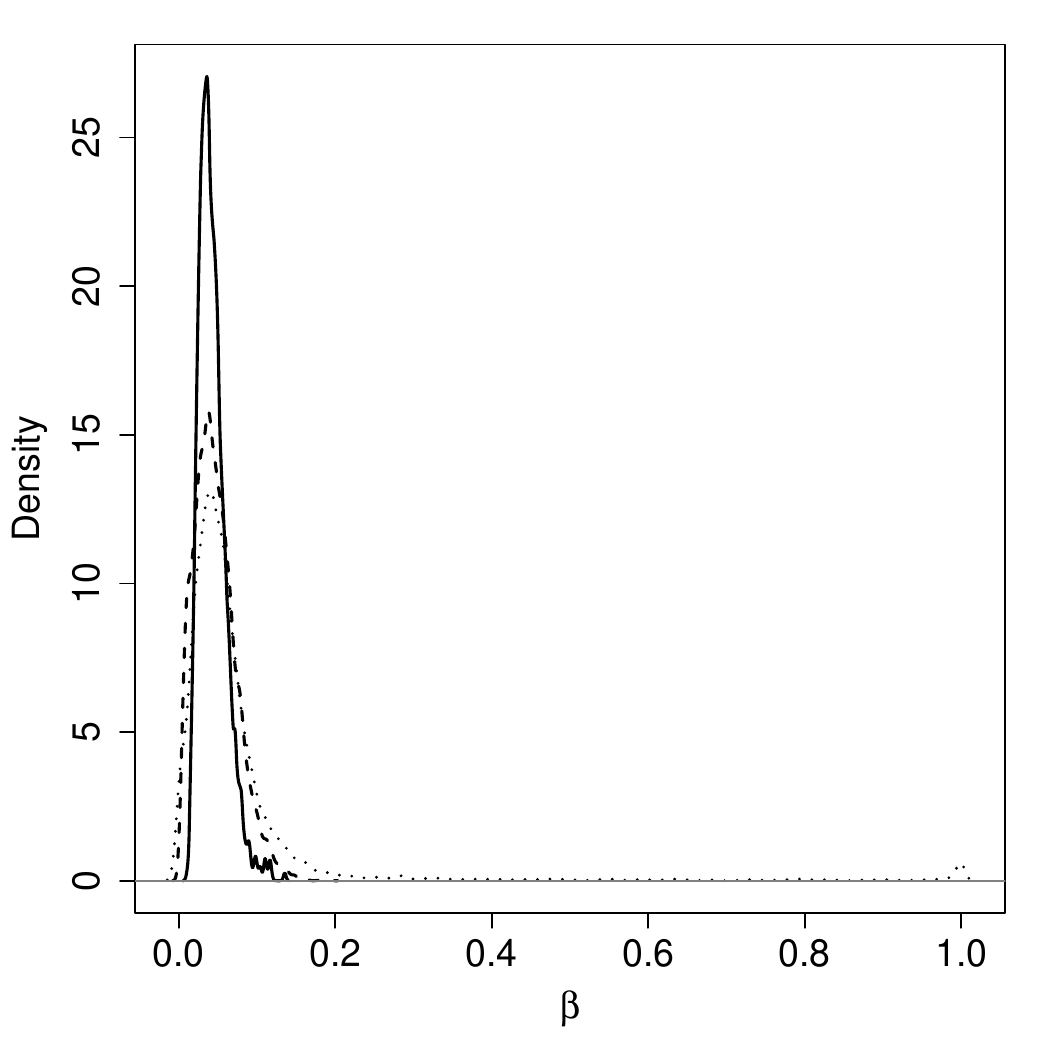}}
    \end{subfigure}
\caption{Estimated marginal posterior densities of parameters $r$, $\kappa$, $\alpha$, $\beta$ in the boom and bust model.  The figures
show kernel density estimates (empirical likelihood ABC (solid), synthetic likelihood (dashed), rejection ABC (dotted)) }
\label{fig:Fbb}
\end{center}
\end{figure}

\subsection{The simple recruitment, boom and bust model}\label{sec:bb}
The simple recruitment, boom and bust model is a discrete stochastic temporal model, primarily used to explain fluctiations in species population over time.  The dynamics is controlled by the parameter vector $\theta=(r,\kappa,\alpha,\beta)$. For $i=o$, $1$, $2$, $\ldots$, $m$, given $X_{ij}(\theta)=x_{ij}$, the distribution of $X_{i(j+1)}(\theta)$ is given by:
\[
X_{i(j+1)}(\theta)\sim\begin{cases}
Poisson(x_{ij}(1+r))+\epsilon_j&\text{if $x_{ij}\le\kappa$}\\
Binomial(x_{ij},\alpha)+\epsilon_j&\text{if $x_{ij}>\kappa$.}
\end{cases}
\]
Here $\epsilon_{j}\sim Poisson(\beta)$ distribution.  The sample paths rapidly cycle between the large and small non-negative integers.

For our simulation study we follow \citet{anNottDrovandi2020} and set $\theta_o=(0.4,50,0.09,0.05)$, and assume a prior of $U(0,1)\times U(30,80)\times U(0,1)\times U(0,1)$ on $\theta$.  We generated observations of length $n=200$, after discarding the first $50$ values to remove the transient phase of the process. 

For each $\theta$ we generated $m=40$ replications from the model.  The summary statistics used were, (a) the proportion of observations in the interval $(0,15)$, (b) the proportion of differences $X_{ij}(\theta)-X_{i(j-1)}(\theta)$ strictly larger than $2$ and (c) the proportion of log-amplitudes of the smoothed periodogram of the data lying in the interval $(5.120,6.278)$.  
The intervals were chosen rather judiciously partly based on the observed data $X_o$.

Our choice of summary statistics is targeted toward specifying the underlying data density rather than any particular parameter.  Clearly, a process can be specified by its probabilities of exceedance of certain thresholds.  The use of lagged differences is also natural for the same reason. The periodogram of the process is connected to its auto-covariance function.  
Thus, the exceedance probabilities of its log amplitude should put constraints on the auto-covariances between the successive observations.  Note that, none of the summary statistics are normally distributed in the case.

The density plots of observations sampled from the proposed ABCel (solid), synthetic likelihood (dashed), and the rejection ABC with a ridge regression adjustment with tolerance $0.001$ (dotted) are presented in Figure \ref{fig:Fbb}.  From the plot, it is clear that the rejection ABC has very heavy tails, which essentially cover the whole of the support of the priors and do not change if different tolerances or rejection methods are used.
The synthetic ABC is not expected to work well in this case. However, they seem to show a lighter tail than the Rejection ABC.  Among the three, the proposed ABCel posterior seems to be the most concentrated around true parameter values and seems to approximate the true posterior well. 
   

\subsection{Stereological data}  

Next, we consider an example concerning the modelling of diameters of inclusions (microscopic particles introduced
in the steel production process) measured from planar cross-sections in a block of steel.  The size of the largest
inclusion in a block is thought to be important for steel strength.  
We focus on an elliptical inclusion model due to \citet{bortot2007inference} here, which is an extension of the spherical model studied by \citet{anderson2002largest}.  Unlike the latter, the elliptical model does not have a tractable likelihood.



It is assumed that the inclusion centres follow a homogeneous Poisson process with a rate $\lambda$. 
For each inclusion, the three principal diameters of the ellipse 
are assumed independent of each other and of the process of inclusion centres.  
Given $V$, the largest diameter for a given inclusion, the two other principal diameters are determined by multiplying $V$ with two independent uniform $U[0,1]$ random variables.
The diameter $V$, conditional on exceeding a threshold value $v_0$ ($5\mu m$ in \citet{bortot2007inference}) is assumed to follow a generalised Pareto distribution:
\begin{displaymath}
\operatorname{pr}(V\leq v|V>v_0)=1-\left\{1+\frac{\xi (v-v_0)}{\sigma}\right\} _{+}^{-\frac{1}{\xi}}.
\end{displaymath}
The parameters of the model are given by $\theta=(\lambda,\sigma,\xi)$.
We assume independent uniform priors 
with ranges $(1,200)$, $(0,10)$ and $(-5,5)$ respectively. 
A detailed implementation of ABC for this example is discussed in \citet{erhardt+s15}.


\begin{figure}[t]
  \begin{center}
  \begin{subfigure}{.3\columnwidth}
    \resizebox{2in}{2.25in}{\rotatebox{90}{\includegraphics{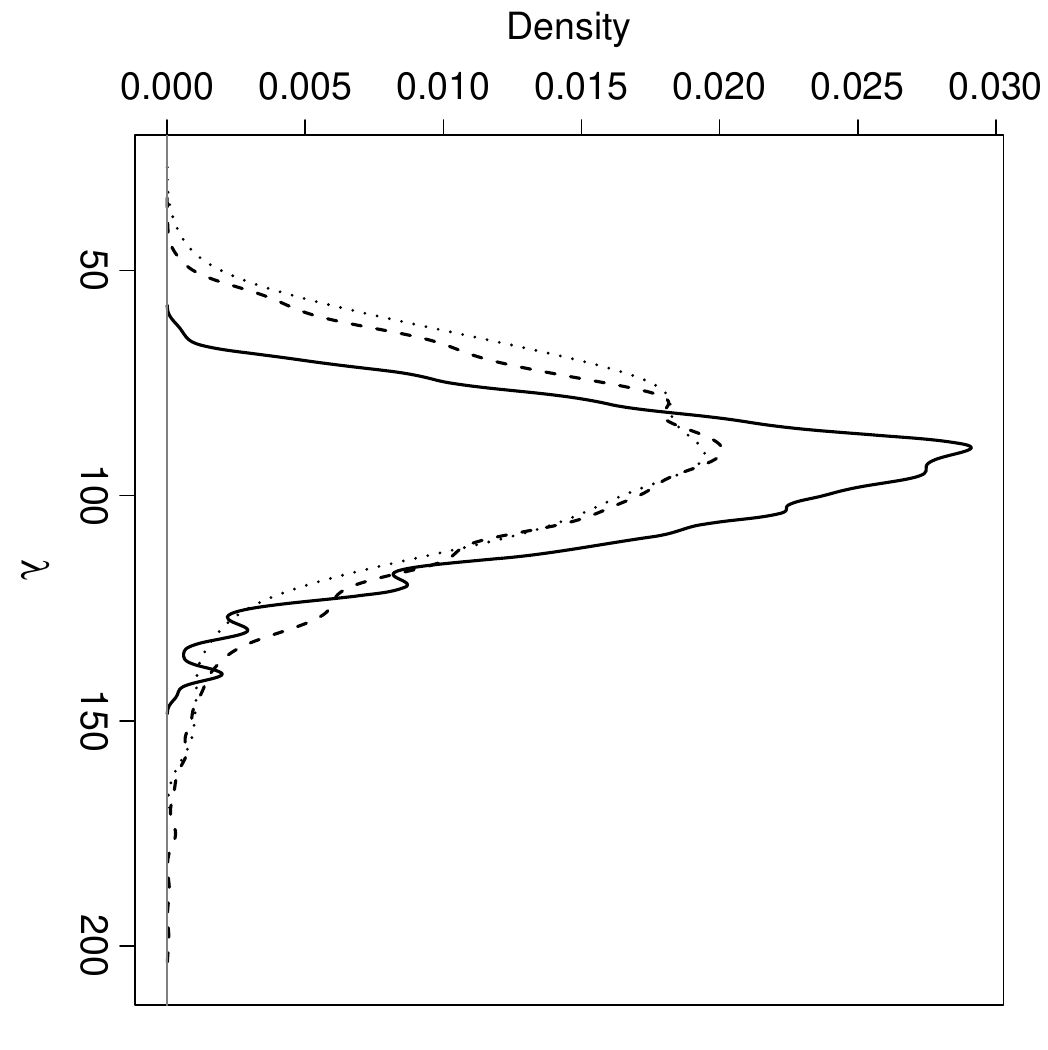}}}
  \end{subfigure}
  \begin{subfigure}{.3\columnwidth}
  \resizebox{2in}{2.25in}{\rotatebox{90}{\includegraphics{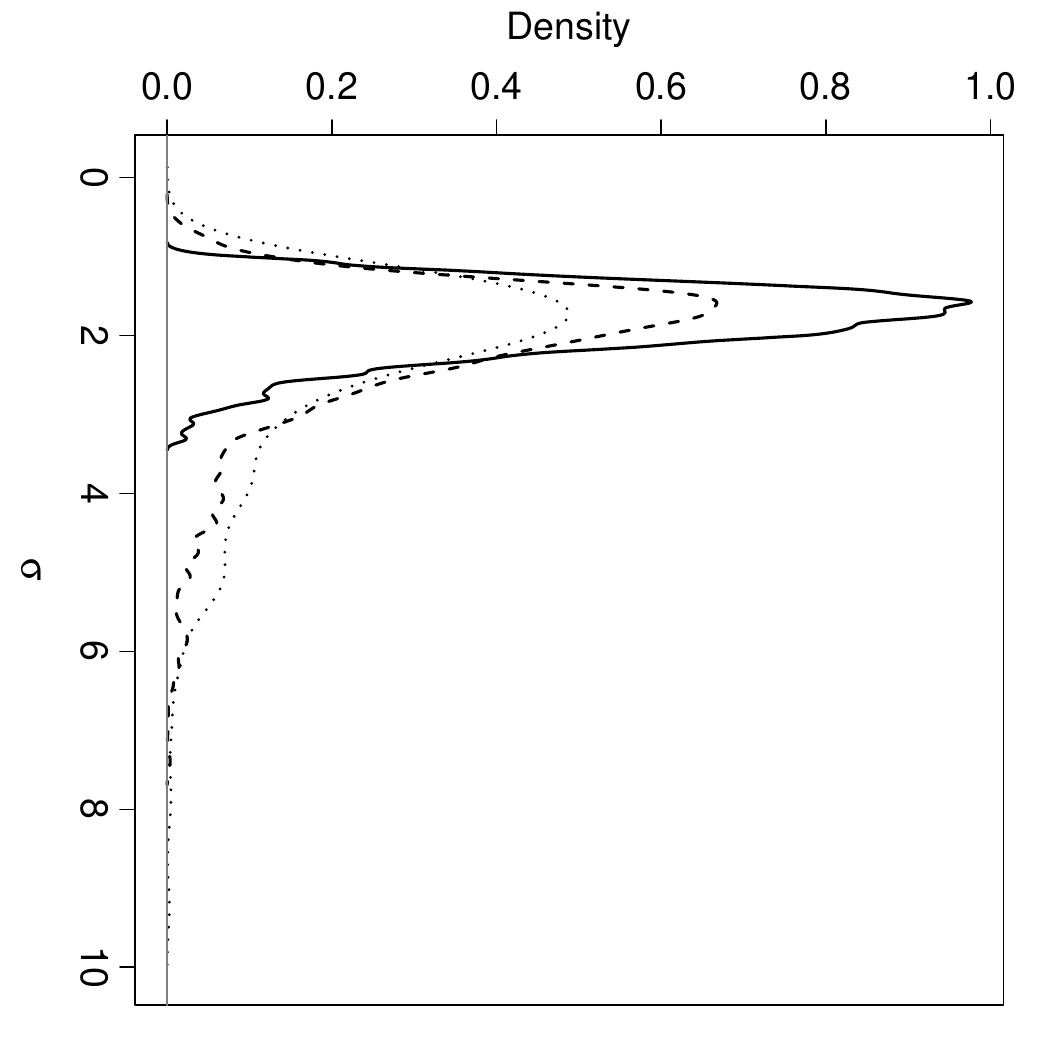}}}
  \end{subfigure}
  \begin{subfigure}{.3\columnwidth}
    \resizebox{2in}{2.25in}{\rotatebox{90}{\includegraphics{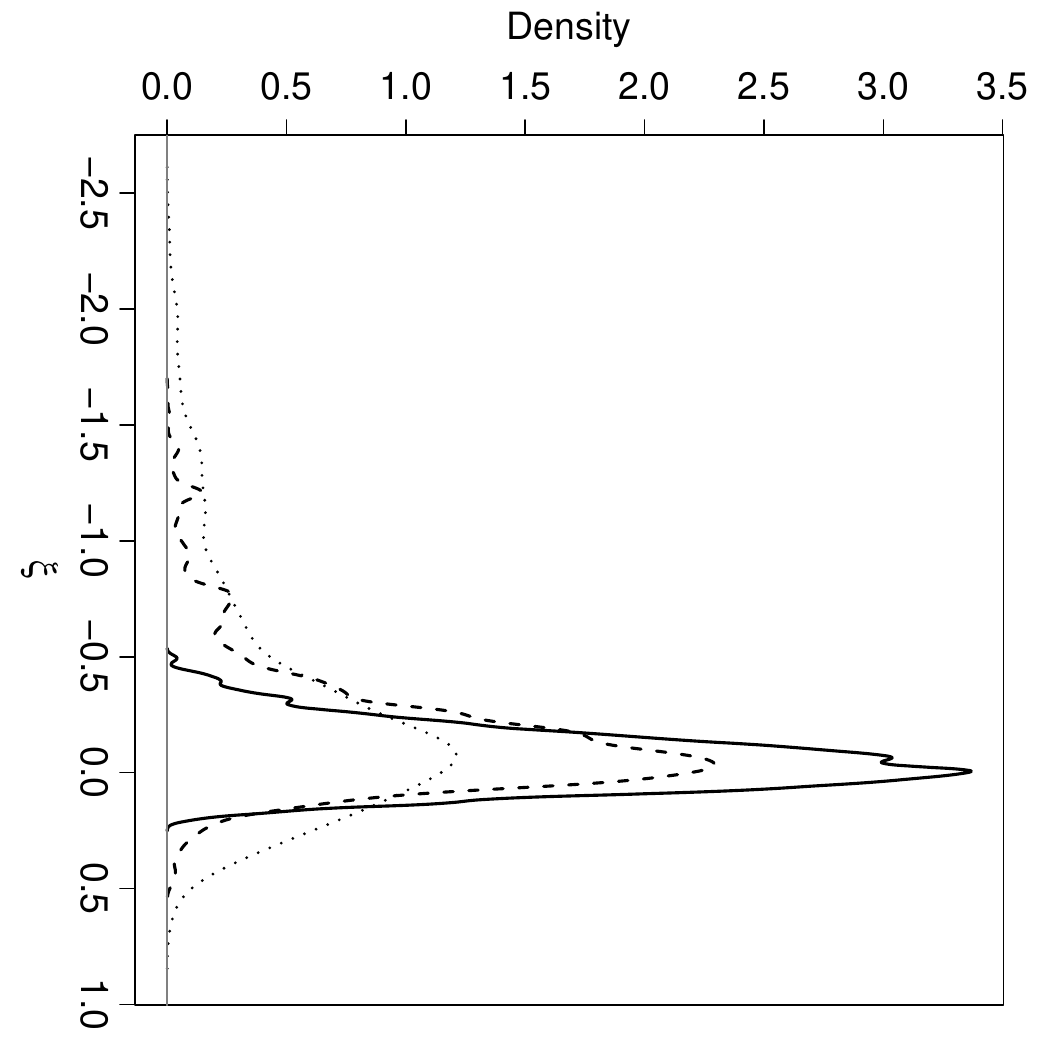}}}
    \end{subfigure}
\caption{Estimated marginal posterior densities of $\lambda$, $\sigma$ and $\xi$ using empirical likelihood ABC (solid), rejection ABC (dotted) and synthetic likelihood (dashed).}
\label{density_inclu}
  \end{center}
  \end{figure}

The observed data has $112$ entries, measuring the largest principal diameters of elliptical cross-sections of inclusions for a planar slice. The number of inclusions $L$ in each dataset generated from the model is random.  
The summary statistics used are $a)$ $(L-112)/100$, $b)$ the mean and $c)$ the median of the observed planar measurements, 
and $d)$ the proportion of planar measurements less than or equal to six (approximately the median for the observed data). 

Using the summary statistics described above, we compare the proposed empirical likelihood-based method with the synthetic likelihood ($m=25$ for both) and a rejection ABC algorithm with small tolerance ($0.00005$) and linear regression adjustment. 
The resulting estimated marginal posterior densities for $\lambda, \sigma, \xi$  are shown in
Figure~\ref{density_inclu}. The results for the proposed empirical likelihood-based method are more concentrated than the rejection ABC or the synthetic likelihood both of whom exhibit quite long tails. 

The chosen summaries mixed faster than those used in \citet{Pham2014note} and were comparable in speed to the synthetic likelihood.

\section{Discussion}

This article develops a new empirical likelihood-based easy-to-use approach to the ABC paradigm called ABCel.  For its implementation, the method only requires a set of summary statistics, their observed values, and the ability to simulate these summary statistics from a given blackbox or a suitable auxiliary model.  We first use a direct information projection to derive an analytic form for an approximation of the target posterior.  
Using this analytic expression, the best approximation to the target posterior is then obtained from a reverse information projection.  The procedure is implemented using a modified empirical likelihood.  By construction, the proposed empirical likelihood estimates the joint distribution of the observed and replicated summaries by minimising a cross-entropy over a large set of distributions.  
Furthermore, for appropriate summaries, at each value of the parameter, the above joint distribution is estimated by approximately equating the marginal densities of the observed and the replicated data.  The construction does not require any specification of a distance function, a tolerance or a bandwidth.  Neither does it assume any asymptotic distribution of the summary statistics.  
No constraints that are functions of the parameter and the data are required either.  We explore the properties of the proposed posterior both analytically and empirically.  The method is posterior consistent under reasonable conditions and shows good performance in simulated and real examples.


The modified empirical likelihood works with user-specified simple summaries like quantiles, moments or proportion of exceedance, that specify the underlying data density.  Summaries based on the spectral density of the data can also be conveniently used.  Even though no specific algorithm is so far available, our experience suggests appropriate simple summary statistics could easily be postulated from basic statistical considerations for almost all problems.

The number of replications depends in principle both on the number and the nature of the summary statistics used.  We make recommendations on the relative magnitudes of the number of replications and the sample size. 

Finally, empirical evidence as seen in the Q-Q plots in Sections \ref{Ssec:graph} and \ref{Ssec:qq} of the supplement, suggest that under suitable conditions, the proposed posteriors would asymptotically converge to a normal density.  The conditions under which such convergences would hold is a natural question for further investigation.

\input{abcelArxivSupp.tex}

\setlength{\bibsep}{0pt plus 0.3ex}


\input{abcelArxiv.bbl}
\end{document}

%% file: abcelArxivSupp.tex
\appendix

In this supplement we present the proofs of the Lemmas, Theorems and Remarks in the main article. We also discuss estimation of differential entropy, and methods to sample observations from an ABCel posterior.  
Illustrative applications on traditional problems like parameter estimation from a g-and-k model are presented.  We also employ the ABCel posterior to estimate the edge probability of a Erd\"{o}s-Renyi random graph. 
Finally, we present the QQ-plots of the samples drawn from the various posteriors with normal densities to demonstrate the possible asymptotic normality of the ABCel posterior.

\section{Proofs of Results in Section \ref{sec:just}.}
\noindent{\it Proof of Theorem 1.}
The proof proceeds by expanding the Kullback-Leibler divergence
\[
D_{KL}\left[q(\theta,\gt)\mid\mid f(\theta,\gt\mid \gn)\right]
\]
when $q(\theta,\gt)=q^{\prime}(\theta)f_0(\gt\mid \theta)$.

For a $f\in\mathcal{F}$, suppose $f(\gn)$ is the marginal distribution of $\gn$.
It is well known that \citep{ormerodWand2010,faesOrmerodWand2011} the so called log evidence i.e. $\log f(\gn)$ can be expressed as:
\begin{equation}\label{eq:fund}
\log f(\gn)=D_{KL}\left[q(\theta,\gt)\mid\mid f(\theta,\gt\mid \gn)\right)]+\int q(\theta,\gt)\log\left(\frac{f(\theta,\gt,\gn)}{q(\theta,\gt)}\right)d\gt~ d\theta.
\end{equation}

For the convenience of notation, for an $f\in\mathcal{F}$ we define:

\begin{align}
f^{\prime\prime}(\theta,\gn)&=\frac{\exp(\et_{\gt\mid\theta}\left[\log f(\theta,\gt,\gn)\right])}{\int\exp(\et_{\gt\mid t}\left[\log f(t,\gt,t^{\prime})\right])dtdt^{\prime}},~~f^{\prime\prime}(\gn)=\int f^{\prime\prime}(\theta,\gn)d\theta~\mbox{and}\nonumber\\
 f^{\prime\prime}(\theta\mid \gn)&=f^{\prime\prime}(\theta,\gn)/f^{\prime\prime}(\gn).\nonumber
\end{align}
 
By substituting the expression of $q(\theta,\gt)\in \mathcal{Q}^{\prime}$ in \eqref{eq:fund} we get: 

\begin{align}
~D_{KL}\left[q(\theta,\gt)\mid\mid f(\theta,\gt\mid \gn)\right)]=\log f(\gn)+\int q^{\prime}(\theta) f_0(\gt\mid \theta)\log f_0(\gt\mid \theta)d\gt d\theta&\nonumber\\
-\hfill\int q^{\prime}(\theta)\left\{\int\log f(\theta,\gt,\gn) f_0(\gt\mid \theta)d\gt-\log q^{\prime}(\theta)\right\}d\theta&\nonumber
\end{align}
\begin{align}
=&\log f^{\prime\prime}(\gn)-\int q^{\prime}(\theta)\log \left(\frac{\exp(\et_{\gt\mid\theta}\left[\log f(\theta,\gt,\gn)\right])}{q^{\prime}(\theta)}\right)d\theta-\int \htr_{\gt\mid\theta}(\theta)q^{\prime}(\theta)d\theta +\log \left(\frac{f(\gn)}{f^{\prime\prime}(\gn)}\right)&\nonumber\\
=&\log f^{\prime\prime}(\gn)-\int q^{\prime}(\theta)\left\{\log \left(\frac{f^{\prime\prime}(\theta,\gn)}{q^{\prime}(\theta)}\right)-\log\int \exp(\et_{\gt\mid t}\left[\log f(t,\gt,t^{\prime})\right])dt dt^{\prime}\right\}d\theta& \nonumber\\
&\hspace{.3\columnwidth}-\int \htr_{\gt\mid\theta}(\theta)q^{\prime}(\theta)d\theta+\log \left(\frac{f(\gn)}{f^{\prime\prime}(\gn)}\right)&\label{eq:varAprx}
\end{align} 

Similar to \eqref{eq:fund} one can show that:
\[
\log f^{\prime\prime}(\gn)=\int q^{\prime}(\theta)\log \left(\frac{f^{\prime\prime}(\theta,\gn)}{q^{\prime}(\theta)}\right)d\theta+D_{KL}\left[q^{\prime}(\theta)\mid\mid  f^{\prime\prime}(\theta\mid \gn)\right],
\]
where second addendum is the Kullback-Leibler divergence between the densities $q^{\prime}(\theta)$ and $f^{\prime\prime}(\theta\mid \gn)$.  Moreover, the third addendum in \eqref{eq:varAprx} depends on the hyper-parameters of $\pi(\theta)$ and thus independent of $\theta$.  Suppose we denote $C^{\prime}=\log\int \exp(\et_{\gt\mid t}\left[\log f(t,\gt,t^{\prime})\right])dt dt^{\prime}$. 

By substituting the above result in \eqref{eq:varAprx} and from \eqref{eq:fund} we get:
\begin{align}
~&D_{KL}\left[q(\theta,\gt)\mid\mid f(\theta,\gt\mid \gn)\right)]=\log f(\gn)-\int q^{\prime}(\theta)f_0(\gt\mid \theta)\log\left(\frac{f(\theta,\gt,\gn)}{q^{\prime}(\theta)f_0(\gt\mid \theta)}\right)d\gt~ d\theta\nonumber\\
=&D_{KL}\left[q^{\prime}(\theta)\mid\mid  f^{\prime\prime}(\theta\mid \gn)\right]-\int \htr_{\gt\mid\theta}(\theta)q^{\prime}(\theta)d\theta-C^{\prime}+\log \left(\frac{f(\gn)}{f^{\prime\prime}(\gn)}\right)\label{eq:varAprx2}
\end{align}

Now by expanding the first two addenda in \eqref{eq:varAprx2} we get:
\begin{align}
~&D_{KL}\left[q^{\prime}(\theta)\mid\mid  f^{\prime\prime}(\theta\mid \gn)\right]-\int \htr_{\gt\mid\theta}(\theta)q^{\prime}(\theta)d\theta=\int q^{\prime}(\theta)\left\{\log\left(\frac{q^{\prime}(\theta)}{f^{\prime\prime}(\theta\mid \gn)}\right)-\htr_{\gt\mid\theta}(\theta)\right\}d\theta\nonumber\\
=&\int q^{\prime}(\theta)\left\{\log\left(\frac{q^{\prime}(\theta)}{f^{\prime\prime}(\theta\mid \gn)exp(\htr_{\gt\mid\theta}(\theta))}\right)\right\}d\theta\nonumber\\
=&\int q^{\prime}(\theta)\left\{\log\left(\frac{q^{\prime}(\theta)}{f^{\prime}(\theta\mid \gn)}\right)-\left(\log\int f^{\prime\prime}(t\mid \gn)\exp(\htr_{\gt\mid t}(t))dt\right) \right\}d\theta\label{eq:varAprx3}
\end{align}
The first addendum in \eqref{eq:varAprx3} is the Kullback-Leibler divergence between $q^{\prime}$ and $f^{\prime}(\theta\mid \gn)$. 
The second addendum is a function of $\gn$ and is independent of $\theta$.  By denoting it by $C(\gn)$ and collecting the terms from \eqref{eq:varAprx2} and \eqref{eq:varAprx3} we get:
\begin{equation}\label{eq:varAprx4}
D_{KL}\left[q(\theta,\gt)\mid\mid f(\theta,\gt\mid \gn)\right]=D_{KL}\left[q^{\prime}(\theta)\mid\mid f^{\prime}(\theta\mid \gn)\right]-C(\gn)-C^{\prime}+\log \left(\frac{f(\gn)}{f^{\prime\prime}(\gn)}\right).
\end{equation}

Note that, the R.H.S. of the equation \eqref{eq:varAprx4} is non-negative for all $q^{\prime}\in\mathcal{Q}_{\Theta}$. Furthermore, only the first addendum depends on $q^{\prime}$, which is also non-negative, with equality holding iff $q^{\prime}(\theta)=f^{\prime}(\theta\mid \gn)$.  
This implies the R.H.S. of \eqref{eq:varAprx4} attains its minimum at $q^{\prime}(\theta)=f^{\prime}(\theta\mid \gn)$.  So, it clearly follows that the information projection of $f(\theta,\gt\mid \gn)$ is given by $f^{\prime}(\theta\mid \gn)f_0(\gt\mid \theta)$.\hfill $\square$

\bigskip
\noindent{\it Proof of Theorem \ref{thm:error}.} From the LHS of \ref{stt:main} we get:
\begin{align}
~&\log f\left(\theta,\gn\right)-\left\{\et_{\gt\mid \theta}\left[\log f\left(\theta,\gt,\gn\right)\right]+\htr_{\gt\mid \theta}(\theta)\right\}\nonumber\\
=&\log f\left(\theta,\gn\right)-\int f_0\left(\gt|\theta\right)\log f\left(\gt|\gn,\theta\right)d\gt-\int f_0\left(\gt|\theta\right)\log f\left(\theta,\gn\right)d\gt+\htr_{\gt\mid \theta}(\theta)\nonumber\\
&=\int f_0\left(\gt\mid\theta\right)\log \left(\frac{f_0\left(\gt\mid\theta\right)}{f\left(\gt\mid \gn,\theta\right)}\right)d\gt=D_{KL}\left[f_0\left(\gt\mid\theta\right)\mid\mid f\left(\gt\mid \gn,\theta\right)\right].\nonumber
\end{align}
Rest of the Theorem follows from above.\hfill$\square$

\section{Proofs of Results in Section \ref{sec:postcons}.}
\noindent{\it Proof of Lemma \ref{lem:1}.} We show that for every $\epsilon>0$, there exists $n_0=n_0(\epsilon)$ such that for any $n\ge n_0$ for all $\theta\in\Theta_n$ the maximisation problem in \eqref{eq:w2} is feasible with probability larger than $1-\epsilon$.  

By assumption, for each $\theta$, random vectors $\xi^{(n)}_i(\theta)$ are i.i.d., put positive mass on each orthant and supremum of their lengths in each orthant diverge to infinity with $n$.  The random vectors $\left\{\xi^{(n)}_i(\theta)-\xi^{(n)}_o(\theta_o)\right\}$ will inherit the same properties.  
That is, there exists integer $n_0$, such that for each $n\ge n_0$, the convex hull of the vectors $\left\{\xi^{(n)}_i(\theta)-\xi^{(n)}_o(\theta_o)\right\}$, $i=1$, $\ldots$, $m(n)$, would contain the unit sphere with probability larger than $1-\epsilon/2$.  


We choose an $n\ge n_0$ and a $\theta\in\Theta_n$. For this choice of $\theta$:
\begin{align}
h^{(n)}_i(\theta,\theta_o)=&b_n\left\{\sm(\theta)-\sm(\theta_o)\right\}+\xi^{(n)}_i(\theta)-\xi^{(n)}_o(\theta_o)=c_n(\theta)+\xi^{(n)}_i(\theta)-\xi^{(n)}_o(\theta_o),\nonumber
\end{align}
where, $\mid\mid\sm(\theta)-\sm(\theta_o)\mid\mid\le b^{-1}_n$.  That is, $\mid\mid c_n(\theta)\mid\mid\le 1$.  
Now, since $-c_n(\theta)$ is in the convex hull of the vectors $\left\{\xi^{(n)}_i(\theta)-\xi^{(n)}_o(\theta_o)\right\}$, $i=1$, $\ldots$, $m(n)$, with probability larger than $1-\epsilon/2$, there exists weights $w\in\Delta_{m(n)-1}$ such that,

\[
-c_n(\theta)=\sum^{m(n)}_{i=1}w_i\left\{\xi^{(n)}_i(\theta)-\xi^{(n)}_o(\theta_o)\right\}.
\]
Now it follows that for the above choice of $w$ that
\[
\sum^{m(n)}_{i=1}w_ih^{(n)}_i(\theta,\theta_o)=c_n(\theta)+\sum^{m(n)}_{i=1}w_i\left\{\xi^{(n)}_i(\theta)-\xi^{(n)}_o(\theta_o)\right\}=0,
\] 
which shows that the problem in \eqref{eq:w2} is feasible.\hfill $\square$

\bigskip

\noindent{\it Proof of Lemma \ref{lem:2}.} Let $\epsilon$ be as in the statement.  By assumption (A1), for some $\delta>0$,  $\mid\mid\sm(\theta)-\sm(\theta_o)\mid\mid>\delta$ for all $\theta$ with $\mid\mid\theta-\theta_o\mid\mid >\epsilon$.

Consider $\eta>0$. We show that there exists $n_0=n_0(\eta)$ such that for any $n\ge n_0$, the constrained maximisation problem in \eqref{eq:w2} is not feasible for all $\mid\mid\theta-\theta_o\mid\mid >\epsilon$, with probability larger than $1-\eta$. 


Let if possible $w\in\Delta_{m(n)-1}$ be a feasible solution.  Hence we get:

\begin{align}
0=&\sum^{m(n)}_{i=1}w_ih^{(n)}_i(\theta,\theta_o)=\sum^{m(n)}_{i=1}w_i\left\{\s{X^{(n)}_i(\theta)}-\s{X^{(n)}_o(\theta_o)}\right\}\nonumber\\
=&\left\{\sm^{(n)}(\theta)-\sm^{(n)}(\theta_o)\right\}+\left\{\sum^{m(n)}_{i=1}w_i\xi^{(n)}_i(\theta)\right\}-\xi^{(n)}_o(\theta_o),\nonumber
\end{align}
so that 
\begin{equation*}
-b_n\left\{\sm(\theta)-\sm(\theta_o)+o(1)\right\}=\sum^{m(n)}_{i=1}w_i\xi^{(n)}_i(\theta)-\xi^{(n)}_o(\theta_o).
\end{equation*}

By dividing both sides by $b_n$ we get:
\begin{equation}\label{eq:cons}
-\left\{\sm(\theta)-\sm(\theta_o)\right\}=\sum^{m(n)}_{i=1}w_i\left\{\frac{\xi^{(n)}_i(\theta)}{b_n}-\frac{\xi^{(n)}_o(\theta_o)}{b_n}\right\}-o(1).
\end{equation}
Now, $\mid\mid\xi^{(n)}_o(\theta_o)\mid\mid/b_n\le \sup_{i\in\{o,1,2\ldots,m(n)\}}\mid\mid\xi^{(n)}_o(\theta_o)\mid\mid/b_n$ and
\begin{align}
\left|\left|\sum^{m(n)}_{i=1}w_i\frac{\xi^{(n)}_i(\theta)}{b_n}\right|\right|\le\sum^{m(n)}_{i=1}w_i\frac{\mid\mid\xi^{(n)}_i(\theta)\mid\mid}{b_n}\le\sup_{i\in\{o,1,2\ldots,m(n)\}}\frac{\mid\mid\xi^{(n)}_i(\theta)\mid\mid}{b_n}.\nonumber
\end{align}
That is, by assumption (A3), there exists $n_0(\eta)$ such that for any $n\ge n_0$, the RHS of \eqref{eq:cons} is less than $\delta$ for all $\theta\in B(\theta_o,\epsilon)$, with probability larger than $1-\eta$.  However, $\mid\mid\sm(\theta)-\sm(\theta_o)\mid\mid>\delta$. 
We arrive at a contradiction.  Thus the problem is infeasible for every $\theta\in B(\theta_o,\epsilon)^C$ with probability larger than $1-\eta$.\hfill $\square$

\medskip

\noindent{\it Proof of Theorem \ref{thm:1}.}  Let $g(\theta)$ be a continuous, bounded function. We choose an $\epsilon>0$. Then by Lemma \ref{lem:2}, there exists $n(\epsilon)$, such that for any $n>n(\epsilon)$ and $\theta\in B\left(\theta_o,\epsilon\right)^C$, 
$l_n(\theta)=0$ and by definition \eqref{eq:mpost} the posterior $\hat{\Pi}_n\left(\theta\mid \s{X_o(\theta_o)}\right)=0$.  That is for any $n>n(\epsilon)$, 
\begin{align}
~&\int_{\Theta}g(\theta)\hat{\Pi}_n\left(\theta\mid \s{X_o(\theta_o)}\right)d\theta=\int_{B\left(\theta_o,\epsilon\right)}g(\theta)\hat{\Pi}_n\left(\theta\mid \s{X_o(\theta_o)}\right)d\theta\nonumber\\
=&\int_{B\left(\theta_o,\epsilon\right)}\left\{g(\theta)-g(\theta_o)\right\}\hat{\Pi}_n\left(\theta\mid \s{X_o(\theta_o)}\right)d\theta +g(\theta_o)\int_{B\left(\theta_o,\epsilon\right)}\hat{\Pi}_n\left(\theta\mid \s{X_o(\theta_o)}\right)d\theta.\nonumber
\end{align}

Since the function $g(\theta)$ is bounded and continuous at $\theta_o$, the first term is negligible.  Furthermore, $\int_{B\left(\theta_o,\epsilon\right)}\hat{\Pi}_n\left(\theta\mid \s{X_o(\theta_o)}\right)d\theta=1$.  This implies the integral converges to $g(\theta_o)$.  This shows, the posterior converges weakly to $\delta_{\theta_o}$.  \hfill $\square$

\section{Details of the Remarks in Section \ref{sec:m}}

Using the notations introduced above, when $r=1$, i.e. there is only one constraint present, 
under conditions similar to those described above, it can be shown that, \citep[Theorem $3.4$]{ghosh2019empirical} for any $\theta\in\Theta$:
\begin{align}\label{eq:ms}
l_m(\theta)&\coloneqq\frac{1}{m}\sum^m_{i=1}\log(\hat{w}(\theta))=-\frac{1}{\mathcal{M}_m(\theta)}\left|E_{\s{X^{(n)}_1(\theta)}\mid\theta}[\s{X^{(n)}_1(\theta)}]-\s{X^{(n)}_o(\theta_o)}]\right|(1+o_p(1)),\nonumber\\
& =-\frac{b_n}{\mathcal{M}_m(\theta)}\left|(\sm(\theta)-\sm(\theta_o)+o(1))-\frac{\xi^{(n)}_o(\theta_o)}{b_n} \right|(1+o_p(1)),  
\end{align}

\noindent{\it Details of Remark \ref{rem:baycon1}.}
In order to ensure the first condition, suppose $\theta\ne\theta_o$, and as  $m,n\rightarrow\infty$, and in \eqref{eq:ms}, $b_n/\mathcal{M}_m(\theta)$ diverges.  Since by assumption (A3), as $m,n\rightarrow\infty$, $\sup_{i\in\{o,1,2,\ldots,m\}}$ $|\xi^{(n)}_o(\theta_o)|/b_n$ $\rightarrow 0$, 
in probability, uniformly over $\theta$, and by assumption (A1), $||\sm(\theta)-\sm(\theta_o)||>0$, for each $\theta\ne\theta_o$, the R.H.S. of \eqref{eq:ms} diverges to $-\infty$.  So $\exp(l_m(\theta))$ converges to zero.  
That is, an upper bound of the rate of growth of $m$ can thus be obtained by inverting the relation $b_n>\mathcal{M}_m(\theta)$.

Depending on the distribution of $\xi^{(n)}_o$, $m$ can be much larger than $n$.
For example, if $\xi^{(n)}_o$ follows a normal distribution with mean zero and variance $\sigma^2_o$, $b_n=\sqrt{n}$ and $\mathcal{M}_m(\theta)$ is of the order $\sigma_o\sqrt{2\log(m)}$, which allows an upper bound of $m$ as large as $\exp(n/(2\sigma^2_o))$.   

\noindent{\it Details of Remark \ref{rem:baycon2}.}
Similar to the argument for the upper bound, for Bayesian consistency $l_m(\theta_o)$ cannot diverge to $-\infty$.  There exists a constant $C_1>0$ such that, $l_m(\theta)>-C_1$ with a high probability.

For \eqref{eq:m}, it follows that when $\theta=\theta_o$:
\begin{equation}\label{eq:thetao}
l_m(\theta_o)=-\frac{\mid\xi^{(n)}_o(\theta_o)\mid}{\mathcal{M}_m(\theta_o)}(1+o_p(1)).
\end{equation}
For simplicity of presentation, we also suppose $\xi^{(n)}_o(\theta_o)$ is a $N(0,\sigma_o^2)$ variable.    

For a fixed $C_1>0$, we first compute $Pr[l_m(\theta_o)\le -C_1]$.  Using the tail bound for a $N(0,\sigma^2_o)$ random variables we get,
\begin{align}
~&Pr[l_m(\theta_o)\le -C_1]= Pr\left[-\frac{\left|\xi^{(n)}_o(\theta_o)\right|}{\mathcal{M}_m(\theta_o)}(1+o_p(1))\le -C_1\right]\nonumber\\
=&Pr\left[\left|\xi^{(n)}_o(\theta_o)\right|\ge C_1\frac{\mathcal{M}_m(\theta_o)}{1+o_p(1)}\right]\le \exp\left(-\frac{1}{2}\left(\frac{C_1\mathcal{M}_m(\theta_o)}{\sigma_o}\right)^2\right).\label{eq:lbound}
\end{align}

Since $\xi^{(n)}_o(\theta_o)$ is normally distributed, $\mathcal{M}_m(\theta_o)=\sigma_o\sqrt{2\log m}$, diverges as $m\rightarrow\infty$.  So the R.H.S. of \eqref{eq:lbound} converges to zero.  That is, for any $C_1>0$, $Pr[l_m(\theta_o)\le -C_1]$ converges to zero.  Furthermore, by substituting the expression for $\mathcal{M}_m(\theta_o)$ in \eqref{eq:lbound} we get:
\begin{equation}\label{eq:lbound2}
Pr[l_m(\theta_o)\le -C_1]\le \exp(-C_1^2\log m)=\frac{1}{m^{C^2_1}}.
\end{equation}

Now as before by setting $p_n=m^{-C^2_1}$, we get $m=p_n^{-1/C^2_1}$.  In particular, if $p_n=n^{-\alpha}$, $m=n^{\alpha/C^2_1}$.  

\noindent{\it Details of Remark \ref{rem:test}.}
The likelihood ratio statistic for testing the null hypothesis of $\theta=\theta_o$ against the unrestricted alternative is given by:
\[
LR(\theta_o)=\frac{\exp(l_m(\theta_o))}{\max_{w\in\Delta_{m-1}}\exp(\sum^m_{i=1}\log(w_i)/m)}.
\]
Clearly, the maximum value the denominator attains is, $1/m$. So the log-likelihood ratio $\log LR(\theta_o)$ turns out to be $l_m(\theta_o)+\log m$.

  The test rejects $H_0$ if $\log LR(\theta_o)$ is smaller than $\log C_0$, for some pre-specified $C_0\in(0,1)$.  Ideally, $C_0$ should be a function of $m$.  However, at this point we assume $C_0$ to be fixed.

  Using \eqref{eq:m}, the probability of rejecting the null hypothesis is given by:
\begin{align}
~&Pr[\log m+l_m(\theta_o)\le \log C_0]=Pr[l_m(\theta_o)\le \log C_0-\log m]\nonumber\\
  =&Pr\left[-\frac{1}{\mathcal{M}_m(\theta_o)}\left|\xi^{(n)}_o(\theta_o)+o(1)\right|(1+o(1))\le \log\left(\frac{C_0}{m}\right)\right]\nonumber\\
  =&Pr\left[\left|\xi^{(n)}_o(\theta_o)+o(1)\right|(1+o(1))\ge -\mathcal{M}_m(\theta_o)\log\left(\frac{C_0}{m}\right)\right]\nonumber
\end{align}

Now Suppose that $\xi^{(n)}(\theta)$ is a $N(0,\sigma^2_0)$ random variable.  Using the tail bounds for a normal distribution, we get:

\begin{align}
  Pr&\left[\left|\xi^{(n)}_o(\theta_o)+o(1)\right|(1+o(1))\ge -\mathcal{M}_m(\theta_o)\log\left(\frac{C_0}{m}\right)\right]\nonumber\\
  &\le exp\left(-\frac{1}{2\sigma^2_o}\left\{\mathcal{M}_m(\theta_o)\log\left(\frac{C_0}{m}\right)\right\}^2\right)
\end{align}
By substituting $\mathcal{M}_m(\theta_o)=\sigma_o\sqrt{2\log m}$ in the exponent of the above expression we get:
\begin{align}
&\frac{1}{2\sigma^2_o}\left\{\mathcal{M}_m(\theta_o)\log\left(\frac{C_0}{m}\right)\right\}^2=(\log m)\left(\log C_0-\log m\right)^2\nonumber\\
=&(\log m)^3-2(\log m)^2\log C_0+(\log m)(\log C_0)^2.\nonumber 
\end{align}

  Clearly, the $(\log m)^3$ term dominates and the probability of rejecting the null hypothesis decreases at the rate of $\exp(-(\log m)^3)$.  This is true even if $C_0$ increases to one with increasing $m$ at a suitable rate.  

  Finally, in order to describe some relationship between $m$ and $n$, suppose we would like to ensure, that the probability of rejecting the null hypothesis reduces at the rate of $p_n$.
Then it follows that the number of replications required to ensure such a rate is of the order $m=\exp((-\log p_n)^{1/3})$.     

\noindent{\it Details of Remark \ref{rem:div}.}
Let us fix $\theta\ne\theta_o$ and suppose $\xi^{(n)}_o(\theta_o)$ follows a $N(0,\sigma^2_o)$ distribution.  Then for a fixed $C_2>0$, it can be shown that:

\begin{align}
Pr[l_m(\theta)\le -C_2]\le& Pr\left[\left|\xi^{(n)}_o(\theta_o)\right|\ge\mathcal{M}_m(\theta)\left\{C_2-\frac{b_n}{\mathcal{M}_m(\theta)}\left|\sm(\theta)-\sm(\theta_o)\right|\right\}\right]\nonumber\\
\le&\exp\left[-\frac{(\mathcal{M}_m(\theta))^2}{2\sigma^2_o}\left\{C_2-\frac{b_n}{\mathcal{M}_m(\theta)}\left|\sm(\theta)-\sm(\theta_o)\right|\right\}^2\right]\nonumber\\
\end{align}
Now by substituting $\mathcal{M}_m(\theta))=\sigma_o\sqrt{2\log m}$ we get:
\begin{equation}\label{eq:lbfs}
Pr[l_m(\theta)\le -C_2]\le \left(\frac{1}{m}\right)^{\left\{C_2-\frac{b_n}{\sigma_o\sqrt{2\log m}}\left|\sm(\theta)-\sm(\theta_o)\right|\right\}^2}.
\end{equation}
Now, if $\mathcal{M}_m(\theta)/b_n=\sigma_o\sqrt{2\log m}/b_n$ diverges with $m$ and $n$, clearly, for large values of $m$ and $n$, $Pr[l_m(\theta)\le -C_2]\approx m^{-C^2_2}$.  That is, for any fixed $C_2>0$ and $\theta\ne\theta_o$,  $l_m(\theta)\ge -C_2$ with a high probability, 
and $\exp(l_m(\theta))$ does not collapse to zero with a high probability. 

Furthermore, for a fixed $n$, R.H.S. of \eqref{eq:lbfs} is a decreasing function in $m$.  That is if the sample size is kept fixed, increasing the number of replications will increase the probability of $l_m(\theta)\ge -C_2$.  As a result, the log-likelihood will be flatter in shape.  
Note that, from \eqref{eq:ms}, it is clear that the variance of the expected log likelihood gets reduced as $m$ increases. This explains a bias-variance trade-off in the choice of $m$.  Such phenomenon is evident from Figure \ref{fig:post}, where the curve joining the means of the 
proposed estimated log posterior progressively flattens with the number of replications. The argument above provides a formal explanation of the phenomenon. 

\section{Differential Entropy Estimation}
Several estimators of differential entropy have been studied in the literature.  The oracle estimator is given by $-\sum^m_{i=1}\log f_0(\s{X_i(\theta)})/m$.  In this article we implement a weighted k-nearest neighbour based Kozachenko-Leonenko estimator \citep{kozLeo87,tsybakovMeulen1996} described in \citet{berrettSamworthMing2019}.

In order to define the estimator, let $||\cdot||$ denote the Euclidean norm on $\mathbb{R}^r$ and we fix an integer $k$ in $\{1,2,\ldots,m-1\}$.  We shall use the replicated summaries $\s{X_i(\theta)}$ to estimate the entropy.  Since the parameter value is fixed, in what follows we drop its explicit mention in the notation.

In the language of \citet{berrettSamworthMing2019}, for each $i=1$, $2$, $\ldots$, $m$, let $\s{X_{(1),i}}$, $\s{X_{(2),i}}$, $\ldots$ $\s{X_{(m-1),i}}$ be a
 permutation of $\{\s{X_1},\s{X_2},\ldots,\s{X_m} \}\setminus\{\s{X_i}\}$ such that $||\s{X_{(1),i}}-\s{X_i}||\le||\s{X_{(2),i}}-\s{X_i}||$ $\le\cdots\le ||\s{X_{(m-1),i}}-\s{X_i}||$.  Suppose we denote, $\rho_{(j),i}\coloneqq ||\s{X_{(j),i}}-\s{X_i}||$, that is  $\rho_{(j),i}$ is the $j$th nearest neighbour of $\s{X_i}$.  Furthermore, for the fixed $k$, define a set of weights $\nu=(\nu_1,\ldots,\nu_k)^T\in \mathbb{R}^k$ as
\begin{align}
\mathcal{V}^{(k)}\coloneqq&\left\{\nu\in\mathbb{R}^k~:~\sum^k_{j=1}\nu_j\frac{\Gamma(j+2l/r)}{\Gamma(j)}=0\text{~for~$l=1$, $\ldots$, $\lfloor r/4\rfloor$,}\right.\nonumber\\
&\left. \sum^k_{j=1}\nu_j=1\text{ and $\nu_j=0$ if $j\not\in\{\lfloor k/r \rfloor,\lfloor 2k/r\rfloor,\ldots,k\}$}\right\}. 
\end{align}   

For a weight vector $\nu\in\mathcal{V}^{(k)}$, \citet{berrettSamworthMing2019} define the weighted Kozachenko-Leonenko estimator of $\htr_{\sa\mid \theta}(\theta)$ as
\begin{equation}\label{eq:hEst}
\hat{H}^0_{\sa\mid \theta}(\theta)=\frac{1}{m}\sum^m_{i=1}\sum^k_{j=1}\nu_j\log\left(\frac{(m-1)\pi^{r/2}\rho^r_{(j),i}}{e^{-\psi(j)}\Gamma(1+r/2)}\right),
\end{equation}
where $\psi$ is the digamma function.

In order to find one entry in $\mathcal{V}^{(k)}$, we solve: 
\begin{equation}\label{eq:euLik}
\hat{\nu}=\arg\min_{\nu\in\mathcal{V}^{(k)}}\sum^k_{j=1}(k\nu_j-1)^2.
\end{equation}
The objective function in \eqref{eq:euLik} is the so called \emph{Euclidean likelihood} (see \citep{owen01}) which has been previously studied by \citet{brown_chen_1998}.

From \citet{berrettSamworthMing2019} it follows that the normalized risk of the proposed estimator converges in a uniform sense to that of the unbiased oracle estimator. Other histogram or kernel-based estimators \citep{hallMorton1993,paninskiYazima2008} can be considered.  Due to the curse of dimensionality, they don't perform well in high dimensions.  
They are also potentially computationally expensive.  

If the summary statistics are approximately normally distributed, it is often sufficient and computationally more efficient to directly use the  expression of differential entropy for a normal random vector, which depends only on the determinant of the covariance matrix.   

Finally, the proposed estimator in \eqref{eq:hEst} only requires a user to specify $k$, i.e. the order of the nearest neighbour.  Ideally, $k$ should depend on $m$.  Our experience suggests any value of $k$ as long as it is not very small or not very large makes little difference.  

Note that, other than the summary statistics and the number of replications $m$ to be generated from the process for each value of $\theta$, this $k$ is the only parameter a user needs to specify in order to compute the proposed posterior.  No other parameters tuning or otherwise are required.

\section{Posterior Sampling}\label{Ssec:impl2}
Similar to all BayesEL procedures the proposed approximate posterior cannot be expressed in a closed form. Thus, one needs to devise Markov Chain Monte Carlo procedures to draw samples from the posterior which can be used to make further inferences.  
Furthermore, it is well known that the support of the posterior may be non-convex \citep{chaudhuri+my17}, which requires a judicious choice of the sampler.  
In the examples below, we use an adaptive Metropolis-Hastings random walk method with a normal proposal 
for the MCMC sampling due to \citet{haarioEtAl2001}. More sophisticated methods could be designed.
The empirical likelihood can generally be computed very fast using the R package {\tt emplik} \citep{zhou+y16}.

Similar to the synthetic likelihood approach \citep{price+dln16}, the sampling procedure is related to the pseudo-marginal Metropolis-Hastings methods \citep{beaumont03,andrieu+r09,doucet+pdk15} in the sense that we sample from a noisy likelihood estimate.  It is well known that such samplers mix slowly if the variance of the likelihood estimate is large.  
For the proposed posterior this can partly be controlled by choosing $m$ appropriately.

\section{Illustrative Examples}


\begin{figure}[t]
  \begin{center}
    \begin{subfigure}{.45\columnwidth}
      \resizebox{2.25in}{2.25in}{\rotatebox{90}{\includegraphics{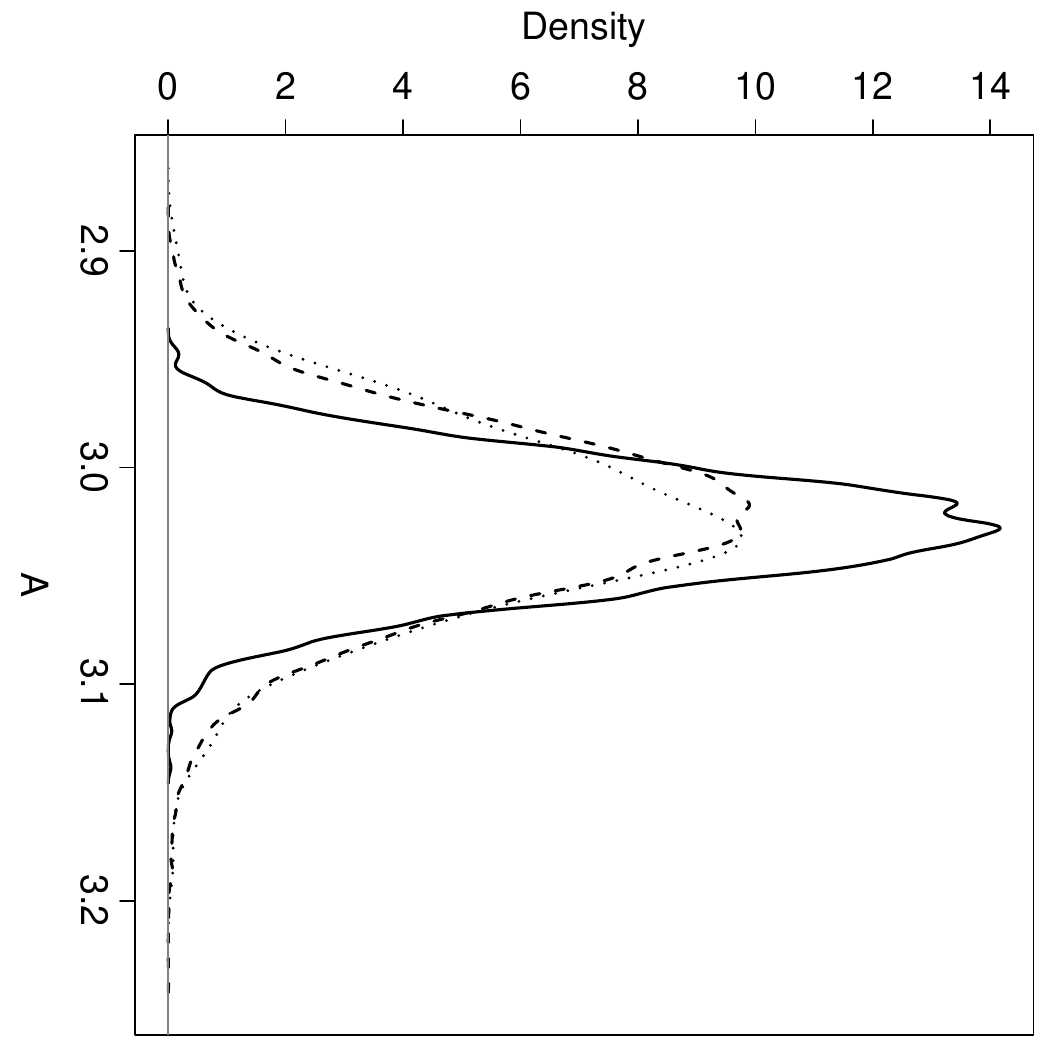}}}
      \end{subfigure}~\begin{subfigure}{.45\columnwidth}
    \resizebox{2.25in}{2.25in}{\rotatebox{90}{\includegraphics{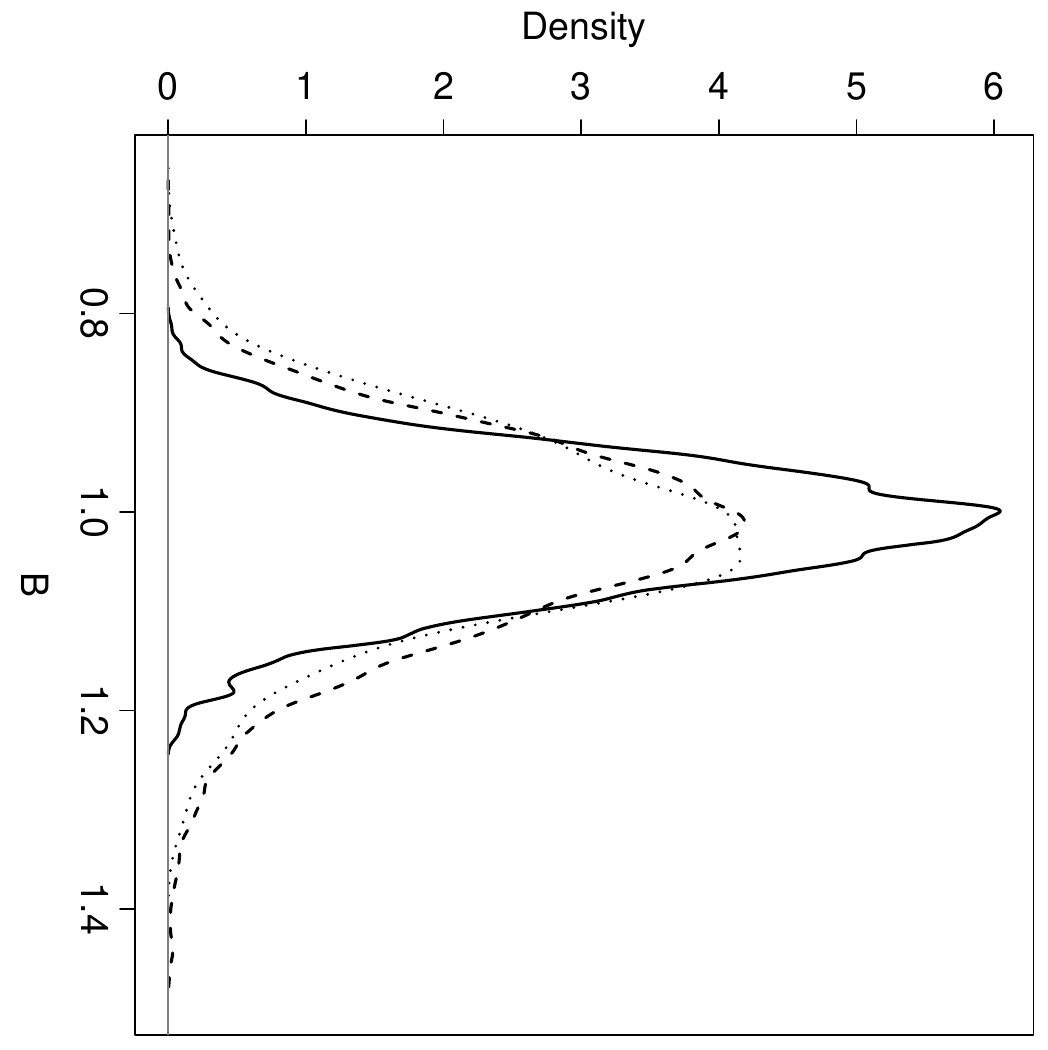}}}
    \end{subfigure}\\
    \begin{subfigure}{.45\columnwidth}
      \resizebox{2.25in}{2.25in}{\rotatebox{90}{\includegraphics{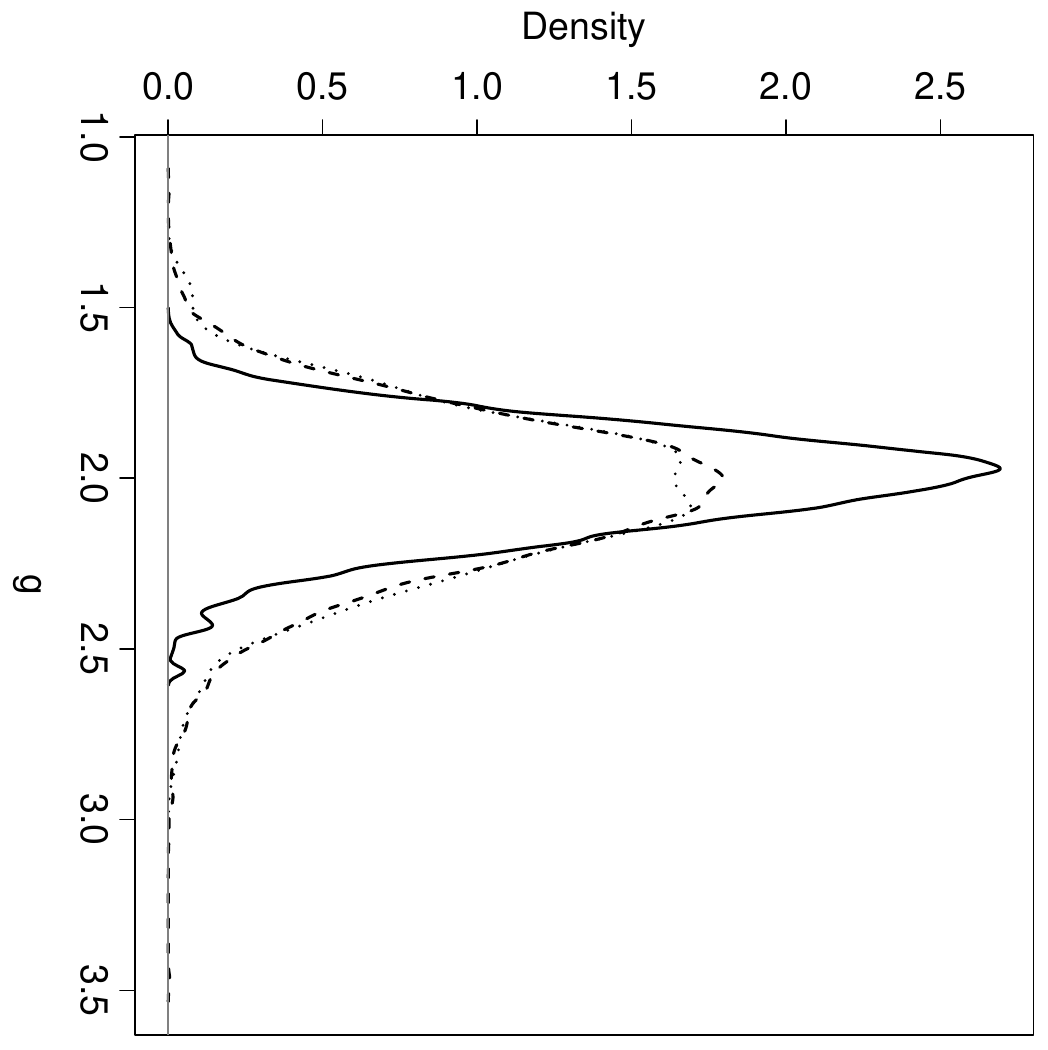}}}
      \end{subfigure}~\begin{subfigure}{.45\columnwidth}
    \resizebox{2.25in}{2.25in}{\includegraphics{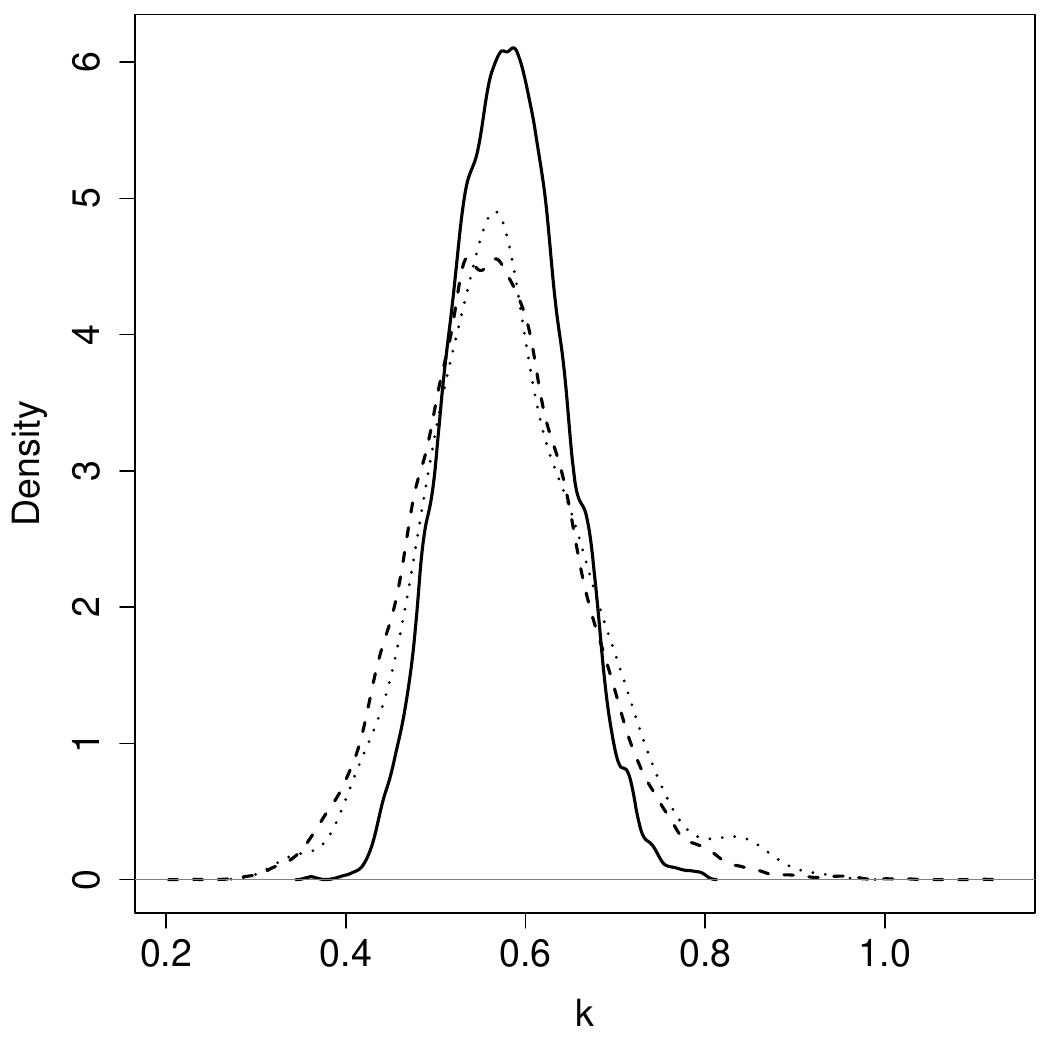}}
    \end{subfigure}
\caption{Estimated marginal posterior densities by proposed method (solid), synthetic likelihood (dashed) and regression ABC (dotted) for parameters of the $g$-and-$k$ model.}
\label{F2}
\end{center}
\end{figure}

\subsection{$g$-and-$k$ distribution}
Our second example concerns inference for the $g$-and-$k$ distribution \citep{haynes1997robustness,peters+s06,allingham2009bayesian}.  The distribution is defined through its quantile function: 
\begin{align}
~&Q(p;A,B,g,k) =A+B \left[ 1+c\times\frac{1-\exp\left\{-gz(p)\right\}}{1+\exp\left\{-gz(p)\right\}} \right] \left\{ 1+{z(p)}^2 \right\} ^kz(p),\nonumber
\end{align} 
where $z(p)$ is the $p$th standard normal quantile and conventionally $c$ is fixed at $0.8$, which results in the constraint $k>-0.5$.  Components of the parameter vector $\theta=(A,B,g,k)$ are respectively related to location, scale, skewness, and kurtosis of the distribution.  There is no closed-form expression for the density function,
however, data can be simulated from this model by transforming uniform random variables on $[0,1]$ by the quantile function.  These features make the g-and-k distribution popular in ABC literature. 


A data set of size $n=1000$ was simulated from the distribution with $(A,B,g,k)=(3,1,2,0.5)$.  A uniform prior $U(0,10)^4$ for $\theta$ was assumed.  We approximate the proposed empirical likelihood and the synthetic
likelihood using $m=40$ data sets each of length $n$ for each value of $\theta$.  The mean and the three quartiles were used as summary statistics.  
Compared to the octile-based summaries used in \citet{drovandi2011likelihood}, these summaries lead to a slightly better estimate for the parameter $k$.
Posterior summaries are based on $100,000$ sampling
iterations after $100,000$ iterations burn in.  
 
The results are presented in Figure~\ref{F2}.  Estimated marginal posterior densities obtained from the synthetic likelihood and proposed empirical likelihood are shown as dashed and solid lines respectively. Also shown is 
an answer based on rejection ABC with small tolerance and linear
regression adjustment \citep{beaumont+zb02}.  For the ABC approach, to improve computational efficiency, we restricted the prior for $\theta$ from $U(0,10)^4$ to $U(2,4)\times U(0,2)\times U(0,4)\times U(0,1)$.
This restricted prior is broad enough to contain the support of the posterior based on the original prior.  The ABC estimated marginal posterior densities (dotted) shown in Figure~\ref{F2} were based on $5,000,000$ samples, choosing the tolerance so that $2000$ samples are kept.
The summary statistics used here are asymptotically normal and $n$ is large, so the synthetic likelihood is expected to work well in this example, which it does. Our proposed method gives comparable results to synthetic likelihood and
the rejection ABC analysis, although there does seem to be some slight underestimation of posterior uncertainty in the empirical likelihood method. 



\subsection{Estimation of Edge Probability of an Erd\"{o}s-Renyi Random Graph}\label{Ssec:graph}

\begin{figure}[ht]
\begin{center}
\begin{subfigure}{.45\columnwidth}
\resizebox{2.25in}{2.25in}{\includegraphics{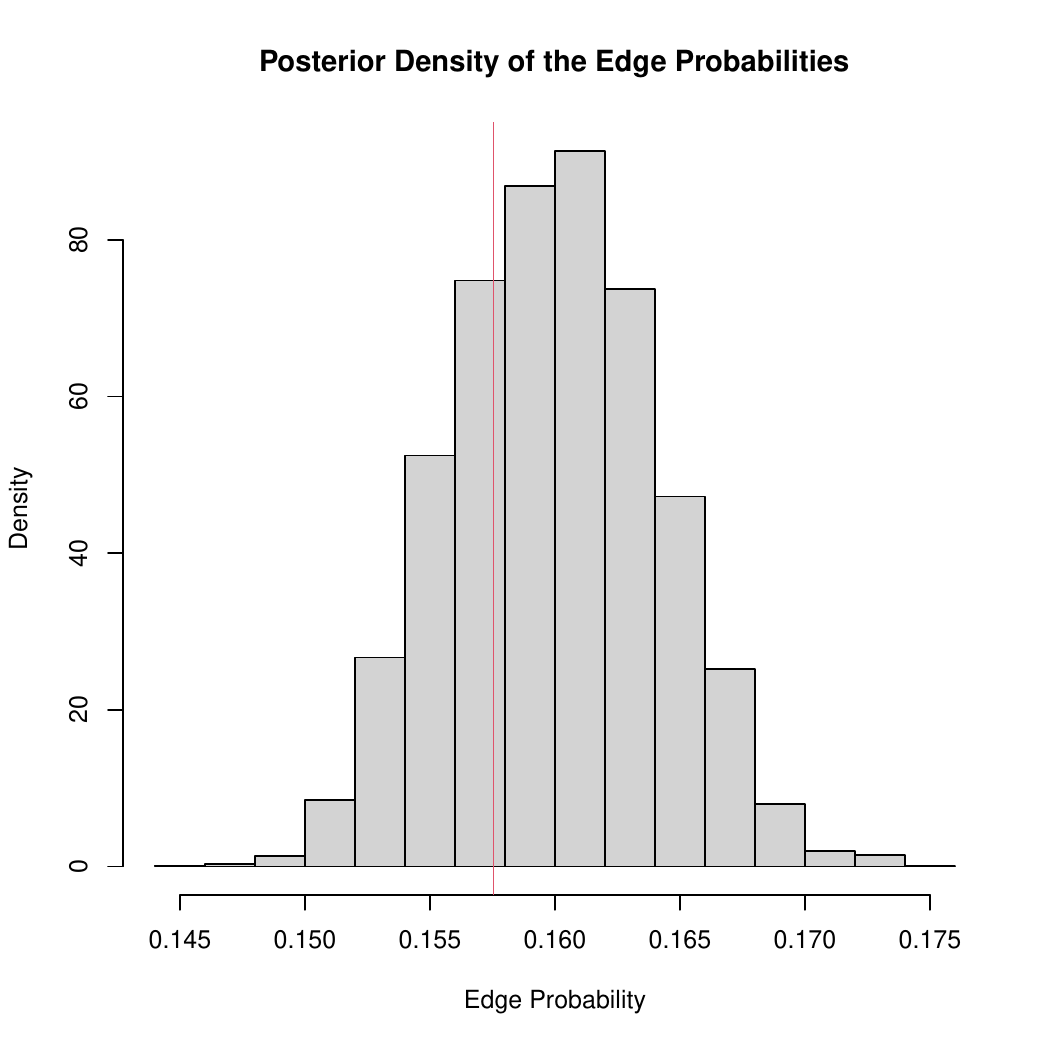}}
\caption{\label{fig:graphA}}
\end{subfigure}
\begin{subfigure}{.45\columnwidth}
\quad\resizebox{2.25in}{2.25in}{\rotatebox{90}{\includegraphics{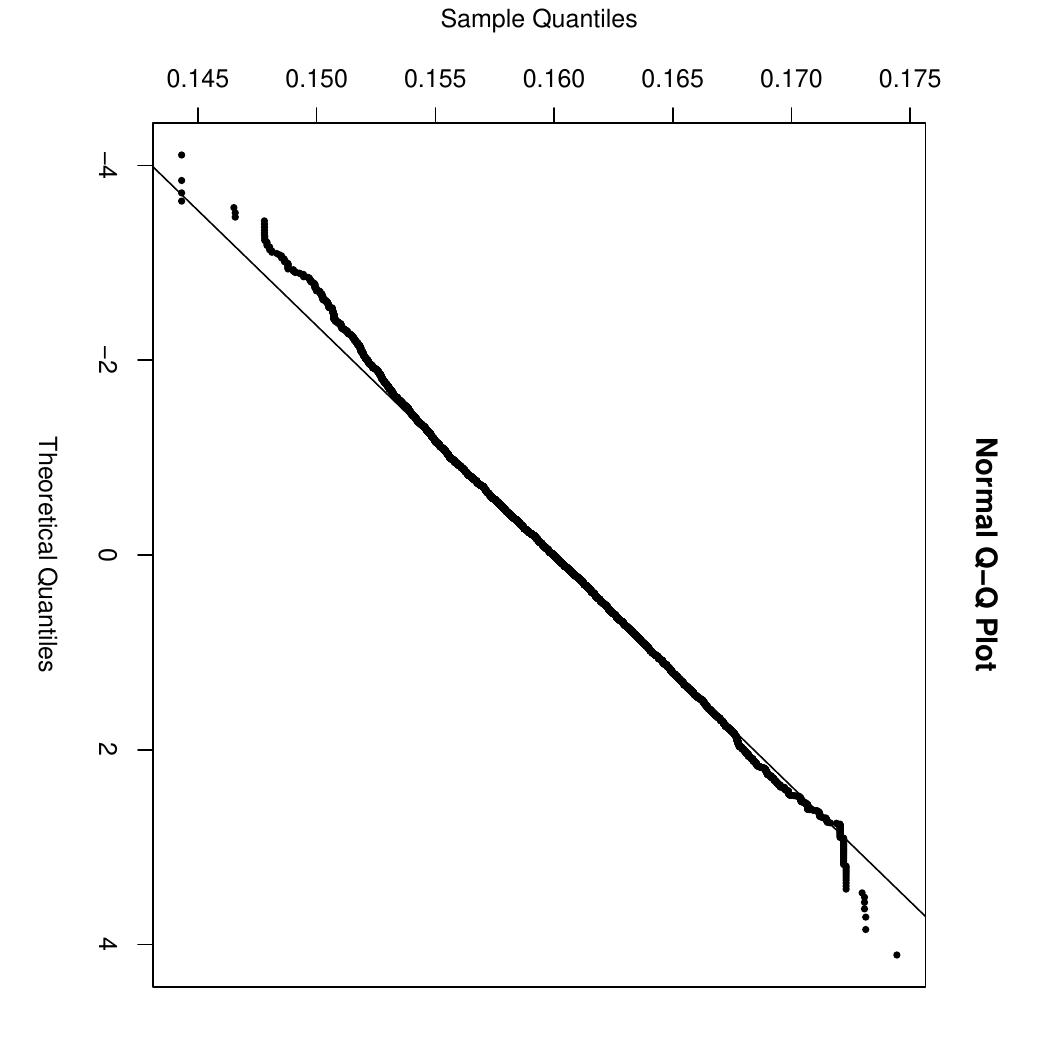}}}
\caption{\label{fig:graphB}}
\end{subfigure}
\end{center}
\caption{The histogram (\ref{fig:graphA}) of the samples drawn from the proposed posterior of the edge probability of an Erd\"{o}s-Renyi graph. The graph had $n=100$ nodes and we use $m=25$ replications.  The Q-Q plot of the sample against normal distribution is presented in Figure \ref{fig:graphB}.}
\label{fig:graph}
\end{figure}

In this example, we estimate the edge probability of an observed Erd\"{o}s-Renyi random graph with $n$ vertices.  Suppose $p$ is the probability of an edge between any two vertices.  We assume that $p$ has a $\text{Beta}(1.5,1.5)$  distribution.
The observed graph had $n=100$ nodes, and the number of edges and the number of triangles were used as two estimating equations.  The posterior was computed using $m=25$ replications.  

The above experiment was repeated $100$ times and the observed coverage of the $95$\% confidence intervals were about $89$\%.  
A typical example of the sampled posterior distribution is presented in Figure \ref{fig:graph}.
In Figure \ref{fig:graphA} the histogram of observations sampled from the posterior in a typical repeat is presented. 
The true value of the edge probability ie. $p_o$ is presented by the vertical red straight line.  From the Q-Q plot (in Figure \ref{fig:graphB}) with the normal distribution, it seems of the posterior has a slightly lighter tail.

The proposed methodology allows an alternative way to estimate the model parameters in a random graph model by avoiding pitfalls of model degeneracies experienced in Exponential Random Graph models (see. e.g. \citet{fellows_handcock_2017}).  The example easily generalizes to more complex models with covariate-dependent node-specific edge probabilities.

      
\section{Q-Q Plots of the Posteriors with Normal Density:}\label{Ssec:qq}
In this section we present the Q-Q plots of the posteriors obtained from the proposed empirical likelihood based methods with normal density and compare those plots with the same plots obtained form the synthetic likelihood and rejection ABC methods.

\begin{figure}[h]
\begin{center}
\resizebox{5in}{5in}{\includegraphics{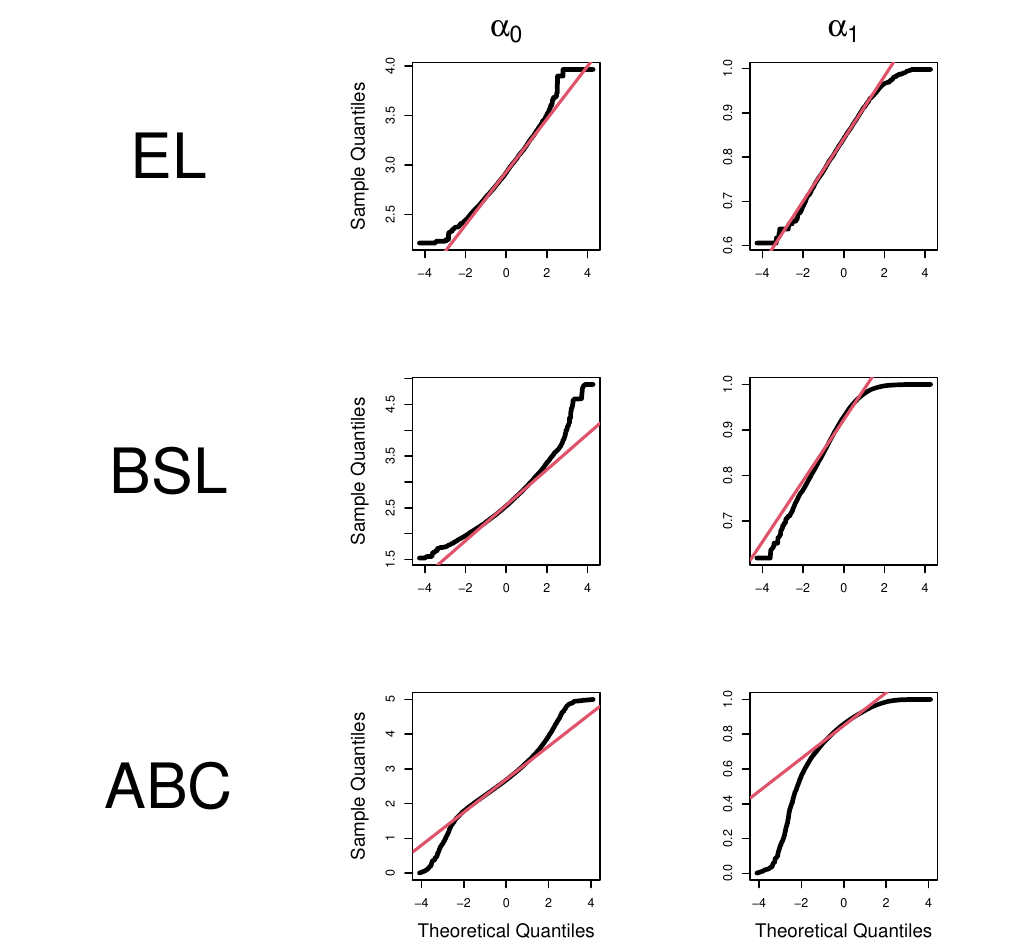}}
\end{center}
\caption{Q-Q plot of the EL, BSL and ABC posteriors for the ARCH(1) model with normal density.}
\label{fig:archqq}
\end{figure}

In each example, for the chosen summaries the proposed modified empirical likelihood based 
\begin{landscape}
\begin{figure}[t]
\begin{center}
\resizebox{9in}{5in}{\includegraphics{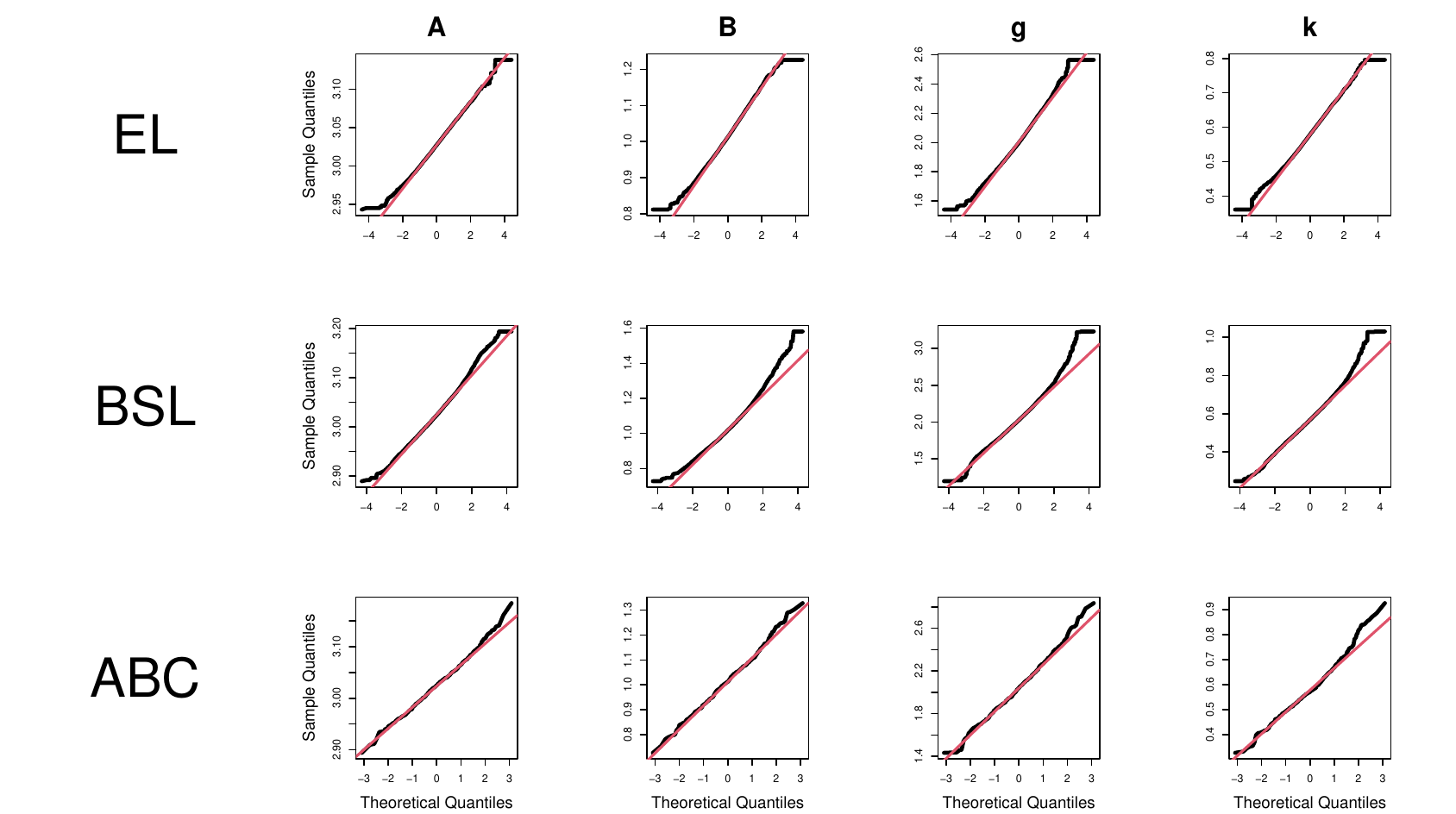}}
\end{center}
\caption{Q-Q plot of the EL, BSL and ABC posteriors for the g-and-k distribution with normal density.}
\label{fig:gkqq}
\end{figure}
\end{landscape}

\begin{landscape}
\begin{figure}[t]
\begin{center}
\resizebox{9in}{5in}{\includegraphics{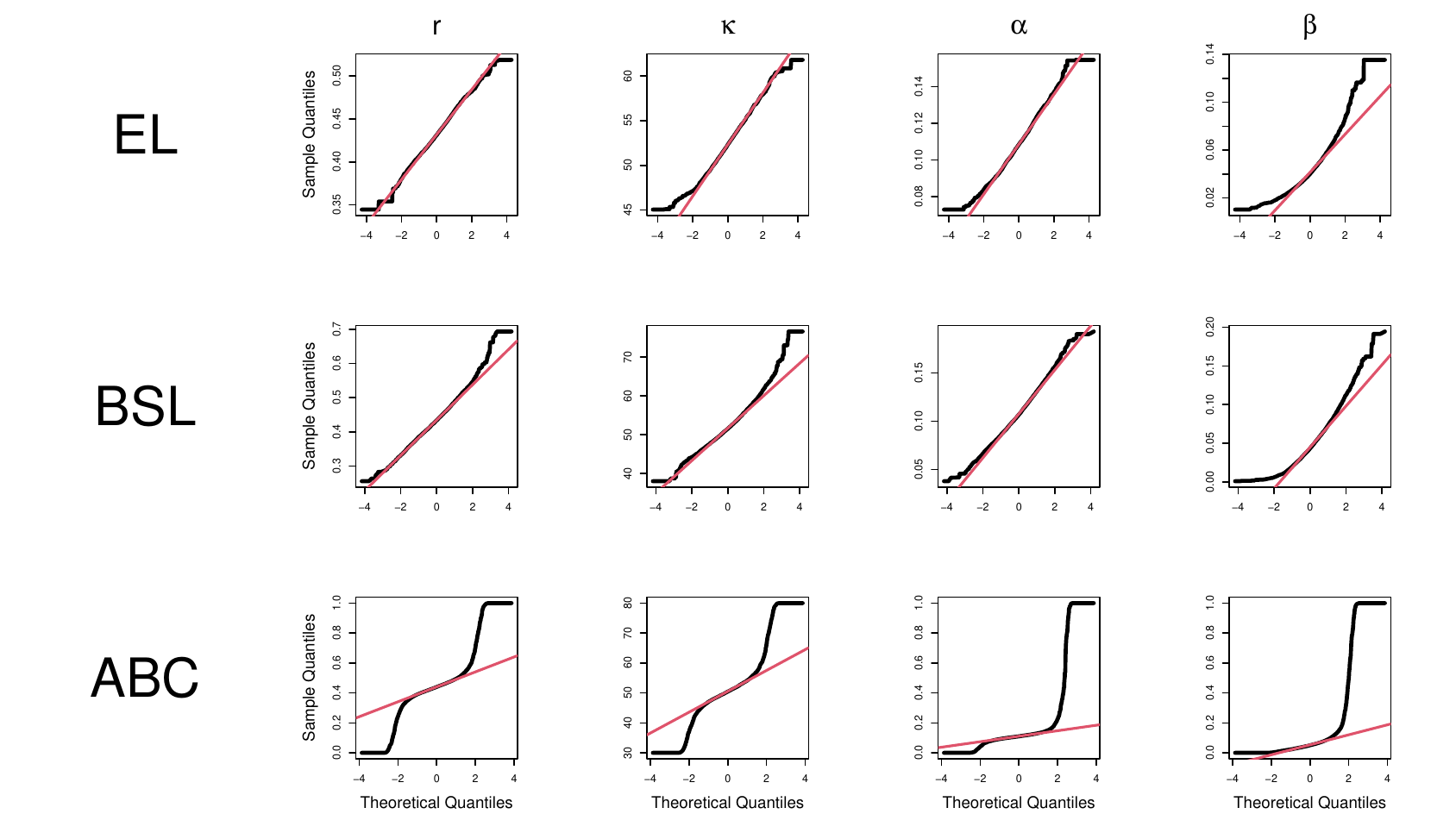}}
\end{center}
\caption{Q-Q plot of the EL, BSL and ABC posteriors for the simple recruitment, boom and bust model with normal density.}
\label{fig:BBqq}
\end{figure}
\end{landscape}

\begin{figure}[h]
\begin{center}
\resizebox{6.5in}{5in}{\includegraphics{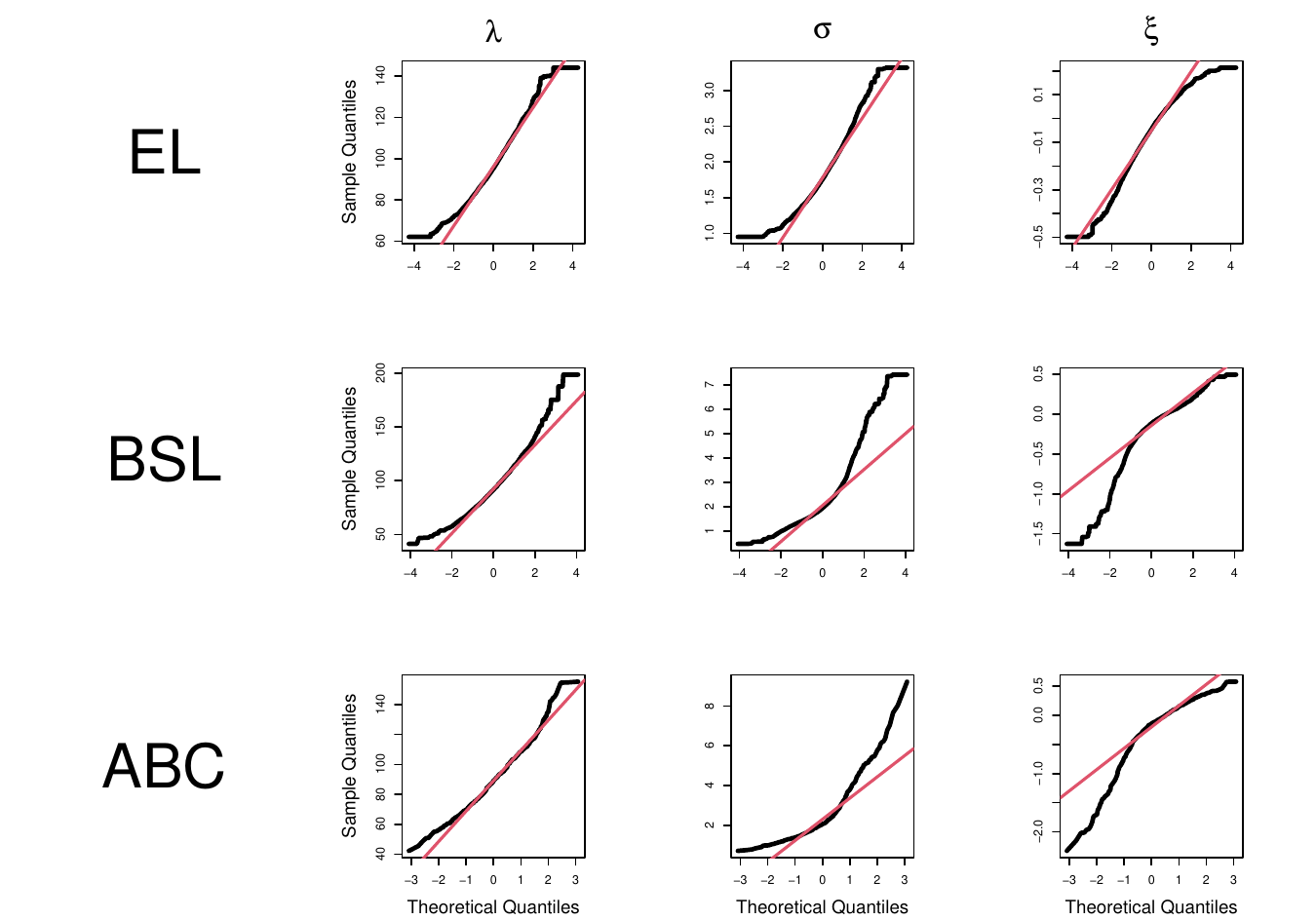}}
\end{center}
\caption{Q-Q plot of the EL, BSL and ABC posteriors for the stereological data with normal density.}
\label{fig:inclqq}
\end{figure}

based method produces posteriors with shapes quite close to a normal density.  This is mostly true even when the summary statistics is not asymptotically normally distributed.

Between Bayesian Synthetic likelihood (BSL) and the rejection ABC (ABC), the former seems to produce posteriors with shapes resembling a normal density than the latter.  For g-and-k (Figure \ref{fig:gkqq}) and simple recruitment, boom and bust model (Figure \ref{fig:BBqq}) the normality of the posterior mostly holds.  For Arch(1) model (Figure \ref{fig:archqq}) the density of $\alpha_1$ is skewed. This happens because of the non-normality of the summary statistics used.  The deviation from normality is evident in stereological data example as well (Figure \ref{fig:inclqq}).   

Other than the g-and-k example in figure \ref{fig:gkqq} in all cases the rejection ABC posteriors show large difference in shape form the normal density.  Such deviations are expected from the results of \citet{frazier+mrr18}.

Recently, \citet{frazierNottDrovandiKohn2022} have proved Bernstein-von-Mises theorem for Bayesian synthetic likelihood posteriors.  Even though similar results are not yet available for the proposed empirical likelihood based method,  The Q-Q plots presented above strongly suggest that under usual regularity conditions the proposed posteriors would asymptotically converge to a normal density.